\documentclass[11pt,letterpaper]{article}

\usepackage[style=alphabetic,natbib=true,maxnames=99,maxalphanames=99,isbn=false,url=false,backref=true,backend=biber,giveninits=false]{biblatex}

\AtEveryBibitem{%
  \ifentrytype{article}{%
    \clearlist{publisher}
    \clearfield{editor}
  }{}
  \ifentrytype{book}{
  }{}
  \ifentrytype{inproceedings}{%
    \clearfield{volume}
    \clearfield{number}
    \clearfield{editor}
    \clearlist{publisher}
    \clearlist{location}
  }{}
  \clearfield{day}
  \clearfield{month}
  \clearfield{isbn}
  \clearfield{url}
}
\usepackage{amsmath,amsthm,thmtools,amsfonts,amssymb}
\usepackage{mathrsfs}
\usepackage{mathtools}
\usepackage{physics}
\usepackage{braket}
\usepackage{array}
\usepackage{booktabs}
\allowdisplaybreaks

\usepackage[margin=1in]{geometry}
\usepackage{xspace}
\usepackage{xcolor}
\usepackage[normalem]{ulem}
\usepackage{subcaption}
\usepackage{enumitem}
\usepackage{microtype}
\usepackage{csquotes}
\usepackage{dsfont}

\usepackage{algorithm}
\usepackage{algpseudocode}

\usepackage{hyperref}
\hypersetup{
  colorlinks=true,
  linkcolor=red,
  citecolor=red,
  filecolor=red,
  urlcolor=red,
}
\usepackage[capitalize]{cleveref}
\AtBeginDocument{%
}
\crefname{step}{step}{steps}
\Crefname{step}{Step}{Steps}
\Crefname{equation}{Eq.}{Eqs.}

\usepackage[skins,many]{tcolorbox}

\makeatletter
\newcounter{HALG@line}
\renewcommand{\theHALG@line}{\thealgorithm.\arabic{ALG@line}}
\makeatother
\newtheorem{theorem}{Theorem}[section]
\newtheorem{lemma}[theorem]{Lemma}
\newtheorem{claim}[theorem]{Claim}
\newtheorem{corollary}[theorem]{Corollary}

\theoremstyle{definition}
\newtheorem{definition}[theorem]{Definition}
\newtheorem{example}[theorem]{Example}
\newtheorem{remark}[theorem]{Remark}

\renewcommand{\theHtheorem}{\thesection.\arabic{theorem}}
\makeatletter
\@for\hmaj@env:=lemma,claim,corollary,proposition,conjecture,definition,example,remark\do{%
  \expandafter\def\csname theH\hmaj@env\endcsname{\theHtheorem}%
}
\makeatother
\newcommand{\calA}{\mathcal{A}}
\newcommand{\calB}{\mathcal{B}}
\newcommand{\calC}{\mathcal{C}}
\newcommand{\calD}{\mathcal{D}}
\newcommand{\calE}{\mathcal{E}}
\newcommand{\calF}{\mathcal{F}}
\newcommand{\calG}{\mathcal{G}}

\newcommand{\calS}{\mathcal{S}}
\newcommand{\calT}{\mathcal{T}}

\newcommand{\calW}{\mathcal{W}}
\newcommand{\calX}{\mathcal{X}}

\newcommand{\bbE}{\mathbb{E}}

\newcommand{\bbN}{\mathbb{N}}

\newcommand{\bbR}{\mathbb{R}}

\newcommand{\bbZ}{\mathbb{Z}}

\newcommand{\bfa}{\mathbf{a}}
\newcommand{\bfb}{\mathbf{b}}

\newcommand{\bfe}{\mathbf{e}}

\newcommand{\bfp}{\mathbf{p}}
\newcommand{\bfq}{\mathbf{q}}
\newcommand{\bfr}{\mathbf{r}}

\newcommand{\bfx}{\mathbf{x}}
\newcommand{\bfy}{\mathbf{y}}
\newcommand{\bfz}{\mathbf{z}}

\newcommand{\bfC}{\mathbf{C}}

\newcommand{\bfU}{\mathbf{U}}
\newcommand{\bfV}{\mathbf{V}}

\newcommand{\bfX}{\mathbf{X}}
\newcommand{\bfY}{\mathbf{Y}}
\newcommand{\bfZ}{\mathbf{Z}}

\DeclarePairedDelimiterXPP{\expect}[1]{\bbE}{[}{]}{}{#1}
\DeclarePairedDelimiterXPP{\pr}[1]{\Pr}{[}{]}{}{#1}
\DeclarePairedDelimiterXPP{\pron}[2]{\Pr_{#1}}{[}{]}{}{#2}
\DeclarePairedDelimiterXPP{\expecton}[2]{\bbE_{#1}}{[}{]}{}{#2}

\DeclarePairedDelimiterXPP{\variance}[1]{\mathrm{Var}}{[}{]}{}{#1}
\DeclarePairedDelimiterXPP{\varianceon}[2]{\mathrm{Var}_{#1}}{[}{]}{}{#2}
\DeclarePairedDelimiterXPP{\covariance}[1]{\mathrm{Cov}}{[}{]}{}{#1}

\DeclareMathOperator{\Multinom}{Multinomial}
\DeclareMathOperator{\Gammadistr}{Gamma}

\newcommand{\ind}[1]{\mathds{1}_{\{#1\}}}

\newcommand{\hmaj}{$h$-\textsf{Majority}\xspace} %
\newcommand{\threemaj}{$3$-\textsf{Majority}\xspace} %
\newcommand{\twochoices}{$2$-\textsf{Choices}\xspace} %
\newcommand{\undstate}{\textsf{UndecidedState}\xspace} %
\newcommand{\dejavu}{\textsf{DéjàVu}\xspace} %
\newcommand{\voter}{\textsf{Voter}\xspace} %
\newcommand{\minority}{\textsf{Minority}\xspace}

\newcommand{\pull}{\textsf{PULL}\xspace} %

\newcommand{\pullmodel}[1]{\ensuremath{\pull(#1)}\xspace} %

\newcommand{\pa}[1]{\left( #1 \right)}
\DeclareMathOperator{\maj}{\textsc{Maj}}

\newcommand{\pmax}{p_{\max}}

\newcommand{\tconsensus}{\tau_{\mathsf{consensus}}}

\newcommand{\pignore}{P_{\mathrm{ignore}}}
\usepackage[style=alphabetic,natbib=true,maxnames=99,maxalphanames=99,isbn=false,url=false,backref=true,backend=biber,giveninits=false]{biblatex}

\AtEveryBibitem{%
  \ifentrytype{article}{%
    \clearlist{publisher}
    \clearfield{editor}
  }{}
  \ifentrytype{book}{
  }{}
  \ifentrytype{inproceedings}{%
    \clearfield{volume}
    \clearfield{number}
    \clearfield{editor}
    \clearlist{publisher}
    \clearlist{location}
  }{}
  \clearfield{day}
  \clearfield{month}
  \clearfield{isbn}
  \clearfield{url}
}
\begin{document}

\begin{flushleft}
  \huge\bf
  Nearly-Optimal Time-Sample Trade-off in Consensus Dynamics
\end{flushleft}
\smallskip

\newcommand{\myemail}[1]{\,$\cdot$\, {\small #1}}
\newcommand{\myaff}[1]{\,$\cdot$\, {\small #1}\par\medskip}
\newenvironment{myabstract}{\list{}{\listparindent 1.5em \itemindent \listparindent \leftmargin 0cm \rightmargin 0cm \parsep 0pt}
\item\relax}{\endlist}
\newenvironment{mycover}{\list{}{\listparindent 0pt \itemindent \listparindent \leftmargin 0cm \rightmargin 1.5cm \parsep 0pt} \raggedright
\item\relax}{\endlist}

\begin{mycover}

  \textbf{Francesco d'Amore}
  \myaff{Gran Sasso Science Institute, Italy}

  \textbf{Joel Rybicki}
  \myaff{Humboldt-Universität zu Berlin, Germany}

  \textbf{Nobutaka Shimizu}
  \myaff{Institute of Science Tokyo, Japan}

  \textbf{Takeharu Shiraga}
  \myaff{Chuo University, Japan}

  \textbf{Robin Vacus}
  \myaff{Sorbonne Université, CNRS, LIP6, France}

  \bigskip
\end{mycover}

\begin{myabstract}
  \noindent\textbf{Abstract.}
How does the number of sampled opinions affect the time required to reach consensus in distributed consensus dynamics? We study the synchronous $h$-\textsf{Majority}, in which every node independently samples $h$ nodes uniformly at random with replacement in each round and adopts a most frequent sampled opinion, breaking ties uniformly at random.
We show that $h$-\textsf{Majority} achieves the nearly-optimal trade-off between the sample size and the consensus time, among a broad class of consensus dynamics.
In particular, we prove the following results.
\begin{itemize}
  \item For every sample size $h \ge 3$ and for every initial configuration, we prove that the process reaches consensus with high probability within $\widetilde{O}\left(1+\min\left\{\frac{1}{\min\{h^2p_{\max},1\}},\frac{\sqrt{n}}{h}\right\}\right)$ rounds, where $n$ is the number of nodes and $p_{\max}$ is the largest initial fraction of nodes supporting any single opinion. This is the first bound on the consensus time for $h$-\textsf{Majority} that holds for every $h\ge3$ and every initial configuration.
  \item When the initially most popular opinion has a sufficiently large lead over every other opinion, we show that all nodes adopt this opinion with high probability within $O(\log n/\min\{h^2p_{\max},1\})$ rounds.
  \item We establish a matching lower bound up to polylogarithmic factors for any fixed update rule that treats all opinions symmetrically and that selects either a sampled opinion or the node's current opinion.
\end{itemize}

 \end{myabstract}

\thispagestyle{empty}
\setcounter{page}{0}
\newpage

\setcounter{tocdepth}{2}
\tableofcontents

\thispagestyle{empty}
\hypersetup{pageanchor=false}
\setcounter{page}{0}
\newpage
\hypersetup{pageanchor=true}
\section{Introduction}

Understanding the computational power of simple local decision rules is key to designing lightweight distributed algorithms and analyzing models of collective behavior \cite{HP01,BCN20}.
\emph{Consensus dynamics}, in which nodes repeatedly update their opinions based on local observations until reaching agreement, provide a basic yet fundamental setting for studying this power.
Such dynamics are attractive in their simplicity: nodes typically retain no state between rounds other than their current opinion, but they can still often reach consensus remarkably fast~\cite{HP01,nearly_tight_analysis,SS25,Doerr11,simple_dynamics,Archivio26,stabilizing_consensus,ignore_or_comply,async_3-majority_soda25,h-Majority-DISC25,DDGGN26,undecided_MFCS}.%

In this setting, the natural computational resource is the amount of information nodes receive about the current state of the system: the more information nodes have, the faster consensus can be reached.
However, for many dynamics, it is still poorly understood how much additional information can speed up the consensus time.
Motivated by this, we address the general question about the trade-off between the total amount of information exchanged and the time to reach consensus:
\begin{quote}
  \begin{center}
    \emph{How much can additional information per update reduce the time to reach consensus?}
  \end{center}
\end{quote}

\paragraph{Our setting.}
To investigate this question, we consider the synchronous \pullmodel{h} communication model~\cite{BKN17} on a fully connected network of~$n$ nodes. In each round, every node in parallel
\begin{enumerate}[label=(\roman*),noitemsep]
  \item samples $h$ nodes chosen independently and uniformly at random with replacement, and
  \item chooses their new opinion by applying a (randomized) rule to the sample obtained (plus, possibly, the node's current opinion).
\end{enumerate}
We further restrict attention to neutral rules that treat all opinion labels symmetrically, that is, there is no prior agreement on a total order (e.g., preferences) on the opinions.

So far, prior work has largely focused on analyzing the consensus time of specific dynamics, often under restrictions on the initial opinion configuration, such as the total number $k$ of initially present opinions or assuming that exactly one opinion is substantially more popular than others, or making specific assumptions about the sample size $h$~\cite{simple_dynamics,stabilizing_consensus,ignore_or_comply,nearly_tight_analysis,quasi-majority,h-Majority-DISC25,SS25,Archivio26,async_3-majority_soda25,DDGGN26,BecchettiCNPS15,fast_convergence_undecided,undecided-soda22,USD_PODC26,BecchettiCPTVZ24}.
In contrast, we impose no such restrictions on the initial configuration, the number $k$ of initial opinions, or the sample size $h$.

\paragraph{Our results in a nutshell.}
As our main result, we prove a new upper bound on the consensus time of the widely studied \hmaj~\cite{simple_dynamics,stabilizing_consensus,ignore_or_comply,nearly_tight_analysis,SS25,Archivio26,h-Majority-DISC25}.
In these dynamics, each node chooses its new opinion by adopting the most frequent opinion in the sample it obtained, breaking ties uniformly at random. %

We prove a new upper bound for any sample size $h\ge3$ and any number $k \ge 2$ of initial opinions that holds for \emph{every} initial configuration. In addition, we establish a general lower bound that holds for \emph{every} dynamics in a large class of neutral dynamics. This lower bound matches our upper bound up to polylogarithmic factors, and gives new lower bounds for other previously studied consensus dynamics~\cite{BecchettiCPTVZ24,DDGGN26}. Both the upper and lower bounds hold with high~probability.

\subsection{Prior work on \texorpdfstring{$h$}{h}-Majority}

Despite its simple definition, analyzing the behavior of \hmaj is often highly non-trivial.
In the special case when $h\in\{1,2\}$, \hmaj has the same transition probabilities as \voter, in which each node samples a single node uniformly at random and adopts its opinion.\footnote{For $h=2$, uniform tie-breaking yields the same transition probabilities as adopting the first sampled opinion, which is exactly the \voter rule.} The expected consensus time of \voter is $O(n)$ from any initial configuration, and can be $\Theta(n)$ on worst-case configurations~\cite{HP01,CEOR13}.

The case $h=3$ has been studied extensively, including under asynchronous updates or communication noise~\cite{simple_dynamics,stabilizing_consensus,ignore_or_comply,nearly_tight_analysis,async_3-majority_soda25,SS25,Archivio26,DZ25}.
Yet, even for the basic setting of synchronous updates, nearly optimal bounds on the consensus time for an arbitrary number $k \ge 2$ of initial opinions were obtained only very recently in 2025 by~\citet{SS25}, followed by a tight characterization in terms of the maximum initial opinion density by~\citet{Archivio26}.

For $h\ge4$, prior work only provides a complete picture for the case of binary opinions ($k=2$).
In particular, consensus is reached in at most $O(\log n/\log h)$ rounds for odd $h=o(n/\log n)$~\cite{quasi-majority}. Moreover,
\citet{hierarchy_Berenbrink} showed that the consensus time of \hmaj for any $h \geq 1$ stochastically dominates that of $(h+1)$-\textsf{Majority}, extending the result from \cite{quasi-majority} to even values of \(h\) as well.
In contrast, much less is known when many opinions are involved ($k\ge 3$), particularly when the number $h$ of samples is super-constant.
For any constant $\varepsilon>0$, if $k/h=O(n^{1/4-\varepsilon})$, then consensus time is at least $\Omega(k/h^2)$ rounds~\cite{simple_dynamics}.
For the case $h=\Omega(k\log n)$,
\citet{h-Majority-DISC25} showed that when the initial gap between the two most popular opinions is sufficiently large, then \emph{plurality consensus} is reached in $O(\log n)$ rounds. See \cref{tab:prior-work} for more details.

\subsection{Our results}

We now summarize our main contributions.
Let $2\le k\le n$ and for~$i \in \{1,\ldots,k\}$, let $p_i$ be the fraction of nodes that initially hold opinion $i$.
Without loss of generality, we assume $p_1\ge \dots \ge p_k$ and write $\pmax=p_1$.
We use $\widetilde{O}$ to hide factors polynomial in $\log n$, and say that an event holds \emph{with high probability} if its probability is at least $1-1/n^c$ for some constant $c>0$. Unless stated otherwise, the sample size is an arbitrary fixed integer $h\ge3$.

\begin{table}[htbp]
  \centering
  \captionsetup{font=small,skip=6pt}
  \caption{Prior work on consensus-time bounds for $h$-Majority.}
  \label{tab:prior-work}
  \renewcommand{\arraystretch}{1.2}
  \setlength{\tabcolsep}{4pt}

  \resizebox{\textwidth}{!}{%
    \begin{tabular}{@{}cccc|cccc@{}}
      \hline
      $\boldsymbol{k}$ & $\boldsymbol{h}$
        & \textbf{Consensus time} & \textbf{Reference}
      & $\boldsymbol{k}$ & $\boldsymbol{h}$
        & \textbf{Consensus time} & \textbf{Reference} \\
      \hline

      Arbitrary
        & $1$
        & $\Theta(n)$ \textnormal{(expected)}
        & \cite{HP01,CEOR13}
      & $2$
        & $3\le h=o(n/\log n)$ (odd)
        & $O(\log n/\log h)$
        & \cite{quasi-majority} \\

      $O((n/\log n)^{1/3})$
        & $3$
        & $O(k\log n)^{\ast}$
        & \cite{simple_dynamics}
      & $2$
        & $\ge3$
        & $O(\log n)$
        & \cite{hierarchy_Berenbrink} \\

      $O((n/\log n)^{1/4})$
        & $3$
        & $\Omega(k\log n)^{\ast}$
        & \cite{simple_dynamics}
      & Arbitrary
        & $3$
        & $\widetilde{\Theta}\!\left(\min\{k,\sqrt{n}\}\right)$
        & \cite{SS25} \\

      $O(hn^{1/4-\varepsilon})$
        & $\ge4$
        & $\Omega(k/h^2)$
        & \cite{simple_dynamics}
      & Arbitrary
        & $\ge Ck\log n$
        & $O(\log n)^{\ast}$
        & \cite{h-Majority-DISC25} \\

      $O(n^{1/3-\varepsilon})$
        & $3$
        & $\widetilde{O}(k^3)$
        & \cite{stabilizing_consensus}
      & Arbitrary
        & $3$
        & $\widetilde{\Theta}\!\left(\min\{1/p_1,\sqrt{n}\}\right)$
        & \cite{Archivio26} \\

      Arbitrary
        & $3$
        & $\widetilde{O}(n^{3/4})$
        & \cite{ignore_or_comply}
      & Arbitrary
        & $\ge3$
        & $\widetilde{\Theta}\!\left(
            1+\min\left\{\frac{1}{h^2p_1},\frac{\sqrt n}{h}\right\}
          \right)$
        & Thms.~\ref{thm:consensus-time-intro} and~\ref{thm:neutral-lower-bound-intro} \\

      $O(n^{1/3}/\sqrt{\log n})$
        & $3$
        & $O(k\log n)$
        & \cite{nearly_tight_analysis}
      & Arbitrary
        & $\ge3$
        & $O\!\left(
            \frac{\log n}{\min\{h^2p_1,1\}}
          \right)^{\ast}$
        & Thm.~\ref{thm:plurality-consensus-intro} \\

      Arbitrary
        & $3$
        & $\widetilde{O}(n^{2/3})$
        & \cite{nearly_tight_analysis}
      & & & & \\

      \hline
    \end{tabular}%
  }

  \smallskip
  \begin{minipage}{\textwidth}
    \footnotesize
    \setlength{\parindent}{0pt}
    Here, $\varepsilon\in(0,1/4)$ is any fixed constant
    and $C>0$ is sufficiently large.

    \smallskip
    ${}^{\ast}$ Plurality consensus under a sufficiently large
    initial gap and, where required, sufficiently large
    plurality support.
  \end{minipage}
\end{table}
\subsubsection{\texorpdfstring
    {A general upper bound for \hmaj}
    {A general upper bound for h-Majority}
}

First, we show a new general upper bound for the consensus time of \hmaj:

\begin{theorem}[Consensus time; see also \cref{thm:proof of the upper bound on the consensus time}]\label{thm:consensus-time-intro}
  For every initial configuration, \emph{\hmaj} reaches consensus within
  $\widetilde{O}\pa{
    1+\min\left\{\frac{1}{\min\{h^2\pmax,1\}},\frac{\sqrt n}{h}\right\}
  }$
  rounds with high probability.
\end{theorem}

The two terms in \cref{thm:consensus-time-intro} can be understood as follows. If $\pmax\ge 1/(h\sqrt n)$ initially, two phases which we call ``symmetry breaking'' and ``amplification'' yield consensus within $\widetilde{O}(1+1/(h^2\pmax))$ rounds. Otherwise, we show that the process reaches a configuration with $\pmax\ge 1/(h\sqrt n)$ within $\widetilde{O}(1+\sqrt{n}/h)$ rounds. Applying the same bound from this configuration yields consensus within another $\widetilde{O}(1+\sqrt{n}/h)$ rounds, since $1/(h^2\pmax)\le\sqrt{n}/h$. Combining the two cases yields the minimum in \cref{thm:consensus-time-intro}. See \cref{sec:proof-overview} for more explanations, and \cref{thm:proof of the upper bound on the consensus time} for an exact statement taking the logarithmic terms into account.

For $h=3$, the bound is $\widetilde{O}(\min\{1/\pmax,\sqrt n\})$, recovering the recent bound of \citet{Archivio26}. Since $\pmax\ge1/k$, it also recovers the worst-case bound $\widetilde{O}(\min\{k,\sqrt n\})$ of \citet{SS25}. However,
\Cref{thm:consensus-time-intro} is the first consensus time bound that holds for an \emph{arbitrary} $h\ge 3$, $k\ge 2$, and an \emph{arbitrary} initial configuration.

If the initial configuration is nearly balanced (i.e., $\pmax=\Theta(1/k)$), our bound is $\widetilde{O}(1+\min\{k/h^2,\sqrt n/h\})$.
\Cref{thm:consensus-time-intro} implies that $h=\Omega(\sqrt k)$ suffices for near-optimal convergence time from every initial configuration with at most $k$ opinions, without any assumption on the initial gap.

\subsubsection{Fast plurality consensus}

We also prove \emph{plurality consensus}, where all nodes eventually adopt the opinion supported by the largest number of nodes in the initial configuration, provided that there is a large enough gap between the initial plurality opinion and any other opinion.
\begin{theorem}[Plurality consensus]\label{thm:plurality-consensus-intro}
  There exists an absolute constant $c>0$ such that, for every initial configuration satisfying
  $\pmax-p_i\ge c\sqrt{
  \frac{\pmax}{\min\{h^2\pmax,1\}}\cdot\frac{\log n}{n}}$ for all $i\ge 2$,
  \emph{\hmaj} reaches consensus on opinion $1$ within $
  O\pa{\frac{\log n}{\min\{h^2\pmax,1\}}}$
  rounds with high probability.
\end{theorem}
For $h=3$, \cref{thm:plurality-consensus-intro} guarantees plurality consensus in $O\left(\frac{\log n}{\pmax}\right)$ rounds under a gap of order $\sqrt{\log n/n}$, which matches the bound of \citet{Archivio26}.

Furthermore, \Cref{thm:plurality-consensus-intro} weakens the requirement on~$h$ to obtain polylogarithmic-time plurality consensus from $h=\widetilde{\Omega}(\pmax^{-1})$ to $h=\Theta(\pmax^{-1/2})$ compared to the state-of-the-art~\cite{h-Majority-DISC25}. In particular, when the initial configuration is approximately balanced ($\pmax \approx 1/k$), the requirement goes from $\widetilde{\Omega}(k)$ to $\Theta(\sqrt{k})$.

Finally, this bound on the plurality consensus time matches that of \dejavu~\cite{DDGGN26}; see \cref{sec:related-works} for a more detailed comparison with prior work.

\subsubsection{\texorpdfstring{Lower bound for neutral $h$-input dynamics}{Lower bound for neutral h-input dynamics}}

Could a different update rule reach consensus faster than \hmaj using the same number of samples?
To make the comparison fair, we consider \emph{neutral $h$-input dynamics} in the \pullmodel{h} model, which contain \hmaj as a special case.
To select an opinion, each node applies the same fixed, possibly randomized rule, which depends only on the $h$ samples and (possibly) its current opinion, and treats all opinion labels symmetrically -- see \cref{def:h-input dynamics} for a formal statement.
The following theorem gives a nearly matching lower bound for this class.

\begin{theorem}[Lower bound; see also \cref{thm:general lower bound on h-input dynamics}]\label{thm:neutral-lower-bound-intro}
  As long as $\pmax < 1$, every neutral $h$-input dynamics requires $\Omega\pa{
    1+\min\left\{\frac{1}{\min\{h^2\pmax,1\}},
    \frac{\sqrt n}{h\sqrt{\log n}}\right\}
  }
  $
  rounds to reach consensus with high probability.
\end{theorem}
\Cref{thm:neutral-lower-bound-intro} matches \cref{thm:consensus-time-intro} up to logarithmic factors, establishing the near-optimality of \hmaj within the class of neutral $h$-input dynamics.
The lower bound from \cref{thm:neutral-lower-bound-intro} shows that the two time scales $1/\min\{h^2\pmax,1\}$ and $\sqrt{n}/h$ cannot be substantially improved by choosing a different neutral update rule.

Since $\pmax\ge1/k$, \cref{thm:consensus-time-intro} implies that \hmaj takes a total of $\widetilde{O}\pa{h+\min\left\{\frac{k}{h},\sqrt n\right\}}$ samples per node until consensus, from every initial configuration with at most $k$ opinions.
Choosing $h=\Theta(\sqrt{k})$ yields polylogarithmic-time consensus using $\widetilde{O}(\sqrt{k})$ samples per node.
Conversely, when $\pmax=\Theta(1/k)$, \cref{thm:neutral-lower-bound-intro} implies that every neutral $h$-input rule requires a total of $\widetilde{\Omega}(\sqrt{k})$ samples per node until consensus, for every sample size $h\ge1$ fixed throughout the execution, since $k\le n$.
Thus, this choice of $h$ minimizes the worst-case total number of samples per node within this class, up to logarithmic factors.

\section{Notation and definitions}\label{sec:preliminaries}
We introduce notations that will be used throughout the paper.
For $m\in\bbN$, write $[m]=\{1,\ldots,m\}$.

For $\bfx\in\bbZ_{\ge0}^k$ with $\abs{\bfx}:=x_1+\dots+x_k=r$, the multinomial coefficient is
\[
  \binom{r}{\bfx}=\binom{r}{x_1,\ldots,x_k}
  =\frac{r!}{x_1!\cdots x_k!}.
\]
For a probability vector $\bfp\in[0,1]^k$ with $\abs{\bfp}=1$, consider $h$ independent samples from $\bfp$.
The count vector is $\bfX=(X_1,\dots,X_k)$ where $X_i$ is the number of occurences of $i$ in the samples.
This vector has the \emph{multinomial distribution} with parameters $(h,\bfp)$, denoted by $\bfX\sim\Multinom(h;\bfp)$.
For every $\bfx\in\bbZ_{\ge0}^k$ with $\abs{\bfx}=h$, its probability mass function is
\begin{equation}\label{eq:density-multinomial}
  \pr{\bfX=\bfx}=\binom{h}{\bfx}\prod_{i=1}^k p_i^{x_i}.
\end{equation}

We formally define \hmaj as follows.
  For a fixed positive integer $h$ and an initial configuration $\bfC(0)\in[k]^n$, \hmaj is a sequence of random variables $(\bfC(t))_{t=0,1,\dots}$, where $\bfC(t)=(C_v(t))_{v\in[n]}\in[k]^n$.
  Given $\bfC(t)\in[k]^n$, the next configuration $\bfC(t+1)$ is sampled according to the following rule:
  \begin{enumerate}
    \item Let $\bfp(t)=(p_i(t))_{i\in[k]}$ be the probability vector defined by $p_i(t)=\frac{|\{v\in[n]:C_v(t)=i\}|}{n}$.
    \item Independently for every node $u\in[n]$, sample $\bfX^{(u)}=(X_1^{(u)},\dots,X_k^{(u)}) \sim \Multinom(h;\bfp(t))$.
      Then set $C_u(t+1)=\maj(\bfX^{(u)})$, where $\maj(\bfX^{(u)})$ is an index attaining the largest coordinate of $\bfX^{(u)}$, chosen uniformly in case of a tie.
  \end{enumerate}
  All sampling and tie-breaking use fresh randomness, independent across nodes and rounds, and all nodes update simultaneously.

We write $\calF_t$ to denote the filtration generated by $(\bfC(s))_{s \leq t}$, and use shorthand notation $\pron{t}{\cdot}$, $\expecton{t}{\cdot}$, and $\varianceon{t}{\cdot}$ for $\pr{\cdot \mid \calF_t}$, $\expect{\cdot \mid \calF_t}$, and $\variance{\cdot \mid \calF_t}$, respectively, unless otherwise specified.
Note that both the processes \(\bfp(t)\) and \(\bfC(t)\) are Markov chains.
Finally, we define $\bfq(t)=\expecton{t}{\bfp(t+1)}$, the one-round expected evolution of $\bfp(t)$.
\section{Proof overview}\label{sec:proof-overview}

A common strategy in the analysis of consensus dynamics is to break symmetry and then amplify the resulting advantage until consensus~\cite{nearly_tight_analysis,async_3-majority_soda25,SS25,USD_PODC26}.
We follow this strategy in three phases, with all guarantees below holding with high probability.
In each case,
round $0$ denotes the start of that phase.
Let \(\pmax(t)\) denote the density of the plurality opinion at the beginning of round \(t\).%

\subsection{The three phases}

 \paragraph{Phase I: Voter.}
    In this phase, we start from an arbitrary initial configuration; however, the phase is only relevant in cases where \(\pmax\) is very small.
    To show that \(\pmax\) grows, we exploit the fact that many opinions vanish.
    We couple \hmaj with \voter (i.e., \(1\)-\textsf{Majority}) to show that $\pmax(T)=\Omega\left(\frac{T}{n\log n}\right)$ for every fixed integer $1\le T\le n\log n$.
    Consequently, $\pmax$ reaches
    \begin{equation}\label{eq:overview-preparation-threshold}
      \pmax(t)\ge c\max\left\{\frac{\log n}{h\sqrt n},\frac{(\log n)^2}{n}\right\}
    \end{equation}
    within $O\left(\max\left\{(\log n)^3,\frac{\sqrt n(\log n)^2}{h}\right\}\right)$ rounds, for a sufficiently large constant $c>0$; see \cref{sec:overview-voter}.
    Ignoring logarithmic factors, the coupling gives $\pmax(T)=\Omega(T/n)$, while the condition for exiting this phase is $\pmax(T)=\Omega(1/(h\sqrt n))$.
    Equating these two scales gives the time scale $T=\widetilde{O}(\sqrt n/h)$ in \cref{thm:consensus-time-intro}.

 \paragraph{Phase II: Symmetry breaking.}
    After reaching the density threshold of \Cref{eq:overview-preparation-threshold}, this phase deals with configurations where the gap between the most popular opinion and other opinions is not large enough.
    We exploit the sampling variance, which creates separations between tied or nearly tied opinions, and a positive expected multiplicative drift, which amplifies the separations that emerge.
    We propose a potential argument that captures these two effects in a single estimate.
    At the end of the phase, the most popular opinion has a constant relative gap: $p_i(t)\le \frac{3}{4}\pmax(t)$ for every $i$ which is not the plurality opinion.
    In particular, this is enough for ensuring plurality consensus.
    This takes $O\left(\frac{\log n}{\min\{h^2\pmax(0),1\}}\right)$ rounds; see \cref{sec:overview-symmetry}.

\paragraph{Phase III: Amplification to consensus.}
    This phase starts when the plurality opinion, relabeled as opinion $1$, satisfies
    \[
      p_1(0)-p_i(0) \ge c\sqrt{
        \frac{p_1(0)}{\min\{h^2p_1(0),1\}}
        \cdot\frac{\log n}{n}}
      \qquad\text{for every }i\ne1
    \]
    for a sufficiently large constant $c>0$.
    The output of Phase II satisfies this condition because its constant relative gap is combined with the density lower bound in \eqref{eq:overview-preparation-threshold}.
    Under this condition, a positive expected multiplicative drift amplifies the existing gaps between the most popular opinion and other competing opinions, while concentration arguments control the variance.
    This yields consensus ($p_1(t)=1$) in at most $O\left(\frac{\log n}{\min\{h^2\pmax(0),1\}}\right)$ additional rounds; see \cref{sec:overview-plurality}.

Phase I can be skipped if the initial maximum density already satisfies \eqref{eq:overview-preparation-threshold}.
Under the assumption on the initial gap in \cref{thm:plurality-consensus-intro}, we can start directly with amplification and obtain plurality consensus.
Intuition for the term \(\min\{h^2\pmax(0),1\}\) in the time scale \(O\left(\frac{\log n}{\min\{h^2\pmax(0),1\}}\right)\) is given later in \cref{sec:overview-expectation}.

\subsection{The main obstacles}

The rest of this section describes the main obstacles, explains the technical ideas used to overcome them, and then outlines the main steps of the proofs.

\paragraph{Obstacle 1: One-round expected evolution.}\label{obs:overview-evolution}
A natural approach is to extend the recent analyses of \threemaj by \citet{SS25} and \citet{Archivio26} to \hmaj.
However, these analyses rely on a simple exact formula for the one-round expected evolution $\bfq$, namely $q_i=p_i(1+p_i-\sum_jp_j^2)$.
For general \hmaj, a direct expression for $q_i$ involves summing over multinomial sample-count vectors and accounting for all possible ties for the maximum.
Deriving useful estimates from this expression is difficult, as also noted by \citet[Section~3]{h-Majority-DISC25}.

A key technique in prior work on \threemaj is to track the gap between two opinion densities and establish conditions under which it grows in expectation~\cite{simple_dynamics,nearly_tight_analysis,async_3-majority_soda25,SS25,Archivio26}.
For a gap $p_i-p_j$, its \emph{drift} is the expected one-round change $(q_i-q_j)-(p_i-p_j)$ conditional on the current configuration; positive drift means that the gap increases in expectation.
Our first main contribution is the following expected relative drift lemma, which bounds the amplification of the relative gap $p_1/p_i-1$.

\begin{lemma}[Expected relative drift]\label{lem:overview-drift}
  Fix a probability vector $\bfp$ with $p_1=\max_jp_j$, and let $q_i$ be the probability that a node adopts opinion $i$ when applying the \hmaj rule to $h$ independent samples from $\bfp$. For every $h\ge3$ and $i\ne1$ with $p_i>0$,
  \[
    \frac{q_1}{q_i}-1\ge
    \left(\frac{p_1}{p_i}-1\right)
    \left(1+\frac1{80}\min\{h^2p_1,1\}\right).
  \]
\end{lemma}
Thus, the relative advantage of the current largest opinion grows at rate $\min\{h^2p_1,1\}$ in the expected update.
The form of this bound is similar to that obtained in \dejavu~\cite{DDGGN26}, though its proof is of different nature (see \cref{sec:related-works} for details about \dejavu and its analysis, which relies on \emph{Poisson races}).
We outline the proof of this lemma in \cref{sec:overview-expectation}.

\paragraph{Obstacle 2: Symmetry breaking.}\label{obs:overview-symmetry}
When $p_1=p_i$, symmetry gives $q_1=q_i$, and the tie is preserved in expectation: a multiplicative estimate such as \cref{lem:overview-drift} does not explain how a gap is created.
The existing plurality-consensus analyses of \hmaj~\cite{h-Majority-DISC25} and \dejavu~\cite{DDGGN26} start from a sufficiently large initial gap and establish its amplification, leaving symmetry breaking from an initially tied configuration untreated.

For \threemaj, conditional on $\bfC(t)$, the gap $p_i(t+1)-p_j(t+1)$ between two fixed opinions $i$ and $j$ is a sum of independent node contributions. This lets us apply variance and anti-concentration estimates, which show that, with constant probability, the gap is of the order of its standard deviation, even when its conditional expectation is zero.

For \hmaj, this gap still admits a representation as a sum of independent contributions, but it need not grow in expectation.
For example, when $h$ is large and there is a unique largest opinion, two non-leading opinions may both vanish, so their gap shrinks to zero. We therefore track amplification through the gap $\pmax(t)-p_j(t)$ between a competitor and the current largest opinion, as in \cref{lem:overview-drift}.
However, this quantity in round~$t+1$ can no longer be written as a sum of independent random variables, even when conditioning on $\bfC(t)$, because the majority opinion in round~$t+1$ may differ from that in round~$t$.
Thus, controlling the expected evolution as in \cref{lem:overview-drift} does not suffice to extend the symmetry-breaking arguments for \threemaj to \hmaj.

Despite these obstacles, we establish the following symmetry-breaking guarantee.

\begin{lemma}[Symmetry breaking, informal]\label{lem:overview-symmetry-informal}
  Suppose that the initial configuration satisfies $\pmax(0)\ge c\max\bigl\{\frac{(\log n)^2}{n},\frac{\log n}{h\sqrt n}\bigr\}$ for a sufficiently large constant $c>0$ (but we do not assume any gap).
  For some $T=O\bigl(\log n/\min\{h^2\pmax(0),1\}\bigr)$, with high probability there is a unique opinion $i$ such that
  \[
    p_j(T)<\frac{3}{4} \cdot p_i(T)\quad\text{for every }j\ne i.
  \]
\end{lemma}
The typical strategy for proving such a result is to analyze the creation of the gap and then its amplification in two separate stages.
It was first used by \citet{Doerr11} and appears in later work on \threemaj and \twochoices~\cite{nearly_tight_analysis,async_3-majority_soda25,SS25}, as well as on \undstate~\cite{fast_convergence_undecided,USD_PODC26}.

In contrast, our proof uses a simple potential argument that captures both phenomena simultaneously.
This technique, the second main contribution of this paper, may therefore substantially simplify symmetry-breaking proofs across several consensus dynamics.
We outline the argument in \cref{sec:overview-symmetry}.

Recall that the Phase I (the ``Voter'' phase) establishes the density condition \eqref{eq:overview-preparation-threshold}.
Together with this density condition, the lower bound makes the resulting gap large enough to start plurality amplification:
\[
  p_i(T)-p_j(T)\ge\frac{\pmax(T)}{4}
  \ge c'\sqrt{\frac{\pmax(T)}{\min\{h^2\pmax(T),1\}}\cdot\frac{\log n}{n}}
  \quad\text{for every }j\ne i,
\]
where $c'>0$ is the constant required for plurality amplification in \cref{sec:overview-plurality}. %
We can therefore relabel opinion $i$ as opinion $1$ and restart the analysis at round $T$.
Plurality amplification followed by the final phase then yields consensus; see \cref{sec:overview-plurality}.

\subsection{Bounding the expected one-round growth of opinions (\texorpdfstring{\cref{sec:expected_growth}}{Section 5})}\label{sec:overview-expectation}

We outline the proof of the expected relative drift bound in \cref{lem:overview-drift}, which addresses \hyperref[obs:overview-evolution]{Obstacle~1}.
Fix a round $t$ and write $\bfp=\bfp(t)$ and $\bfq=\bfq(t)$.
Relabel opinions so that $p_1=\max_j p_j$.
Recall from \cref{sec:preliminaries} that each node updates to $\maj(\bfX)$ with $\bfX\sim\Multinom(h;\bfp)$, so it suffices to analyze a single node.
For $i\in[k]$, let $\calW_i=\{\maj(\bfX)=i\}$, so that $q_i=\pr{\calW_i}$.
Fix an opinion $i\ne1$ with $p_i>0$; If \(p_i=0\), opinion \(i\) has zero probability of winning and there is nothing to prove.
We compare $\pr{\calW_1}$ with $\pr{\calW_i}$ directly, without separately estimating either term.
\Cref{lem:overview-drift} gives
\begin{align}
  \frac{\pr{\calW_1}}{\pr{\calW_i}} \ge 1 + \left(\frac{p_1}{p_i} - 1\right)\left(1 + \frac{1}{80}\min\{h^2p_1,1\}\right).\label{eq:overview:multinomial}
\end{align}
The term \(\min\{h^2p_1,1\}\) is associated with
collisions in the sample.
Indeed, by a union bound,
\[
  \pr{\text{the sample has a collision}}
  \le
  \min\left\{1,\binom{h}{2}\sum_{j=1}^k p_j^2\right\}
  \le
  \min\left\{1,\binom{h}{2}p_1\right\}
  = O\bigl(\min\{h^2p_1,1\}\bigr).
\]
To understand why this term shows up in the bound, we need to connect \hmaj with \voter.
Generate \(h\) samples
and let the voter update copy a uniformly random position in the
sample; this is equivalent to sampling one uniformly
random node from the graph.
If the \(h\) sampled opinions are all
distinct, then \hmaj uniformly breaks the tie and hence has
exactly the same update rule.
Thus, under this coupling, \hmaj can
differ from the voter update only when the sample contains a
collision.
Such collisions are therefore responsible for the
amplification of relative gaps.

\paragraph{Probability transport arguments.}
If \(p_i=p_1\), the inequality \eqref{eq:overview:multinomial} is trivially true.
For the rest of this subsection, assume \(0<p_i<p_1\).
For \(j\in\{1,i\}\) and any sample-count vector \(\bfx\) with
nonnegative integer entries summing to \(h\), define
\(\mu_j(\bfx)=\pr{\calW_j,\bfX=\bfx}\), the probability that opinion
\(j\) wins the majority rule \emph{and} the sampled configuration is \(\bfx\).
Let \(\calX= \{\bfx\in\{0,\ldots,h\}^k:\sum_j x_j=h\}\).
We define a bijection over \(\calX\) as follows: \(\bfx\mapsto \bfx^{1 \leftrightarrow i}\), where \(\bfx^{1 \leftrightarrow i}\) is obtained by exchanging the entries of~\(x_1\) and~\(x_i\).
We prove that \(\mu_i(\bfx)\) and \(\mu_1(\bfx^{1 \leftrightarrow i})\) are related as follows (\cref{lemma:score-aux}):
\begin{align}
  \mu_1(\bfx^{1 \leftrightarrow i}) = \mu_i(\bfx) + (\rho - 1)\Psi(\bfx)\mu_i(\bfx), \label{eq:overview:swapping-prob}
\end{align}
where \(\rho = p_1/p_i\) and \(\Psi(\bfx) = (\rho^{x_i - x_1} - 1)/(\rho - 1)\).
To see this, consider a configuration \(\bfx\) that contributes to \(\pr{\calW_i}\), that is, where \(x_i\) is maximum (possibly, with ties).
The map \(\bfx\mapsto \bfx^{1 \leftrightarrow i}\) restricted to such configurations is also a bijection between configurations that contribute to \(\pr{\calW_i}\) and those that contribute to \(\pr{\calW_1}\).
While moving from \(\bfx\) (for \(\pr{\calW_i}\)) to \( \bfx^{1 \leftrightarrow i}\) (for \(\pr{\calW_1}\)), the tie-winning probabilities are the same, while the sample probability gets multiplied by \(\rho^{x_i - x_1}\).
Hence, \(\mu_1(\bfx^{1 \leftrightarrow i}) = \rho^{x_i - x_1}\mu_i(\bfx)\), and
\(\mu_1(\bfx^{1 \leftrightarrow i}) - \mu_i(\bfx) = (\rho^{x_i - x_1} - 1)\mu_i(\bfx)\), proving \cref{eq:overview:swapping-prob}.

Let \(S = \sum_{\bfx\in \calX} \Psi(\bfx) \mu_i(\bfx)\).
Since \(\bfx\mapsto \bfx^{1 \leftrightarrow i}\) is a bijection in \(\calX\), we can express the difference \(\pr{\calW_1} - \pr{\calW_i}\) as a function of \(S\) and \(\rho\) as follows (\cref{lemma:score-identity}):
\begin{align*}
  \pr{\calW_1} & = \sum_{\bfx\in \calX} \mu_1(\bfx^{1 \leftrightarrow i})
   = \sum_{\bfx\in \calX} \mu_i(\bfx)\left(1 + (\rho-1) \Psi(\bfx)\right)
   = \pr{\calW_i} + (\rho-1)S.
\end{align*}
Proving \cref{eq:overview:multinomial} reduces to showing that \(S \ge \pr{\calW_i}(1 + \min\{h^2p_1,1\}/80)\).
Let \(\varepsilon = \min\{h^2p_1,1\}/80.\)
If \(\Psi(\bfx) \ge (1 + \varepsilon)\) for all configurations where \(x_i\) is a maximum, then we would immediately get \(S \ge (1+\varepsilon) \sum_{\bfx\in \calX} \mu_i(\bfx) = (1+\varepsilon) \pr{\calW_i}\).
Unfortunately, this is not the case.

To give intuition behind the definition of \(\Psi\) (and, hence, \(S\)), consider a configuration \(\bfx\) where opinion \(i\) is maximum (possibly, not unique).
Now, depending on the entries \(x_i\) and \(x_1\) in \(\bfx\), we have
\begin{align*}
  \Psi(\bfx) =
  \begin{cases}
    0, \quad & x_i = x_1; \\
    1, \quad &  x_i = x_1 + 1; \\
    1 + \rho + \ldots + \rho^{x_i - x_1 - 1} \ge 1 + \rho, \quad &  x_i > x_1 + 1.
  \end{cases}
\end{align*}
Therefore, for some configurations \(\Psi(\bfx) < 1 + \varepsilon\), while for other configurations \(\Psi(\bfx) \ge 1 + \rho \ge 1 + \varepsilon\).
In our analysis, we compare the total \emph{probability mass} (according to \(\mu_i(\bfx)\)) of configurations \(\bfx\) that have \(x_i > x_1 + 1\) with that of configurations \(\bfx\) that have \(x_i \le x_1 + 1\).
We do that via what we call \emph{transport lemmas}, that is, we compare probabilities of configurations \(\bfx\) with \(\mu_i(\bfx) > 0\) with configurations obtained by moving units from \(x_1\) to other entries of \(\bfx\), thus with larger values of \(\Psi(\bfx)\).
We call \(\Psi(\bfx)\) the ``score'' of configuration \(\bfx\).

The transport arguments differ according to whether \(hp_1\le1/4\) or \(hp_1>1/4\).
In the small-sample case, at most \(h\) opinions appear in any sample, so the absent opinions carry total probability mass at least \(1-hp_1\ge3/4\).
This provides enough destinations to compare configurations with \(x_1>0\) to configurations with smaller \(x_1\): we move one occurrence of opinion \(1\) to an absent opinion, keeping \(x_i\) unchanged.
For configurations with \(\mu_i(\bfx)>0\) and \(x_i\ge2\), iterating this transport reaches \(x_1=0\), where the score is at least \(1+\rho\), giving ``surplus'' with respect to the required score \(1+\varepsilon\).
The resulting mass comparisons, together with a separate transport for cases in which we have all distinct samples in a configuration, ensure that this surplus compensates for the tied configurations; see \cref{lemma:transport-to-empty,lemma:witness-lemma}.
When \(hp_1>1/4\), absent opinions might not carry enough probability mass for the same argument to work.
We instead move an occurrence of opinion \(1\) directly to opinion \(i\), increasing \(x_i-x_1\) by two.
This transport relates tied configurations to configurations with surplus score; see \cref{lemma:magical-transport-combination,lemma:witness-lemma}.

We remark that
\cite{h-Majority-DISC25} also uses a transport argument, which inspired us to look for more general techniques based on probability transport arguments.
We highlight the difference between our work and \cite{h-Majority-DISC25}.
There, the authors prove that, if \(hp_1 \ge C \log n\), \emph{ties do not really matter}.
Their approach crucially relies on the assumption \(hp_1 \ge C \log n\).
It is needed that each opinion \(j\) such that \(p_j = \Theta(p_1)\) is sampled roughly \(\Omega(\log n)\) times with high probability in the multinomial sample, and the crucial ingredient is a lemma stating that the probability that opinion \(1\) wins (possibly with ties occurring) is comparable to the probability that opinion \(1\) wins without ties.
This latter fact is not true for \(hp_1 \ll \log n\).
Our transport lemmas are of different nature, and we do not employ the same argument here (even for the case \(h p_1 \ge C \log n\)).

\subsection{Preparation via coupling with Voter (\texorpdfstring{\cref{sec:majorization}}{Section 5.2})}\label{sec:overview-voter}

The preparation phase raises the maximum density from $1/n$ (in the worst case) to the threshold in~\eqref{eq:overview-preparation-threshold}.
We extend the \emph{majorization coupling argument} for \threemaj of \citet{ignore_or_comply} to \hmaj for every $h\ge2$.
Specifically, we construct a coupling under which $\bfp^{(h)}(t)$ \emph{majorizes} $\bfp^{(1)}(t)$ almost surely for every $t=0,1,\ldots$, provided $\bfp^{(h)}(0)=\bfp^{(1)}(0)$.
Here $\bfp^{(r)}(t)$ denotes the vector of opinion densities of $r$-\textsc{majority} at round $t$, with $r=1$ corresponding to \voter.
We refer to \cref{def:majorization} for the formal definition of majorization.
The key consequence is that, at every round $t$, the number of surviving opinions in \hmaj (i.e., the number of indices $i$ with $p_i^{(h)}(t)>0$) is at most that in \voter.

For every fixed integer $1\le t\le n\log n$, \voter has at most $\kappa(t)=O\left(\frac{n\log n}{t}\right)$ surviving opinions with high probability~\cite{ignore_or_comply}.
The coupling gives the same upper bound for \hmaj.
Since the opinion densities sum to one, on this event $\pmax^{(h)}(t)\ge\frac{1}{\kappa(t)}=\Omega\left(\frac{t}{n\log n}\right)$.
This yields \eqref{eq:overview-preparation-threshold} within the stated preparation time.

For $k=2$, \citet{hierarchy_Berenbrink} constructed a majorization coupling of \hmaj and \((h-1)\)-\textsf{Majority} throughout their executions from the same initial configuration.
This shows that increasing the sample size from $h-1$ to $h$ cannot increase the consensus time in the sense of stochastic dominance; they left open whether this conclusion extends to $k>2$.
Our analysis gives partial progress: for arbitrary $k$, we obtain a majorization coupling of \emph{one-round} update of \hmaj and \((h-1)\)-\textsf{Majority} from the \emph{same} configuration.
Such a coupling can fail for different configurations; see \cref{rem:majorization-scope}.
Thus, the one-round comparison alone does not establish the corresponding comparison of consensus times.

\subsection{Symmetry breaking via potential (\texorpdfstring{\cref{sec:symmetry_breaking}}{Section 7})}\label{sec:overview-symmetry}

We now explain how the potential argument overcomes \hyperref[obs:overview-symmetry]{Obstacle~2}, allowing the largest opinion to change throughout the phase.
For each opinion $j$, define the normalized squared gap $G_j(t)$ and the decreasing function $\Phi_a$ by
\[
  G_j(t)=\frac{(\pmax(t)-p_j(t))^2}{\pmax(t)+p_j(t)} \text{ and }\Phi_a(x)=\frac{1}{1+ax},
\]
where $a>0$ is a carefully chosen parameter.
As the normalized gap $\frac{\pmax(t)-p_j(t)}{\sqrt{\pmax(t)+p_j(t)}}$ grows, $G_j(t)$ increases and the potential $\Phi_a(G_j(t))$ decreases.
Thus, showing that the potential decreases provides a route to proving symmetry breaking.
The key estimate is the following contraction in conditional expectation.

\begin{lemma}[Potential contraction; \cref{lem:drift for hmaj}]\label{lem:overview-potential}
  Fix $t$, an opinion $i$ with $p_i(t)=\pmax(t)$, and a parameter $0<\rho\le\frac{1}{80}\min\{h^2\pmax(t),1\}$. Set $a=\frac{n\rho}{28}$. If $j\ne i$ and $p_j(t)\ge\frac{\pmax(t)}{2}$, then
  \[
    \expecton{t}{\Phi_a(G_j(t+1))}
    \le \left(1-\frac{\rho}{420}\right)\Phi_a(G_j(t)).
  \]
\end{lemma}
The proof of this contraction combines the expected relative drift bound in \cref{lem:overview-drift} with moment estimates for the gap between a currently largest opinion and a fixed competitor.

Assuming \cref{lem:overview-potential}, the main argument is to iterate it and apply Markov's inequality.
Indeed, $p_j(t)\ge\frac{\pmax(t)}{2}$ implies $G_j(t)\le\frac{\pmax(t)}{6}\le\frac{1}{6}$, so $\Phi_a(G_j(t))\ge\frac{6}{6+a}$.
Thus, a potential below this threshold already certifies that $p_j(t)<\frac{\pmax(t)}{2}$.
The contraction holds even when $G_j(t)=0$, so the same argument applies without an initial gap.

A fixed opinion may itself become the leader, so we apply the argument to the pair potential $\Psi_{ij}(t)=\Phi_a(\max\{G_i(t),G_j(t)\})$ for $i\ne j$.
It inherits the same contraction while both densities are at least $\frac{\pmax(t)}{2}$: apply the lemma to an opinion attaining $\max\{G_i(t),G_j(t)\}$.
If both gaps are zero, choose $i$ as a largest opinion and apply the bound to $j$.
Fix $\rho=\frac{1}{160}\min\{h^2\pmax(0),1\}$ and $a=\frac{n\rho}{28}$.
Let $\tau=\inf\bigl\{t\ge0\colon \min\{p_i(t),p_j(t)\}\le\frac{\pmax(t)}{2}\text{ or }\pmax(t)\le\frac{\pmax(0)}{2}\bigr\}$ be the first round when either member has density at most half the current maximum or the maximum is at most half its initial value.
Iteration and Markov's inequality, applied with this stopping rule, give $\pr{\tau>T}\le(1+\frac{a}{6})\exp(-\frac{\rho T}{420})$.
Consequently, for a suitable $T=O\left(\frac{\log n}{\rho}\right)$, with high probability there is a round $t\le T$ such that
$\min\{p_i(t),p_j(t)\}\le\frac{\pmax(t)}{2}$ or
$\pmax(t)\le\frac{\pmax(0)}{2}$.
By a martingale concentration argument (\cref{lem:taumaxdown}), we can show that the latter event is unlikely.
Therefore, with high probability, the potential cannot keep contracting throughout these $T$ rounds while both opinions remain above this threshold.

Further, for each opinion $j$, with high probability,
if $p_j(s)\le\frac{\pmax(s)}{2}$ at some round $s\le T$,
then $p_j(T)<\frac{3}{4}\pmax(T)$.
This also follows from a martingale concentration argument (\cref{lem:weak cannot be strong}).
Combining this with the pair-potential bound and taking a union bound over all $i\ne j$ gives
$\min\{p_i(T),p_j(T)\}<\frac{3}{4}\pmax(T)$ for every pair, with high probability.
Choosing an opinion $i$ with $p_i(T)=\pmax(T)$ gives
$p_j(T)<\frac{3}{4}p_i(T)$ for every $j\ne i$,
yielding the symmetry-breaking guarantee in \cref{lem:overview-symmetry-informal}.

\subsection{Plurality consensus: amplification and the final phase (\texorpdfstring{\cref{sec:plurality-consensus}}{Section 6})}\label{sec:overview-plurality}

With the multinomial comparison in \cref{lem:overview-drift} at hand, we can derive one-round expected growth of the relative gaps and the leading density, provided that opinion $1$ is the most frequent opinion at the current round. If the initial gap is large enough, concentration arguments turn these estimates into multi-round high-probability statements. In particular, $p_1(t)$ remains the largest opinion and grows until it has no serious competitor.

Write $\Delta_i(t)=p_1(t)-p_i(t)$ and $\Delta(t)=\min_{i>1}\Delta_i(t)$, keeping the opinion labels fixed throughout this phase. The phase starts with $\Delta(0)\ge c\sqrt{\frac{p_1(0)}{\min\{h^2p_1(0),1\}}\cdot\frac{\log n}{n}}$. Its central guarantee is as follows.
\begin{lemma}[One amplification phase]\label{lem:overview-amplification-phase}
  For a sufficiently large constant $c>0$, suppose opinion $1$ is initially the largest and $\Delta(0)\ge c\sqrt{\frac{p_1(0)}{\min\{h^2p_1(0),1\}}\cdot\frac{\log n}{n}}$. With probability at least $1-1/n^2$, at some round $T=O(\log(2p_1(0)/\Delta(0))/\min\{h^2p_1(0),1\})$, either $p_1(T)\ge3/5$, or all of the following hold: $p_1(T)\ge2p_1(0)$, opinion $1$ is still the largest, and
  \[
    \Delta(T)\ge\max\left\{
      c\sqrt{\frac{p_1(T)}{\min\{h^2p_1(T),1\}}\cdot\frac{\log n}{n}},
    \frac{p_1(T)}{32}\right\}.
  \]
\end{lemma}

The point of the second outcome is that the next phase starts with both a valid gap condition and a constant relative gap. The process is Markovian, so the same argument can be restarted at each successful phase. At most $O(\log n)$ phases suffice to reach density $3/5$. The first phase costs $O(\log n/\min\{h^2p_1(0),1\})$ rounds; in every later phase, the logarithm of the density-to-gap ratio is bounded by a constant. Consequently, their total duration is still $O(\log n/\min\{h^2p_1(0),1\})$, rather than incurring an extra logarithmic factor.

The concentration argument must track the accumulated noise, rather than demand that every round closely follows its expectation.
Following the Freedman-based martingale framework developed by \citet{async_3-majority_soda25} and \citet{SS25}, we expose the new opinions of the nodes one by one within each round.
Since Freedman's inequality applies to martingales, we normalize densities and gaps by the accumulated expected growth.
Freedman's inequality then controls the normalized density and gap martingales up to the relevant stopping times.

\begin{lemma}[Final phase]\label{lem:overview-final-phase}
  If $p_1(0)\ge3/5$, then \hmaj reaches consensus on opinion $1$ within $O(\log n)$ rounds with probability at least $1-1/n^2$.
\end{lemma}
This final phase follows from the contraction of the total support of the other opinions once opinion $1$ has a sufficiently large constant majority. Combining it with the amplification phases proves \cref{thm:plurality-consensus-intro}; adding preparation and symmetry breaking proves \cref{thm:consensus-time-intro}.

\subsection{The lower bound (\texorpdfstring{\cref{sec:lowerbound on h-input dynamics}}{Section 8})}
Our lower bound starts from a consequence of neutrality: no opinion can grow too quickly in expectation. Specifically, for every opinion $i\in[k]$, we prove that
\begin{align}
  \expecton{t}{p_i(t+1)} - p_i(t) \le 2h^2\pmax(t)^2.
  \label{eq:overview-lower-drift}
\end{align}
To see why, fix the current configuration and write $p_j=p_j(t)$.
As a baseline, consider the voter rule, which simply copies the first sampled opinion and hence selects opinion $i$ with probability $p_i$.
When all sampled opinions are distinct, neutrality allows the rule to favor sample positions but not opinion labels, so any systematic advantage over this baseline (i.e., $\expecton{t}{p_i(t+1)}- p_i(t)$) must come from repetitions.
We bound the additional selection probability by separating two cases:
\begin{itemize}
  \item \emph{Opinion $i$ appears at least twice.}
    This event has probability at most $\binom{h}{2}p_i^2$, so its contribution is $O(h^2p_i^2)$.
  \item \emph{Opinion $i$ appears only once.}
    A general neutral rule may prefer opinions that appear only once.
    Let $S$ be the number of opinions appearing exactly once among the $h$ sampled opinions.
    By neutrality, these $S$ opinions are interchangeable after averaging over sample positions, so the contribution from such samples is at most the expectation of $1/S$ over them.
    To see how this compares with $p_i$, fix the position of the unique $i$ and let $S'$ count the opinions appearing exactly once among the other $h-1$ sampled opinions.
    When none of these sampled opinions equals $i$, we have $S=1+S'$.
    Summing over the $h$ possible positions of $i$, and dropping the restriction that the other sampled opinions avoid $i$, bounds this contribution by
    \[
      p_i\cdot \expect*{\frac{h}{1+S'}}
      \le p_i\expect{h-S'}
      = p_i\bigl(1+\expect{h-1-S'}\bigr),
    \]
    where we use $h/(1+s)\le h-s$ for $0\le s\le h-1$.
    Here $h-1-S'$ counts the sample positions among the other $h-1$ sampled opinions occupied by repeated opinions.
    In expectation, this number is at most $2\binom{h-1}{2}\sum_jp_j^2$.
    Hence, the contribution to the right-hand side of \eqref{eq:overview-lower-drift} in this case is at most $O(h^2p_i\sum_jp_j^2)$.
\end{itemize}
Combining these two contributions and using $p_i\le\pmax(t)$ and $\sum_jp_j^2\le\pmax(t)$ bounds the expected increase by $O(h^2\pmax(t)^2)$.
Keeping track of the constants gives \eqref{eq:overview-lower-drift}.

The drift bound \eqref{eq:overview-lower-drift} suggests a natural
time scale for the maximum density to double. Indeed,
\[
  \max_{i\in[k]}\expecton{t}{p_i(t+1)}
  \le \pmax(t)\bigl(1+2h^2\pmax(t)\bigr).
\]
As long as $\pmax(t)<2\pmax(0)$, the multiplicative factor on the
right-hand side is at most $1+4h^2\pmax(0)$. When
$h^2\pmax(0)$ is small, growth at this rate would require
$\Omega(1/(h^2\pmax(0)))$ rounds to double the maximum density.
This explains the time scale arising from the conditional expected
increments.

To obtain a lower bound for the actual process, we must also
control deviations arising from the variance of the process.
Conditional on the current
configuration, the support of each opinion is a sum of independent
Bernoulli random variables. Together with
\eqref{eq:overview-lower-drift}, this gives
\begin{equation}\label{eq:overview-lower-variance}
  \varianceon{t}{p_i(t+1)}
  \le \frac{\expecton{t}{p_i(t+1)}}{n}
  =O\left(\frac{\pmax(t)}{n}\right)
\end{equation}
when $h^2\pmax(t)$ is small. Before the maximum density doubles,
the sum of these conditional variances over
$T:=O(1/(h^2\pmax(0)))$ rounds is $O(T \cdot \pmax(t)/n)=O(1/(h^2n))$, corresponding to
a standard-deviation scale of $O(1/(h\sqrt{n}))$.
We use Freedman's inequality to control the accumulated variance and resulting deviations
simultaneously for all opinions.
A smaller one-round
variance allows the drift-based lower bound to remain effective
for smaller initial maximum densities.

Thus, neutrality limits growth that comes from collisions to the time scale $1/\min\{h^2\pmax(0),1\}$, while the variance of the process limits how long this argument can be maintained uniformly to the time scale $\widetilde{\Omega}(\sqrt n/h)$.
This explains why the lower bound contains the minimum of the same two scales as the upper bound.

\paragraph{Extension to rules that use the node's current opinion.}
So far, we have considered dynamics in which nodes always \emph{comply} with their samples by adopting one of the $h$ sampled opinions.
Our argument also extends to dynamics in which nodes may instead \emph{ignore} the sampled opinions and retain their current opinion, even when it is absent from the sample.
For example, \twochoices ignores its samples when the two sampled opinions disagree, and \dejavu capped at $h$ samples ignores them when no opinion is sampled twice.
This distinction between complying with and ignoring the samples is central to the comparison between \threemaj and \twochoices in \citet{ignore_or_comply}.
For our lower bound, the relevant additional condition is that the rule retains the current opinion whenever it and the $h$ sampled opinions are all distinct.
Both \twochoices and \dejavu capped at $h$ samples satisfy this condition.
Such rules have smaller one-round density variance than the general bound in \eqref{eq:overview-lower-variance}, which allows us to obtain a sharper consensus-time lower bound.
In particular, for \twochoices, we recover the known $\widetilde{\Omega}(1/\pmax(0))$ lower bound of \citet{Archivio26}.
See \cref{sec:lowerbound on h-input dynamics} for the precise variance estimate and lower bound.

\subsection{Outline of the paper}

We use the notation and definitions introduced in Section~\ref{sec:preliminaries} throughout the paper.
\Cref{sec:expected_growth} analyzes the expected one-round growth of the opinions using an analysis of the multinomial distribution.
\Cref{sec:plurality-consensus} then gives the time bound on plurality consensus.
\Cref{sec:symmetry_breaking} analyzes the symmetry-breaking phase.
\Cref{sec:lowerbound on h-input dynamics} gives the time lower bound for uniform $h$-sampling dynamics.
\section{Further related work}\label{sec:related-works}

\paragraph{3-Majority and 2-Choices.}
The \threemaj and \twochoices are canonical examples of how a few samples can amplify plurality advantages and accelerate consensus without additional states~\cite{BCN20}.
In \twochoices, first studied implicitly for two opinions in~\cite{Doerr11}, each node adopts the common opinion of two independent uniform samples, retaining its own if they disagree.
For two opinions, the analysis of~\cite{Doerr11} gives an $O(\log n)$ worst-case consensus time with high probability for both dynamics, compared with \voter's $\Theta(n)$ worst-case expected time~\cite{HP01}.
\threemaj with many opinions was first studied in~\cite{simple_dynamics} and further analyzed in~\cite{stabilizing_consensus,ignore_or_comply,nearly_tight_analysis}.
The consensus times of both dynamics were determined up to polylogarithmic factors in the worst case with arbitrary $k$ in~\cite{SS25}, and for each initial configuration in terms of $\pmax$ in~\cite{Archivio26}.
For every nonconsensus initial configuration, the stronger $\widetilde{\Omega}(1/\pmax)$ lower bound for \twochoices (while $\widetilde{\Omega}(\min\{1/\pmax,\sqrt n\})$ for \threemaj) holds with high probability.

In the asynchronous model,
only one node selected uniformly at random in each time step updates its opinion.
For \threemaj, the worst-case consensus time is $\widetilde{\Theta}(\min\{nk,n^{3/2}\})$ steps with high probability~\cite{async_3-majority_soda25}.
For two opinions, exact formulas for the probability of consensus on each opinion are given for both dynamics in~\cite[Corollaries~3 and~4]{order_statistics}.

\paragraph{DéjàVu.}
In \dejavu~\cite{DDGGN26}, each node samples opinions until one appears twice and then adopts the repeated opinion.
Interestingly, this setting is equivalent to a \emph{Poisson race}.
Given a configuration with opinion densities \(p_1 \ge \ldots \ge p_k\), consider independent Poisson processes \((N_i(t))_{t\ge0}\), one for each \(i \in [k]\), where \(N_i\) has rate \(p_i\).
Let \(T_i = \min\{t \ge 0 : N_i(t) = 2\}\), with the convention that \(T_i = \infty\) if \(p_i = 0\).
Note that \(T_i \sim \Gammadistr(2,p_i)\) for every \(i\) such that \(p_i > 0\).
The probability that a node selects opinion \(i\) in \dejavu is exactly \(\pr{T_i = \min_{j \in [k]}\{T_j\}}\).
This probability can be computed by exploiting properties of Gamma distributions.
In its \pullmodel{h} version, sampling is capped at $h$; if no repetition occurs, the node retains its current opinion.
The Poisson race setting is linked to \pullmodel{h} via the generalized birthday paradox: the probability that a collision occurs within \(h\) samples is equal to the probability that \dejavu encounters a repeated opinion within the first \(h\) samples.

For $h\ge2$, this dynamics reaches plurality consensus within $O(\log n/\min\{h^2\pmax,1\})$ rounds with high probability, provided that $n\pmax=\omega(\log n)$ and $p_1-p_2$ is sufficiently large.
To the best of our knowledge, this is the only other plurality consensus dynamics studied in the \pullmodel{h} model.
Note that our lower bound from
\cref{thm:general lower bound on h-input dynamics} applies to \dejavu as well.
Combined with our upper bound for \hmaj, this shows that \hmaj is at least as efficient as \dejavu in
\pullmodel{h}, up to polylogarithmic factors, both in terms of convergence time and in terms of sample complexity.
Moreover, the bounds give a polynomial separation in favor of \hmaj whenever \(\pmax\) is polynomially smaller than \(1/(h\sqrt n)\).

However, if we modify the definition of \pullmodel{h} to allow \emph{adaptive stopping}, \dejavu can beat \hmaj in terms of sample complexity when \(h \norm{\bfp}_2 \) is sufficiently large.
Adaptive stopping in \pullmodel{h} is defined as follows: in each round, a node samples (sequentially) \emph{at most} \(h\) opinions, but stops at the \(i\)-th sample if the update rule is already determined by the first \(i\) samples.
A more detailed comparison is given in \cref{sec:lowerbound on h-input dynamics}.

\paragraph{Undecided-state dynamics.}
\undstate uses an auxiliary undecided state: a decided node becomes undecided upon observing a different decided opinion, while an undecided node adopts any decided opinion it observes; otherwise, its state is unchanged.
The rule was introduced for two opinions in the population protocol model, where uniformly random pairs interact sequentially~\cite{undecided_Angluin}.
For two opinions, the synchronous version reaches consensus within $O(\log n)$ rounds with high probability from any initial configuration with a decided node, and this bound is tight~\cite{undecided_MFCS}.
The binary synchronous dynamics has also been studied under uniform communication noise~\cite{DAmoreCN22}.
For many opinions, early synchronous results required a sufficiently large initial plurality advantage~\cite{BecchettiCNPS15}, whereas asynchronous consensus was proved within $O(kn\log n)$ interactions with high probability for $k=O(\sqrt n/\log^2 n)$ from any initial configuration without undecided nodes~\cite{fast_convergence_undecided}.
Recent work covers arbitrary $k$ in both models, accounting for possible absorption in the all-undecided state under synchronous updates~\cite{USD_PODC26}.
A variant that alternates phases of becoming undecided and adopting opinions achieves consensus within $O(\log^2 n)$ synchronous rounds with high probability for arbitrary $k$~\cite{undecided-soda22}.

\paragraph{Minority.}
In \minority, each node adopts a least frequent opinion among those present in its sample, breaking ties uniformly at random.
For two opinions, \citet{BecchettiCPTVZ24} proved polylogarithmic-time consensus under synchronous updates for a sufficiently large sample size $h = \Omega(\sqrt{n \log n})$, whereas the expected consensus time can be exponential under asynchronous updates.
D'Archivio and Vacus~\cite{DArchivioV24} established an almost-linear worst-case time lower bound for memory-less protocols with constant sample size in the related bit-dissemination problem. \minority is another interesting example of neutral $h$-input dynamics, so our general lower bound applies to it as well.

\paragraph{Voter.}
The \voter rule, in which each node adopts the opinion of one uniformly sampled node, has been studied as a probabilistic polling mechanism for proportionate agreement: an opinion wins with probability equal to its initial fraction of supporters~\cite{HP01}.
The linear voting framework generalizes such rules and gives winning probabilities and convergence-time bounds~\cite{linear_voting}.
In our analysis, comparison with \voter provides a bound on the consensus time for \hmaj, which is useful when many opinions are present; see \cref{sec:majorization}.
\section{Expected one-round growth} \label{sec:expected_growth}

In this section, we focus on a single round of the process.
We omit the round index from the notation and assume that the entries of $\bfp= \bfp(t)$ are sorted in decreasing order, $p_1 \geq \cdots \geq p_k$.
We write $\bfX= (X_1, \dots, X_k)$ to denote a sample from the multinomial distribution with parameter~$(h,\bfp)$.
Recall that $\bfq= \bfq(\bfp) \coloneq  \expecton{t}{\bfp(t+1)}$, and that~$\maj(\bfX)$ denotes the winning opinion (accounting for uniform tie-breaking). Finally, we define an event $\calW_i=\{\maj(\bfX)=i\}$, so that~$q_i=\pr{\calW_i}$.

\subsection{\texorpdfstring
    {Gap between 1 and $i$}
    {Gap between 1 and i}
}

In this section, we study how the ratio between the majority opinion and any other opinion changes from round~$t$ to round~$t+1$.

\begin{theorem}\label{thm:multinomial:growth}
  Let \(h \ge 3\) and \(i \in [k]\) such that \(p_i > 0\).
  Then
  \begin{align*}%
    \frac{q_1}{q_i} = \frac{\pr*{\calW_1}}{\pr*{\calW_i}}
    \ge 1 +
    \left(\frac{p_1}{p_i}-1\right)
    \left(1+(1/80) \cdot \min\{h^2 p_1, 1\}\right).
  \end{align*}
\end{theorem}
The theorem is trivial for \(i = 1\).
We will focus only on opinions \(i \neq 1\).
To prove the theorem, we will consider two differently behaving regimes:
\begin{itemize}
  \item if $hp_1 \le 1/4$, then the samples are ``sparse'' and likely have few collisions, and
  \item if $hp_1 \ge 1/4$, then the samples are ``dense'' and tend to have more collisions.
\end{itemize}

\subsubsection{Multinomial sample spaces} \label{sec:multinomial_utils}

Let $\calX\coloneq  \{ \bfx\in \mathbb{Z}_{\ge 0}^k : x_1+\dots+x_k = h\}$ be the support of the multinomial distribution $\Multinom(h;\bfp)$. For $\bfx\in \calX$, let
\[
  M(\bfx)\coloneq \{j\in[k] : x_j= \norm{\bfx}_\infty \}
\]
be the set of opinions that have the highest frequency in $\bfx$.
For each $i \in [k]$, we define the measure
\[
  \mu_i(\bfx)
  \coloneq \pr{\calW_i \cap \{\bfX= \bfx\}}.
\]
This is the probability that $\bfx$ contributes to $i$ being the ``winning opinion'' in the sample.
Clearly,
\begin{equation}\label{eq:wi-identity}
  \pr*{\calW_i}=\sum_{\bfx}\mu_i(\bfx) = \mu_i(\calX).
\end{equation}

\begin{figure} [htbp]
  \centering
  \includegraphics[width=0.6\linewidth]{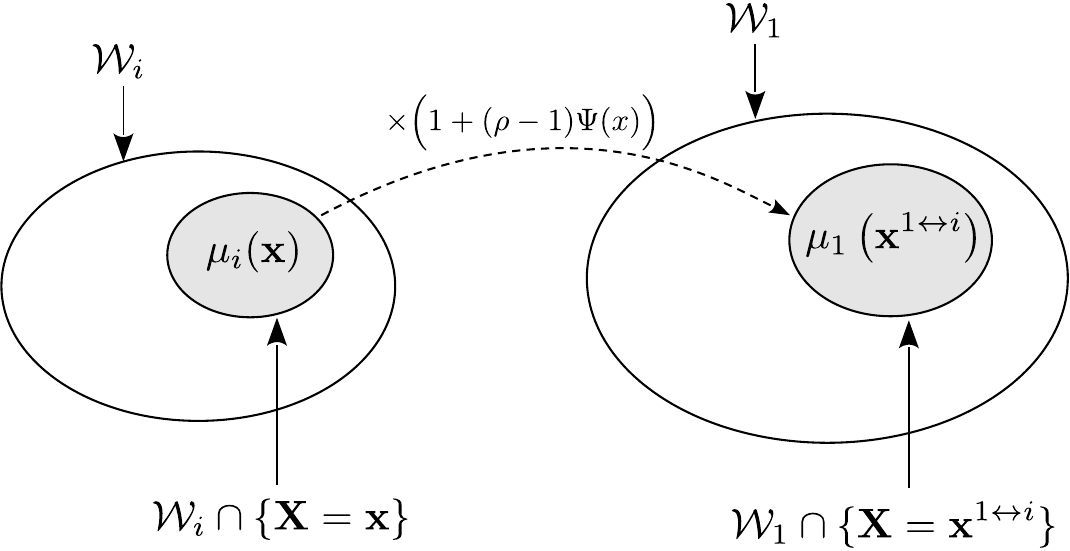}
  \caption{{\bf Intuition behind \cref{lemma:score-aux}:} for any~$\bfx$ such that opinion~$i$ has a non-zero probability to win, there is a corresponding $\bfx^{1 \leftrightarrow i}$ that is more likely to be sampled while granting opinion 1 the same probability to win.
  \label{fig:score}}
\end{figure}

\paragraph{Scoring functions.}
If $p_i = p_1$, then the claim of \Cref{thm:multinomial:growth} is trivially true.
From now on, we fix  $i \in \{2, \ldots, k\}$ such that $0 < p_i < p_1$.
We write $\rho = p_1/p_i > 1$,
and define scoring functions that will be heavily used in the analysis:
\[
  \Psi(\bfx)\coloneq \frac{\rho^{x_i-x_1}-1}{\rho-1} \quad \text{ and } \quad  S\coloneq \sum_{\bfx\in \calX}\Psi(\bfx) \mu_i(\bfx).
\]
Note that $\Psi$ is well-defined, since $\rho > 1$.
These definitions rely on the following idea: for every~$\bfx\in \calX$, let~$\bfx^{1 \leftrightarrow i}$ be the vector obtained from $\bfx$ by exchanging $x_1$ and $x_i$.
As it turns out, $\mu_1(\bfx^{1 \leftrightarrow i})$ can be easily expressed as a function of $\rho$, $\mu_i(\bfx)$ and $\Psi(\bfx)$. We formalize this relationship in \cref{lemma:score-aux} and illustrate it in \cref{fig:score}.

\begin{lemma} \label{lemma:score-aux}
  For every~$\bfx\in \calX$ such that $\mu_i(\bfx) > 0$,
  \begin{equation*}
    \mu_1(\bfx^{1 \leftrightarrow i}) = \mu_i(\bfx) + (\rho-1) \, \Psi(\bfx) \,  \mu_i(\bfx).
  \end{equation*}
\end{lemma}
\begin{proof}
  Since $|M(\bfx)| = |M(\bfx^{1 \leftrightarrow i})|$, we get from \Cref{eq:density-multinomial}:
  \begin{equation*}
    \frac{\pr{\bfX= \bfx^{1 \leftrightarrow i}}}{\pr{\bfX= \bfx}} = \rho^{x_i-x_1}.
  \end{equation*}
  Therefore,
  \begin{equation*}
    \mu_1(\bfx^{1 \leftrightarrow i}) = \mu_i(\bfx) \cdot \rho^{x_i-x_1} = \mu_i(\bfx) \cdot \left[ 1 + (\rho-1) \, \Psi(\bfx)  \right]. \qedhere %
  \end{equation*}
\end{proof}

This identity translates immediately to a nice relationship between $S$ and $\pr*{\calW_1}-\pr*{\calW_i}$.

\begin{corollary} \label{lemma:score-identity}
  We have that
  \begin{equation*}
    \pr*{\calW_1}-\pr*{\calW_i} = (\rho-1)S.
  \end{equation*}
\end{corollary}
\begin{proof}
  Since the map: $\bfx\rightarrow \bfx^{1 \leftrightarrow i}$ is a bijection from $\calX$ to $\calX$,
  \begin{align*}
    \pr*{\calW_1} &= \sum_{\bfx\in \calX}\mu_1(\bfx^{1 \leftrightarrow i}) \\
    &= \sum_{\bfx\in \calX} \left[ \mu_i(\bfx) + (\rho-1) \, \Psi(\bfx) \,  \mu_i(\bfx) \right] & \text{(by \Cref{lemma:score-aux})} \\
    &=  \pr*{\calW_i} + (\rho-1)S. \qedhere
  \end{align*}
\end{proof}

Equipped with \Cref{lemma:score-identity}, it now suffices to lower bound $S$ to prove \Cref{thm:multinomial:growth}.
More precisely, our target is $S\ge (1+\min\{h^2p_1,1\}/80)\mu_i(\calX)$.
On the support of $\mu_i$, the score $\Psi$ is zero when $x_i=x_1$, one when $x_i=x_1+1$, and at least $1+\rho$ when $x_i\ge x_1+2$.
Thus, relative to the needed score of \(1 + \min\{h^2 p_1,1\}/80\), configurations \(\bfx\) with \(\mu_i(\bfx) > 0\) such that \(x_1 = x_i\) create a deficit larger than \(1\), while if \(x_1 = x_i -1\) the deficit is \(\min\{h^2 p_1,1\}/80\), and if \(x_1 \le x_i - 2\) create a surplus.
We compare the probability masses of these configurations through transport maps: the surplus must compensate for the deficit and still leave a margin proportional to $\min\{h^2p_1,1\}\mu_i(\calX)$.

As already mentioned in \cref{sec:overview-expectation}, we have two regimes:
the case \(h p_1 \le 1/4\) and the case \(hp_1 > 1/4\).
To analyze both regimes, the following sample sets will be useful.

\paragraph{Partitioning samples in which $i$ can win.}
For all $1 \le t \le h$ and $0 \le s \le t$, we define
\begin{equation} \label{eq:C_def}
  \calC_{t,s} \coloneq  \{\bfx: x_i=t \text{ and } x_1=s\} \subseteq \calX.
\end{equation}
Observe that any $\bfx$ with $\mu_i(\bfx)>0$ has any opinion at most $x_i$ times.
Hence, the disjoint sets $\calC_{t,s}$, where $1 \le t \le h$ and $0 \le s \le t$, cover the support of $\mu_i$.
We also consider the following set
\[
  \calE\coloneq \{\bfx\in \calX: x_i=2, x_1=0 \text{ and } x_j \le 1 \text{ for } j \notin \{1,i\}\}.
\]
The mass of the set $\calE$ is used to bound the mass of $\calC_{1,0}$ and \(\calC_{1,1}\) sets.
The reason to single out $\calE$ is that the configurations in $\calC_{1,1}$ and $\calC_{1,0}$ on the support of $\mu_i$ have all sampled opinions distinct.
Moving the occurrence of opinion $1$ in a configuration of $\calC_{1,1}$ to opinion $i$ produces a configuration in $\calE$: opinion $i$ becomes the unique winner, and its score increases from zero to $1+\rho$.
Starting from $\calC_{1,0}$, moving an occurrence of another sampled opinion to $i$ also produces a configuration in $\calE$, now increasing the score from one to $1+\rho$.
The following lemma quantifies these two transports.

\begin{lemma}\label{lemma:witness-lemma}
  We have
  \[
    \mu_i(\calC_{1,0}) \le \frac{2 \, \mu_i(\calE)}{h(h-1)p_i}
    \qquad \text{ and } \qquad
    \mu_i(\calC_{1,1})=\frac{2 \, \rho \, \mu_i(\calE)}{h}.
  \]
\end{lemma}
\begin{proof}
  Let $Z_1, \ldots, Z_r$ be values sampled independently from the distribution $\bfp$.
  Define
  \begin{equation}
    \theta_r \coloneq  \pr*{Z_1, \ldots, Z_r \in [k] \setminus \{1,i\} \text{ and they are pairwise distinct}}.
  \end{equation}
  Note that $\theta_r \le \theta_{r-1}$ for any $r > 1$. We get that
  \begin{equation}\label{eq:e-prob}
    \pr*{\bfX\in \calE} = \binom{h}{2} p_i^2 \theta_{h-2},
  \end{equation}
  because if $\calE$ happens, exactly two samples take the value $i$, the value $1$ is never present, and the remaining $h-2$ samples are pairwise distinct within $[k] \setminus \{1,i\}$.
  Observe that $\mu_i(\calE) = \pr*{\bfX\in \calE}$, because for any $\bfx\in\calE$ we have $M(\bfx) = \{i\}$. Hence, \Cref{eq:e-prob} can be rewritten as
  \begin{equation}\label{eq:cale-prob}
    \theta_{h-2} = \frac{\pr*{\bfX\in \calE}}{\binom{h}{2} p_i^2} = \frac{2 \mu_i(\calE)}{h(h-1)p_i^2}.
  \end{equation}

  To show the first bound of the lemma, consider a sample $\bfx\in \calC_{1,0}$.
  By definition, $x_i = 1$. Moreover, if $\mu_i(\bfx)>0$, then $i \in M(\bfx)$ and no other coordinate can exceed $1$. Therefore, all sampled opinions are tied and $|M(\bfx)|=h$.
  Since $i$ is chosen uniformly at random to win from $M(\bfx)$, we get %
  \begin{align*}
    \mu_i(\calC_{1,0}) &= \pr{\bfX\in \calC_{1,0},\ i\in M(\bfX)} \cdot \frac{1}{h} \\
    &= h \, p_i \, \theta_{h-1} \cdot \frac{1}{h} \\
    &= p_i \, \theta_{h-1} \le p_i \, \theta_{h-2}.
  \end{align*}
  A similar reasoning in the case of $\calC_{1,1}$, where there are $h$ choices for the position of 1 and $h-1$ choices for the position of $i$, gives
  \begin{align*}
    \mu_i(\calC_{1,1}) &=  h(h-1) \, p_1 \, p_i \, \theta_{h-2} \cdot \frac{1}{h} = (h-1) \, p_1 \, p_i \, \theta_{h-2}.
  \end{align*}
  Using \Cref{eq:cale-prob} on both equations yields the desired statement.
\end{proof}

\subsubsection{\texorpdfstring
    {Case of sparse samples: $hp_1$ is small}
    {Case of sparse samples: h p1 is small}
}\label{sec:multinomial:small-sample}

In this section, we consider the case when $hp_1 \le 1/4$.
This roughly means the samples tend to have fairly few collisions.
Recall the definition of~$\calC_{t,s}$ in \Cref{eq:C_def}.
To analyze this case, we further define
\[
  \calT\coloneq  \bigcup_{t =1}^h \calC_{t,t} \qquad  \calT^+ \coloneq  \calT\setminus \calC_{1,1}
  \qquad
  \calB\coloneq   \bigcup_{t=2}^h \calC_{t,0},
\]
Here, $\calT$ consists of samples, where opinion $1$ and $i$ are tied.
The set $\calB$ supplies surplus score to compensate for these ties: on the support of $\mu_i$, its configurations have $x_1=0$ and $x_i\ge2$, hence score at least $1+\rho$.

\paragraph{Outline.}
Since all configurations outside $\calT$ on the support of $\mu_i$ have score at least one, we obtain
\begin{align*}
  S &\ge \mu_i(\calX\setminus\calT)+\rho\mu_i(\calB) \\
    &= \pr{\calW_i}-\mu_i(\calT)+\rho\mu_i(\calB).
\end{align*}
Thus, we must show that the surplus $\rho\mu_i(\calB)$ pays for the tied configurations and leaves a positive margin relative to $\pr{\calW_i}$.
Expanding \(\calT\), we obtain
\[
  S-\pr{\calW_i}
  \ge -\mu_i(\calT^+)
       +\underbrace{\rho\mu_i(\calB)-\mu_i(\calC_{1,1})}_{\zeta}.
\]
The transport bounds below show that $\mu_i(\calT^+)\le\zeta/3$ and $\zeta\ge\min\{h^2p_1,1\}\pr{\calW_i}/24$.
Together, these yield the required surplus:
\[
  S-\pr{\calW_i}\ge\frac{2}{3}\zeta
  \ge\frac{\min\{h^2p_1,1\}}{36}\pr{\calW_i}.
\]

\paragraph{Analysis.}
The small-sample assumption makes it possible to reach configurations in \(\calB\) by repeatedly moving occurrences of opinion $1$ to absent opinions.
We have ``enough'' absent configurations because at most $h$ opinions appear in a sample, so absent opinions carry total probability mass at least $1-hp_1\ge3/4$.
This provides enough ``destination mass''.
The next lemma makes this comparison precise.
Iterating it moves from $\calC_{t,s}$ to $\calC_{t,0}$, and in particular from the configurations in $\calC_{t,t}$ to configurations with score at least $1+\rho$ when $t\ge2$.

\begin{lemma} \label{lemma:transport-to-empty}
  Let $1 \le s \le t \le h$.
  If $hp_1 \le 1/4$, then
  \begin{equation}
    \mu_i(\calC_{t,s}) \le \frac{1}{3s} \cdot \mu_i(\calC_{t,s-1}). \label{eq:transport-to-empty}
  \end{equation}
\end{lemma}
\begin{proof}
  For $\bfx\in \calC_{t,s}$, define $Z(\bfx) = \{ j \in [k] : x_j = 0\}$ to be the indices of $\bfx$ that contain the value zero. Since $|\bfx|=h$ and $p_1$ is the largest entry in $\bfp$, we have
  \begin{equation*}
    \sum_{j \in Z(\bfx)} p_j = 1 - \sum_{j \notin Z(\bfx)} p_j \ge 1-hp_1.
  \end{equation*}
  In particular, $Z(\bfx) \neq \emptyset$.
  For each $j \in Z(\bfx)$,
  define $\bfx^{1 \to j} \coloneq  \bfx- \bfe_1 + \bfe_j =
  (x_1-1, \ldots, x_j+1, \ldots, x_k)$  to be the vector, where we move one unit from the first coordinate to the $j$th coordinate. Note that $x_1 = s > 0$ and $\bfx^{1 \to j} \in \calC_{t,s-1}$.
  For any $\bfx$ such that $\mu_i(\bfx)>0$, the definition of $\mu_i$ and applying \Cref{eq:density-multinomial} yields
  \begin{align*}
    \frac{\mu_i(\bfx^{1 \to j} )}{\mu_i(\bfx)} = \frac{\pr*{\bfX= \bfx^{1 \to j}}}{\pr*{\bfX= \bfx}} \cdot \frac{|M(\bfx)|}{|M(\bfx^{1 \to j} )|} %
    = \frac{s \, p_j}{p_1} \cdot \frac{|M(\bfx)|}{|M(\bfx^{1 \to j} )|}%
    \ge \frac{s \, p_j}{p_1},
  \end{align*}
  where the last line follows from the fact $|M(\bfx^{1 \to j})| \le |M(\bfx)| $.
  In particular, we get that
  \begin{equation}\label{eq:bfx1jlb}
    \mu_i(\bfx^{1 \to j}) \ge \mu_i(\bfx) \frac{s \, p_j}{p_1}.
  \end{equation}
  Now \Cref{eq:bfx1jlb} implies
  \begin{equation}\label{eq:xjtransport}
    \sum_{j \in Z(\bfx)} \mu_i(\bfx^{1 \to j}) \ge \mu_i(\bfx) \frac{s}{p_1} \sum_{j \in Z(\bfx)} p_j \ge \mu_i(\bfx) \frac{s(1-hp_1)}{p_1}.
  \end{equation}
  Finally, for $\bfy\in \calC_{t,s-1}$, define
  \begin{equation*}%
    N(\bfy) \coloneq  |\{(\bfx,j) : \bfx^{1 \to j} = \bfy, \bfx\in \calC_{t,s}, j \in Z(\bfx)\}|.
  \end{equation*}
  Observe that $N(\bfy)<h$.
  Indeed, for every pair $(\bfx,j)$ counted by $N(\bfy)$, we have $y_j=1$ and $j\ne i$.
  Moreover, once $\bfy$ and $j$ are fixed, the vector $\bfx$ is uniquely determined.
  Since $y_i=t\ge1$ and $\abs{\bfy}=h$, there are at most $h-1$ possible indices $j \neq i$ such that \(y_j = 1\).
  Now we have
  \begin{align*}
    \mu_i(\calC_{t,s}) &= \sum_{\bfx\in \calC_{t,s}} \mu_i(\bfx) \\
    &\le \frac{p_1}{s(1-hp_1)} \sum_{\bfx\in \calC_{t,s}} \sum_{j \in Z(\bfx)} \mu_i(\bfx^{1 \to j}) && \text{(by \Cref{eq:xjtransport})} \\
    &= \frac{p_1}{s(1-hp_1)}  \sum_{\bfy\in \calC_{t,s-1}} N(\bfy) \mu_i(\bfy) && \text{(by regrouping the summands)} \\
    &\le \frac{hp_1}{s(1-hp_1)} \sum_{\bfy\in \calC_{t,s-1}}\mu_i(\bfy) && \text{(\(N(\bfy) < h\))} \\
    &\le \frac{1}{3s} \mu_i(\calC_{t,s-1}). && \text{($hp_1 \le 1/4 \implies hp_1/(1-hp_1) \le 1/3$)}\qedhere
  \end{align*}
\end{proof}

A corollary of the above lemma relates the probability mass of \(\calT^+\) to that of \(\calB\).
\begin{corollary}\label{cor:t-bound}
  We have
  \(
    \mu_i(\calT^+) \le \frac19 \, \mu_i(\calB).
  \)
\end{corollary}
\begin{proof}
  We have that
  \begin{align*}
    \mu_i(\calT^+) &= \sum_{t=2}^{h} \mu_i(\calC_{t,t}) && \text{(by definition of $\calT^+$)}\\
    &\le \sum_{t=2}^{h} \frac{1}{3^t t!} \cdot \mu_i(\calC_{t,0}) && \text{(by iterating \Cref{eq:transport-to-empty})} \\
    &\le \pa{\sum_{r=2}^{h} \frac{1}{3^r r!}} \cdot  \pa{\sum_{t=2}^{h}  \mu_i(\calC_{t,0})} \\
    &\le  \sum_{r=2}^{h}\frac{1}{3^r r!} \cdot \mu_i(\calB) && \text{(by definition of $\calB$)} \\
    &\le  (e^{1/3}-1-\tfrac{1}{3}) \cdot \mu_i(\calB) && \text{(because $\sum_{r=0}^{\infty}\frac{1}{3^r r!} =e^{1/3}$)}\\
    &\le  \tfrac{1}{9} \, \mu_i(\calB). \qedhere
  \end{align*}
\end{proof}

The same transport argument also shows that the mass of configurations with $x_1=0$, namely $\calC_{1,0}\cup\calB$, controls the entire winning probability of opinion $i$.

\begin{corollary}\label{cor:wi-bound}
  We have $\pr*{\calW_i} \le e^{1/3} \left( \mu_i(\calC_{1,0}) + \mu_i(\calB) \right)$.
\end{corollary}
\begin{proof}
  Note that  $\{\calC_{t,s} : 1 \le t \le h, 0 \le s \le t\}$ is a partition of the support of $\mu_i$. Hence,
  \begin{align*}
    \pr*{\calW_i} &= \sum_{\bfx}\mu_i(\bfx) && \text{(by \Cref{eq:wi-identity})} \\
    &= \sum_{t=1}^h \sum_{s=0}^t \mu_i(\calC_{t,s}) && \text{(by partition of the support)}\\
    &\le \sum_{t=1}^h \sum_{s=0}^t \frac{1}{3^s s!} \cdot \mu_i(\calC_{t,0}) && \text{(by iterating \Cref{eq:transport-to-empty})} \\
    & \le \pa{\sum_{s=0}^\infty \frac{1}{3^s s!}} \cdot \pa{\sum_{t=1}^h  \mu_i(\calC_{t,0})} \\
    & = e^{1/3} \sum_{t=1}^h  \mu_i(\calC_{t,0}) && \text{(because $\sum_{t=0}^{\infty}\frac{1}{3^t t!} =e^{1/3}$)} \\
    &= e^{1/3} ( \mu_i(\calC_{1,0}) + \mu_i(\calB) ). && \text{(by definition of $\calB$)} \qedhere
  \end{align*}
\end{proof}

The remaining mass to bound is that of $\calC_{1,1}$: transport to an absent opinion only reaches $\calC_{1,0}$, whose score is one and provides no surplus.
Here we use the transport to $\calE$ from \cref{lemma:witness-lemma}.
Since $\mu_i(\calC_{1,1})=2\rho\mu_i(\calE)/h$, we can use only a fraction $2/h$ of the surplus $\rho\mu_i(\calE)$ to compensate for $\calC_{1,1}$.
Since $\calE\subseteq\calB$, this payment leaves ``free'' at least a fraction $1-2/h$ of the total surplus $\rho\mu_i(\calB)$; denote the remaining surplus by $\zeta$.
Moreover, $\zeta\ge\rho(1-2/h)\mu_i(\calE)$, so \cref{lemma:witness-lemma} also bounds $\zeta$ from below in terms of $\mu_i(\calC_{1,0})$.
Together with \cref{cor:wi-bound}, this converts the remaining surplus into a lower bound proportional to the entire winning probability $\pr{\calW_i}$.

\begin{lemma}\label{lemma:surplus-bounds}
  Define $\zeta \coloneq  \rho \, \mu_i(\calB) - \mu_i(\calC_{1,1})$ and $m \coloneq  \min\{h^2p_1, 1\}$.
  Then
  \begin{enumerate}[label=(\alph*)]
    \item $3\zeta \ge \mu_i(\calB)$, and
    \item $\zeta \ge \tfrac{m}{24} \pr*{\calW_i}$.
  \end{enumerate}
\end{lemma}
\begin{proof}
  By \Cref{lemma:witness-lemma}, we have
  \begin{equation}\label{eq:witness}
    \zeta = \rho \mu_i(\calB) - \frac{2\rho}{h} \mu_i(\calE).
  \end{equation}
  The first bound (a) of the lemma now follows from \Cref{eq:witness}, since
  $\calE\subseteq \calB$ and $h \ge 3$, and so
  \begin{align*}
    \zeta \ge \rho \mu_i(\calB) - \frac{2\rho}{h} \mu_i(\calB) = \rho\left(1-\frac{2}{h}\right) \mu_i(\calB) \ge \frac13 \mu_i({\calB}).
  \end{align*}
  Next, we show  an auxiliary bound
  \begin{equation}\label{eq:frac-surplus}
    \frac{9}{m} \zeta \ge \mu_i(\calC_{1,0}).
  \end{equation}
  To see why this holds, observe that
  \begin{align*}
    \zeta &= \rho \mu_i(\calB) - \frac{2\rho}{h} \mu_i(\calE) && \text{(by \Cref{eq:witness})}\\
    &\ge \left(1-\frac{2}{h} \right)\rho \mu_i(\calE) && \text{(by $\calE\subseteq \calB$)} \\
    &\ge  \left(1-\frac{2}{h} \right) \frac{\rho h(h-1)p_i}{2} \mu_i(\calC_{1,0}) && \text{(by \Cref{lemma:witness-lemma})} \\
    &= \frac{p_1(h-1)(h-2)}{2} \mu_i(\calC_{1,0})
    \ge \frac{m}{9} \mu_i(\calC_{1,0}),
  \end{align*}
  where the last inequality follows from the fact that $h^2/9 \le (h-1)(h-2)/2$ for $h \ge 3$.
  The second bound (b) of the lemma now follows, as
  \begin{align*}
    \pr*{\calW_i} &\le e^{1/3} \left( \mu_i(\calC_{1,0}) + \mu_i(\calB) \right) && \text{(by \Cref{cor:wi-bound})} \\
    &\le \zeta e^{1/3} \left( \frac{9}{m} + 3 \right) && \text{(by \Cref{eq:frac-surplus} and (a))} \\
    &\le \zeta \frac{24}{m}. && \text{(because $m \le 1$ and $e^{1/3} \le 2$)} \qedhere
  \end{align*}
\end{proof}

\begin{lemma}\label{lemma:sparse-core}
  If $hp_1 \le 1/4$, then
  \[
    S\ge \bigl(1+ 1/36 \cdot \min\{h^2p_1,1\}\bigr)\cdot \pr*{\calW_i}.
  \]
\end{lemma}
\begin{proof}
  By definition of $\Psi$, observe that for any $\bfx$ such that $\mu_i(\bfx)>0$ we have the following properties:
  \begin{enumerate}[label=(\roman*)]
    \item $\Psi(\bfx) = 0$ for any $\bfx\in \calT$,
    \item $\Psi(\bfx) \ge 1$ for any $\bfx\notin \calT$, and
    \item $\Psi(\bfx) \ge 1+\rho$ for any $\bfx\in \calB$.
  \end{enumerate}
  Using the above observations, we get that
  \begin{align*}
    S &= \sum_{\bfx}\Psi(\bfx) \mu_i(\bfx) \\
    &\ge \sum_{\bfx} \mu_i(\bfx) - \sum_{\bfx\in \calT} \mu_i(\bfx) + \sum_{\bfx\in \calB} \rho \mu_i(\bfx) && \text{(by properties (i)--(iii))}\\
    &=  \pr*{\calW_i} - \mu_i(\calT) + \rho \mu_i(\calB) && \text{(by \Cref{eq:wi-identity})} \\
    &= \pr*{\calW_i} - \mu_i(\calT^+) - \mu_i(\calC_{1,1}) + \rho \mu_i(\calB) && \text{(by definition of $\calT^+$ )} \\
    &= \pr*{\calW_i} - \mu_i(\calT^+) + \zeta && \text{(by definition of $\zeta$)} \\
    &\ge \pr*{\calW_i} - \frac{1}{3} \zeta + \zeta && \text{(by \Cref{cor:t-bound} and \Cref{lemma:surplus-bounds}(a))} \\
    &\ge \left( 1 + \frac{\min\{h^2p_1,1\}}{24} \cdot \frac{2}{3}\right) \cdot \pr*{\calW_i}. && \text{(by \Cref{lemma:surplus-bounds}(b) and rearranging)} \qedhere
  \end{align*}
\end{proof}

\subsubsection{\texorpdfstring{Case of dense samples: $hp_1$ is large}{Case of dense samples: h p1 is large}}\label{sec:multinomial:large-sample}

In this regime, the bound $1-hp_1$ no longer guarantees a positive constant mass of absent opinions, so the previous transport argument mapping to empty coordinates cannot be applied.
Instead, we move an occurrence of opinion $1$ directly to opinion $i$.
This increases $x_i-x_1$ by two, suggesting that we group configurations by this difference rather than by the two counts separately.
In particular, a configuration with a tie is transported to a configuration with score at least $1+\rho$.

Let $hp_1 \ge 1/4$. We now consider the following sample sets
\begin{align*}
  \calD_0=\{\bfx\in \calX: x_i = x_1 \}
  \qquad
  \calD_1=\{\bfx\in \calX: x_i = x_1 +1 \}
  \qquad
  \calD_{\ge2}=\{\bfx\in \calX: x_i \ge x_1 + 2 \}.
\end{align*}
Because $\mu_i(\bfx) > 0$ implies $i \in M(\bfx)$,  we have that
\begin{equation}\label{eq:prwi-mu-identity}
  \pr*{\calW_i} = \mu_i(\calD_0) +  \mu_i(\calD_1) +  \mu_i(\calD_{\ge 2}).
\end{equation}
Since $\Psi$ is 0 on $\calD_0$, $\Psi$ is always 1 on $\calD_1$, and $\Psi$ is at least $(1+\rho)$ on $\calD_{\ge2}$, we get that
\begin{align*}
  S &= \sum_{\bfx}\Psi(\bfx) \, \mu_i(\bfx) \\
  &\ge \mu_i(\calD_1) + (1+\rho) \mu_i(\calD_{\ge2}) = \pr*{\calW_i} + \rho \, \mu_i(\calD_{\ge2}) - \mu_i(\calD_0).
\end{align*}
The challenge is now to bound $\mu_i(\calD_{0})$ and $\mu_i(\calD_{\ge2})$.

\paragraph{Outline.}
We first prove the transport lemma stating that
\[
  \rho\mu_i(\calD_{\ge2})
  \ge \frac43\mu_i(\calD_0)+\frac1{40}\mu_i(\calD_1).
\]
Using the lower bound on \(S\), this gives two bounds on the score:
\begin{align*}
  S-\pr{\calW_i}
  &\ge \frac13\mu_i(\calD_0)+\frac1{40}\mu_i(\calD_1), \\
  S-\pr{\calW_i}
  &\ge \frac14\rho\mu_i(\calD_{\ge2}),
\end{align*}
where the second follows because $\mu_i(\calD_0)\le\frac34\rho\mu_i(\calD_{\ge2})$.
Adding these bounds and using $\rho>1$, we obtain
\begin{align*}
  2\bigl(S-\pr{\calW_i}\bigr)
  &\ge \frac13\mu_i(\calD_0)+\frac1{40}\mu_i(\calD_1)
       +\frac14\rho\mu_i(\calD_{\ge2}) \\
  &\ge \frac1{40}\bigl(\mu_i(\calD_0)+\mu_i(\calD_1)+\mu_i(\calD_{\ge2})\bigr)
   =\frac1{40}\pr{\calW_i}.
\end{align*}
Thus, $S\ge(1+1/80)\pr{\calW_i}$, which is the desired bound in this regime.

\paragraph{Analysis.}
As the first step toward this, we prove the following transport lemma.

\begin{lemma}\label{lemma:magical-transport-combination}
  If $hp_1 \ge 1/4$, then
  \[
    \rho \, \mu_i(\calD_{\ge2})
    \ge
    \frac43 \mu_i(\calD_0)+ \frac{1}{40} \mu_i(\calD_1).
  \]
\end{lemma}
\begin{proof}
  Consider the following sets
  \begin{align*}
    \calD_0^+ = \{\bfx\in \calD_0 : x_1 \ge 2 \} \quad \text{ and } \quad \calD_1^+ = \{\bfx\in \calD_1 : x_1 \ge 1 \}.
  \end{align*}
  Let $\bfx\in \calX$ such that $x_1 > 0$. Then define $\tau(\bfx) = (x_1-1, \ldots, x_i + 1, \ldots, x_k)$, i.e., one opinion 1 sample is switched to opinion $i$.
  Observe that if $i \in M(\bfx)$, then by \Cref{eq:density-multinomial} we get
  \begin{equation}\label{eq:sigma-transport}
    \rho \cdot \mu_i(\tau(\bfx))
    = \mu_i(\bfx) \cdot \frac{x_1}{x_i+1} \cdot \frac{|M(\bfx)|}{|M(\tau(\bfx))|}.
  \end{equation}
  By \Cref{eq:sigma-transport}, we get the following bounds:
  \begin{enumerate}[label=(\alph*)]
    \item If $\bfx\in \calD_0^+$, then  $\rho \mu_i(\tau(\bfx)) \ge \tfrac43 \mu_i(\bfx)$. Hence, $\rho \mu_i(\tau(\calD_0^+)) \ge \frac43 \mu_i(\calD_0^+)$.
    \item If $\bfx\in \calD_1^+$, then $\rho \mu_i(\tau(\bfx)) \ge \tfrac13 \mu_i(\bfx)$. Hence, $\rho \mu_i(\tau(\calD_1^+)) \ge \frac13 \mu_i(\calD_1^+)$.
  \end{enumerate}
  Moreover, we have that $\tau(\calD_0^+) \subseteq \calD_{\ge2}$ and $\tau(\calD_1^+) \subseteq \calD_{\ge2}$.
  Since $\tau$ is injective on $\calD_0^+$ and $\calD_1^+$, we get that
  $\tau(\calD_0^+)$, $\tau(\calD_1^+)$, $\calE$ are disjoint subsets of $\calD_{\ge2}$.
  Hence, we get that
  \begin{align*}
    \rho \mu_i(\calD_{\ge2}) &\ge \rho \left( \mu_i(\tau(\calD_0^+)) + \mu_i(\tau(\calD_1^+)) + \mu_i(\calE) \right) \\
    &\ge \frac43 \mu_i(\calD_0^+) + \frac13 \mu_i(\calD_1^+) + \rho \mu_i(\calE). && \text{(by  (a) and (b))}
  \end{align*}
  Next, we lower bound $\rho \mu_i(\calE)$.
  Observe that
  \begin{equation}\label{eq:c11-c10-identity}
    \mu_i(\calC_{1,1}) = \mu_i(\calD_0 \setminus \calD_0^+) \quad \text{ and } \quad \mu_i(\calC_{1,0}) = \mu_i(\calD_1\setminus\calD_1^+).
  \end{equation}
  Applying \Cref{lemma:witness-lemma}, we get that
  \begin{align*}
    \frac{1}{40} \mu_i(\calC_{1,0}) + \frac43 \mu_i(\calC_{1,1}) &\le \mu_i(\calE) \cdot \left( \frac{2}{40h(h-1)p_i} + \frac{8\rho}{3h}  \right) && \text{(by \Cref{lemma:witness-lemma})} \\
    &\le  \mu_i(\calE) \cdot \left( \frac{8hp_1}{40h(h-1)p_i} + \frac{8\rho}{3h} \right) && \text{(by $hp_1 \ge 1/4$)} \\
    &=  \mu_i(\calE) \rho \cdot \left( \frac{1}{5(h-1)} + \frac{8}{3h} \right) && \text{(by $\rho = p_1/p_i$)} \\
    &\le  \mu_i(\calE) \rho  \left( \frac{1}{10} + \frac{8}{9} \right) \le \mu_i(\calE) \rho && \text{(by $h \ge 3$)}.
  \end{align*}
  The claim of the lemma now follows by combining the above, as
  \begin{align*}
    \rho \mu_i(\calD_{\ge2}) &\ge  \frac43 \mu_i(\calD_0^+) + \frac13 \mu_i(\calD_1^+) + \rho \mu_i(\calE) \\
    &\ge \frac43 \mu_i(\calD_0^+) + \frac13 \mu_i(\calD_1^+) +  \frac{1}{40} \mu_i(\calC_{1,0}) + \frac43 \mu_i(\calC_{1,1}) &&  \text{(by the above)} \\
    &\ge \frac43 \mu_i(\calD_0^+) + \frac13 \mu_i(\calD_1^+) +  \frac{1}{40} \mu_i(\calD_1 \setminus \calD_1^+) + \frac43 \mu_i(\calD_0 \setminus \calD_0^+) &&  \text{(by \Cref{eq:c11-c10-identity})} \\
    &\ge \frac43 \mu_i(\calD_0) +  \frac{1}{40} \mu_i(\calD_1).  \qedhere
  \end{align*}
\end{proof}

\begin{lemma} \label{lemma:dense-score}
  If $hp_1 \ge 1/4$, then
  \[
    S\ge \bigl(1+1/80\bigr)\cdot \pr*{\calW_i}.
  \]
\end{lemma}
\begin{proof}
  Recall that
  \begin{equation*}
    S = \sum_{\bfx\in \calX}\Psi(\bfx) \mu_i(\bfx) \ge \pr*{\calW_i} + \rho \mu_i(\calD_{\ge2}) - \mu_i(\calD_0).
  \end{equation*}
  Thus, by rearranging, we have that
  \begin{equation}\label{eq:s-wi-dif}
    S - \pr*{\calW_i} \ge \rho\mu_i(\calD_{\ge2}) - \mu_i(\calD_0)
  \end{equation}
  Applying \Cref{lemma:magical-transport-combination} to \Cref{eq:s-wi-dif}, we have
  \begin{equation}\label{eq:m1}
    S - \pr*{\calW_i} \ge \frac{1}{3} \mu_i(\calD_0) + \frac{1}{40} \mu_i(\calD_1)
  \end{equation}
  By \Cref{lemma:magical-transport-combination}, we also have
  $    \frac34
  \rho \mu_i(\calD_{\ge2})
  \ge
  \mu_i(\calD_0)$, so
  \begin{equation}\label{eq:m2}
    S - \pr*{\calW_i} \ge \rho\mu_i(\calD_{\ge2}) - \mu_i(\calD_0) \ge \frac14 \rho \mu_i(\calD_{\ge2})
  \end{equation}
  Hence, combining the above
  \begin{align*}
    2(S - \pr*{\calW_i}) &\ge  \frac{1}{3} \mu_i(\calD_0) + \frac{1}{40} \mu_i(\calD_1) + \frac14 \rho \mu_i(\calD_{\ge2}) && \text{(by \Cref{eq:m1,eq:m2})} \\
    &\ge
    1/40\Big(  \mu_i(\calD_0) +  \mu_i(\calD_1) + \mu_i(\calD_{\ge2}) \Big) && \text{(because $\rho > 1$)} \\
    &= 1/40 \cdot \pr*{\calW_i} && \text{(by \Cref{eq:prwi-mu-identity})}.
  \end{align*}
  The claim now follows by rearranging, as
  \[
    S \ge \pr*{\calW_i} + 1/80 \cdot \pr*{\calW_i} = \pr*{\calW_i} \cdot (1 + 1/80). \qedhere
  \]
\end{proof}

\subsubsection{\texorpdfstring{Proof of \Cref{thm:multinomial:growth}}{Proof of the expected-growth theorem}}

Combining \Cref{lemma:sparse-core} and \Cref{lemma:dense-score} yields
\[
  S \ge (1+1/80 \cdot \min\{h^2 p_1, 1\}) \cdot \pr*{\calW_i}.
\]
By \Cref{lemma:score-identity}, we have
\begin{align*}
  \pr*{\calW_1}-\pr*{\calW_i} &= (\rho-1)S \\
  &\ge \left(\frac{p_1}{p_i} -1\right) (1+1/80 \cdot \min\{h^2p_1, 1\}) \cdot \pr*{\calW_i}
\end{align*}
and rearranging gives the claimed bound.

\subsubsection{A small corollary}

\begin{corollary}\label{cor:multinomial:weak-comparison}
  Suppose $p_1=\max_{i \in [k]} p_i$.
  Then, for every $i \in [k]$,
  \[
    \pr*{\calW_i}
    \le \frac{p_i}{p_1}\pr{\calW_1}
    \le \pr*{\calW_1}.
  \]
  In particular, it holds that $\pr*{\calW_1}\ge p_1$.
\end{corollary}

\begin{proof}
  If \(p_i > 0\), the first inequality follows
  from \Cref{thm:multinomial:growth}.
  For $p_i=0$, it follows from $\pr{\calW_i}=0$;
  for $i=1$, it is an equality.
  The second inequality follows from $p_i\le p_1$.
  Summing the first inequality gives
  \[
    1=\sum_i\pr{\calW_i}
    \le \frac{\pr{\calW_1}}{p_1}\sum_i p_i
    =\frac{\pr{\calW_1}}{p_1}.
  \]
\end{proof}

\subsection{Majorization}
\label{sec:majorization}
We construct a coupling of synchronous \hmaj and \voter that
preserves the majorization order at every round. The key ingredient compares
one round of \hmaj and \((h-1)\)-\textsf{Majority} starting from the same opinion distribution.

Recall that $\bfp$ is the
current opinion distribution, $\bfX\sim\Multinom(h;\bfp)$ is a multinomial sample count vector,
and $\maj(\bfX)$ is its winning opinion, with uniform tie-breaking.
For sample count vectors, uppercase letters such as $\bfX$ denote random
vectors, whereas lowercase letters such as $\bfx$ denote their fixed values.

We retain the sample-space and mode notation from
\Cref{sec:multinomial_utils}. When the sample size varies, write
\[
 \calX_r=\{\bfx\in\bbZ_{\ge0}^k:\abs{\bfx}=r\},\qquad
 \nu_r(\bfx)=\pr{\bfX=\bfx}
 =\binom{r}{\bfx}\prod_{i=1}^k p_i^{x_i},
\]
where $\bfX$ has distribution $\Multinom(r;\bfp)$ and $0^0=1$.
Thus $\calX_h=\calX$. Throughout this subsection, $h\ge2$;
we allow $r\ge1$ to include the comparison with $h-1$ when $h=2$.
Using the existing mode set $M(\bfx)$, define the second layer by
\[
 L(\bfx)=\{i\in[k]:x_i=\norm{\bfx}_\infty-1\}.
\]
We add a sample-size superscript to the measures and winner probabilities
of \Cref{sec:multinomial_utils}:
\[
 \mu_i^{(r)}(\bfx)=\pr{\calW_i\cap\{\bfX=\bfx\}}
 =\nu_r(\bfx)\frac{\ind{i\in M(\bfx)}}{|M(\bfx)|},\qquad
 q_i^{(r)}(\bfp)=\pr{\calW_i}.
\]
Here $\calW_i$ is understood for sample size $r$. By the same identity as
\eqref{eq:wi-identity},
\[
 q_i^{(r)}(\bfp)=\sum_{\bfx\in\calX_r}\mu_i^{(r)}(\bfx),\qquad
 \bfq^{(r)}(\bfp)=(q_i^{(r)}(\bfp))_{i\in[k]}.
\]
In particular, $\mu_i^{(h)}=\mu_i$ and $\bfq^{(h)}(\bfp)=\bfq(\bfp)$.
In the proofs below, we fix $\bfp$ and abbreviate
$q_i^{(r)}(\bfp)$ and $\bfq^{(r)}(\bfp)$ as
$q_i^{(r)}$ and $\bfq^{(r)}$, respectively.
For the synchronous process on $n$ vertices,
\[
 n\bfp(t+1)\mid\bfp(t)=\bfp\ \sim\Multinom(n;\bfq^{(r)}(\bfp)),
 \qquad \expecton{t}{\bfp(t+1)}=\bfq^{(r)}(\bfp).
\]
All samples and tie-breaking choices are independent across vertices.

\begin{definition}[Majorization] \label{def:majorization}
For nonnegative vectors $\bfx,\bfy\in\bbR^k$ satisfying
$\sum_{i=1}^k x_i=\sum_{i=1}^k y_i$,
let $\bfx^\downarrow,\bfy^\downarrow$ be their nonincreasing rearrangements.
We write $\bfx\succeq\bfy$ if
\[
 \sum_{i=1}^b x_i^\downarrow\ge\sum_{i=1}^b y_i^\downarrow
 \quad\text{for every }b\in[k].
\]
\end{definition}

\paragraph{The two processes.}
Write $\bfp^{(h)}(t)$ and $\bfp^{(1)}(t)$ for the empirical
opinion distributions of synchronous \hmaj and \voter on the same
number $n$ of vertices with opinion set $[k]$. \voter is exactly
\(1\)-\textsc{majority}, so we denote its empirical distribution by $\bfp^{(1)}(t)$.
In \voter, each vertex
independently copies the opinion of a uniformly sampled vertex, with
replacement. Thus its next-round opinion count vector, given current
distribution $\bfr$, has distribution $\Multinom(n;\bfr)$.
The corresponding distribution for \hmaj from $\bfp$ is
$\Multinom(n;\bfq^{(h)}(\bfp))$.

\begin{theorem}[Coupling of \hmaj and \voter]
\label{thm:coupling-voter}
Let $h\ge2$, and let the initial empirical distributions satisfy
$\bfp^{(h)}(0)\succeq\bfp^{(1)}(0)$.
There exists a coupling of the two processes, with their correct marginal
laws, such that
\[
 \pr{\bfp^{(h)}(t)\succeq\bfp^{(1)}(t)
       \text{ for every integer }t\ge0}=1.
\]
In particular, this holds when the two processes start from the same
configuration.
\end{theorem}

The next lemma shows that increasing the sample size makes the winner
distribution larger in the majorization order; this is the key comparison
used to prove \Cref{thm:coupling-voter}.

\begin{lemma}
\label{lem:sample-size-majorization}
For every probability vector $\bfp$ and every integer $h\ge2$, it holds that $\bfq^{(h)}(\bfp)\succeq\bfq^{(h-1)}(\bfp)$.
More precisely, if $p_1\ge\cdots\ge p_k$, then
\[
 q_1^{(h)}(\bfp)\ge\cdots\ge q_k^{(h)}(\bfp),
 \qquad
 q_1^{(h-1)}(\bfp)\ge\cdots\ge q_k^{(h-1)}(\bfp),
\]
and for every $b\in[k]$,
\[
 \sum_{i=1}^b q_i^{(h)}(\bfp)\ge
 \sum_{i=1}^b q_i^{(h-1)}(\bfp)
\]
\end{lemma}

To compare prefix sums in the original opinion order, we first show that
winner probabilities preserve the ordering of the current opinion frequencies.

\begin{lemma}
\label{lem:majorization-order}
Let $h\ge2$ and $r\in\{h-1,h\}$. If $p_i\ge p_j$, then
$q_i^{(r)}(\bfp)\ge q_j^{(r)}(\bfp)$.
\end{lemma}
\begin{proof}
As in \Cref{sec:multinomial_utils}, let $\bfx^{i\leftrightarrow j}$ denote
$\bfx$ with coordinates $i$ and $j$ exchanged, and set
\[
 g(\bfx)=\frac{\ind{i\in M(\bfx)}-\ind{j\in M(\bfx)}}{|M(\bfx)|}.
\]
Swapping coordinates negates $g$, so $g(\bfx^{i\leftrightarrow j})=-g(\bfx)$.
Using \eqref{eq:wi-identity} with sample size $r$,
\[
 q_i^{(r)}-q_j^{(r)}
 =\sum_{\bfx\in\calX_r}(\mu_i^{(r)}(\bfx)-\mu_j^{(r)}(\bfx))
 =\sum_{\bfx\in\calX_r}\nu_r(\bfx)g(\bfx).
\]
Since $\bfx\mapsto\bfx^{i\leftrightarrow j}$ is a bijection of
$\calX_r$, averaging this sum with its reindexed version and using
$g(\bfx^{i\leftrightarrow j})=-g(\bfx)$ gives
\begin{align*}
 q_i^{(r)}-q_j^{(r)}
 &=\frac12\sum_{\bfx\in\calX_r}
 \left(\nu_r(\bfx)g(\bfx)
 +\nu_r(\bfx^{i\leftrightarrow j})g(\bfx^{i\leftrightarrow j})\right) \\
 &=
 \frac12\sum_{\bfx\in\calX_r}
 (\nu_r(\bfx)-\nu_r(\bfx^{i\leftrightarrow j}))g(\bfx).
\end{align*}
If $g(\bfx)>0$, then $x_i>x_j$. Put $d=x_i-x_j>0$.
Using $p_i\ge p_j$, we have
\[
 \nu_r(\bfx)-\nu_r(\bfx^{i\leftrightarrow j})=
 \binom{r}{\bfx}
 \prod_{\ell\ne i,j}p_\ell^{x_\ell}
 (p_i p_j)^{x_j}(p_i^d-p_j^d)\ge0.
\]
In particular, we have
$(\nu_r(\bfx)-\nu_r(\bfx^{i\leftrightarrow j}))g(\bfx)\ge0$.
If $g(\bfx)<0$, the reversed argument makes the probability difference
nonpositive, so again
$(\nu_r(\bfx)-\nu_r(\bfx^{i\leftrightarrow j}))g(\bfx)\ge0$.
Thus every summand
in the preceding expression for the winner-probability difference is
nonnegative, and hence
\[
 q_i^{(r)}-q_j^{(r)}=
 \frac12\sum_{\bfx\in\calX_r}
 (\nu_r(\bfx)-\nu_r(\bfx^{i\leftrightarrow j}))g(\bfx)\ge0.
\]
\end{proof}

The next lemma relates sample sizes $h$ and $h-1$ by deleting one uniformly
chosen sample, allowing us to compare their winner probabilities.

\begin{lemma}[Deletion construction]
\label{lem:majorization-deletion}
Let $h\ge2$ and $\bfX\sim\Multinom(h;\bfp)$. Given $\bfX$, choose $I$
with $\pr{I=i\mid\bfX}=X_i/h$, and let $\bfY=\bfX-\bfe_I$,
where $\bfe_i$ is the $i$th coordinate vector. Then the distribution of
$\bfY$ is identical to $\Multinom(h-1;\bfp)$.
\end{lemma}
\begin{proof}
Generate $h$ independent opinions with distribution $\bfp$ and delete a
uniformly selected sample position, independently of the sampled opinions.
Given their counts, the deleted opinion has distribution $X_i/h$.
Given the deleted position, the remaining $h-1$ opinions are still
independent with distribution $\bfp$. Their count vector is $\bfY$.
\end{proof}

Now we are ready to prove the key result (\cref{lem:sample-size-majorization}).
\begin{proof}[Proof of \Cref{lem:sample-size-majorization}]
By relabeling, assume $p_1\ge\cdots\ge p_k$. By
\Cref{lem:majorization-order}, it suffices to compare prefix sums.
Fix $b\in[k]$ and put $F_b(\bfx)=\tfrac{|M(\bfx)\cap[b]|}{|M(\bfx)|}$.
In view of the deletion construction of \Cref{lem:majorization-deletion}, define $D_b(\bfx)=\sum_{i:x_i>0}\frac{x_i}{h}
    (F_b(\bfx)-F_b(\bfx-\bfe_i))$.
Then, \cref{lem:majorization-deletion} yields
\begin{equation}\label{eq:majorization-difference}
 \sum_{i=1}^b(q_i^{(h)}-q_i^{(h-1)})=\expect{D_b(\bfX)}.
\end{equation}
We first evaluate $D_b$ for a fixed $\bfx\in\calX_h$.
Deleting a non-top opinion leaves $M(\bfx)$ unchanged.
If $a=|M(\bfx)|\ge2$, deleting a top opinion $i$ leaves
$M(\bfx)\setminus\{i\}$ as the top set. We obtain
\[
 D_b(\bfx)=\frac{\norm{\bfx}_\infty}{h}\sum_{i\in M(\bfx)}
 \left(\frac{|M(\bfx)\cap[b]|}{a}
 -\frac{|M(\bfx)\cap[b]|-\ind{i\le b}}{a-1}\right)=0.
\]
If there is a unique top opinion $i$ and $L(\bfx)=\varnothing$,
deleting $i$ also leaves the top set unchanged, so $D_b(\bfx)=0$.

In the remainder of the proof, we consider count vectors $\bfx$ satisfying
$M(\bfx)=\{i\}$ and $L(\bfx)\ne\varnothing$, since all other count vectors
contribute zero to $\expect{D_b(\bfX)}$. For such a vector, deleting $i$
creates the top set $\{i\}\cup L(\bfx)$. Consequently,
\begin{equation}\label{eq:majorization-local}
 D_b(\bfx)=\frac{x_i}{h(|L(\bfx)|+1)}
 \sum_{j\in L(\bfx)}(\ind{i\le b}-\ind{j\le b}).
\end{equation}
Consider the finite set
\[
 \calT=\{(\bfx,i,j):\bfx\in\calX_h,\ M(\bfx)=\{i\},\ j\in L(\bfx)\}.
\]
For each triple, $x_i=x_j+1$, and the map $\Phi\colon (\bfx,i,j)\mapsto (\bfx',j,i)$ for $\bfx'=\bfx-\bfe_i+\bfe_j$
is an involution of $\calT$. It exchanges the counts of $i$ and $j$.
In particular, $x'_j=x_i$ and $|L(\bfx')|=|L(\bfx)|$.
Pairing terms in \eqref{eq:majorization-local} gives
\begin{equation}\label{eq:majorization-pairing}
 2\expect{D_b(\bfX)}=
 \sum_{(\bfx,i,j)\in\calT}
 \frac{x_i(\nu_h(\bfx)-\nu_h(\bfx'))
       (\ind{i\le b}-\ind{j\le b})}{h(|L(\bfx)|+1)}.
\end{equation}
This sum runs over all triples, so each pair is counted twice: once at
$(\bfx,i,j)$ and once at its image $(\bfx',j,i)=\Phi(\bfx,i,j)$.
The factor two on the left accounts for these two occurrences of the
same paired contribution.
Since the multinomial coefficient is unchanged by the swap,
\[
 \nu_h(\bfx)-\nu_h(\bfx')=
 \binom{h}{\bfx}
 \prod_{\ell\ne i,j}p_\ell^{x_\ell}
 (p_i p_j)^{x_j}(p_i-p_j).
\]
If $i\le b<j$, then $p_i\ge p_j$ and both factors in the numerator
of \eqref{eq:majorization-pairing} are nonnegative.
If $j\le b<i$, both factors are nonpositive.
Otherwise the indicator difference is zero. Hence every summand is
nonnegative, and \eqref{eq:majorization-difference} proves the required
prefix inequality. Both winner distributions have total mass one,
which completes the majorization assertion.
\end{proof}

\begin{corollary}\label{cor:majorization-voter}
For integers $h\ge r\ge2$,
$\bfq^{(h)}(\bfp)\succeq\bfq^{(r)}(\bfp)\succeq\bfp$.
Moreover, $\bfq^{(2)}(\bfp)=\bfp$.
\end{corollary}
\begin{proof}
For two samples, choosing the more frequent opinion and breaking
a tie uniformly is equivalent in distribution to choosing a uniformly
random sample position; hence $\bfq^{(2)}(\bfp)=\bfp$.
Iterating \Cref{lem:sample-size-majorization} from sample size two proves the claim.
\end{proof}

The next lemma turns a majorization comparison of probability vectors into
a coupling of their multinomial count vectors that preserves the same order,
as in Berenbrink et al.~\cite{ignore_or_comply}.

\begin{lemma}[Multinomial coupling {\cite[Section 2.2 and Appendix A.3]{ignore_or_comply}}]\label{lem:multinomial-coupling}
Let $\bfa,\bfb\in[0,1]^k$ be probability vectors with
$\bfa\succeq\bfb$, and let $n\ge1$. There exists a jointly distributed
pair $(\bfU,\bfV)$ whose marginal distributions are
$\Multinom(n;\bfa)$ and $\Multinom(n;\bfb)$, respectively, such that
$\pr{\bfU\succeq\bfV}=1$.
\end{lemma}

We now prove the coupling theorem, \Cref{thm:coupling-voter}.

\begin{proof}[Proof of \Cref{thm:coupling-voter}]
For any two current empirical distributions $\bfp\succeq\bfr$,
\Cref{cor:majorization-voter} gives
\[
 \bfq^{(h)}(\bfp)\succeq\bfp\succeq\bfr.
\]
By \Cref{lem:multinomial-coupling}, we can therefore choose a joint
next-round count vector $(\bfU,\bfV)$ with respective marginal
 distributions $\Multinom(n;\bfq^{(h)}(\bfp))$ and
$\Multinom(n;\bfr)$, supported on $\bfU\succeq\bfV$.
Set the next empirical distributions to $\bfU/n$ and $\bfV/n$.
Since scaling both vectors by $1/n$ preserves majorization, this is a
one-round coupling with the correct transition laws and with
$\bfU/n\succeq\bfV/n$ almost surely.

For each ordered pair of current states, fix one such joint transition
law. The empirical state space is finite, so these choices define a
joint Markov transition rule. Starting from the prescribed initial
states, apply this rule at every round, using fresh randomness
conditional on the current pair of states. Each marginal process has
its original transition law. Induction gives
$\pr{\bfp^{(h)}(t)\succeq\bfp^{(1)}(t)}=1$ for every
$t\ge0$. The union of the exceptional events over the countably many
rounds has probability zero, proving the simultaneous assertion.

If vertex-level configurations are desired, for each coupled count
vector choose a uniform assignment of its opinions to the
$n$ vertices, separately for each process. Independent identically
distributed vertex updates are uniform conditional on their count
vector, so this also recovers the correct configuration-level laws.
\end{proof}

\begin{remark}[Scope of the comparison] \label{rem:majorization-scope}
The sample-size comparison in \Cref{lem:sample-size-majorization} compares
expected next-round distributions from the same $\bfp$. Although it yields
the \voter coupling in \Cref{thm:coupling-voter}, it does not by itself yield
a trajectory coupling between \hmaj and \((h-1)\)-\textsc{majority}. In particular, different starting distributions
cannot be substituted into the statement merely because they are ordered
by majorization. For example, with
$\bfp=(1/2,1/2,0)$ and $\bfr=(1/2,1/4,1/4)$, we have
$\bfp\succeq\bfr$, whereas
\[
 \bfq^{(4)}(\bfp)=(1/2,1/2,0)\not\succeq
 \bfq^{(3)}(\bfr)=(9/16,7/32,7/32).
\]
\end{remark}

The coupling transfers the known reduction in the number of opinions in
\voter to \hmaj.

\begin{corollary}[Number of remaining opinions]
\label{cor:majorization-largest-opinion}
There exists an absolute constant $C>0$ such that the following holds.
Consider \hmaj with $h\ge2$ on $n\ge2$ vertices, starting from an arbitrary
initial configuration. For every fixed integer $1\le T\le n\log n$,
with high probability,
\[
 |\{i\in[k]:p_i^{(h)}(T)>0\}|\le\frac{C n\log n}{T}.
\]
In particular, $\pmax^{(h)}(T)\ge T/(C n\log n)$ with high probability,
where $\pmax^{(h)}(T)=\max_{i\in[k]}p_i^{(h)}(T)$.
\end{corollary}
\begin{proof}
Couple \hmaj and \voter from the same initial configuration using
\Cref{thm:coupling-voter}. If \voter has $m$ remaining opinions at round
$T$, then
\[
 1=\sum_{i=1}^m\bigl(\bfp^{(1)}(T)^\downarrow\bigr)_i
 \le\sum_{i=1}^m\bigl(\bfp^{(h)}(T)^\downarrow\bigr)_i\le1.
\]
Hence, almost surely,
\[
 |\{i\in[k]:p_i^{(h)}(T)>0\}|
 \le |\{i\in[k]:p_i^{(1)}(T)>0\}|.
\]
By \cite[Lemma 3]{ignore_or_comply}, \voter reaches at most
$m$ opinions within $O((n/m)\log n)$ rounds with probability at least
$1-1/n$, from any initial configuration. Taking
$m=\min\{n,\lceil C n\log n/(2T)\rceil\}$, choosing the constant $C$
in the statement sufficiently large, gives at most $C n\log n/T$
opinions at round $T$ with high
probability (the case $m=n$ is deterministic). The coupling comparison
proves the claim.
\end{proof}

\section{Plurality consensus}\label{sec:plurality-consensus}

In this section, we study the consensus time starting from a configuration in which symmetry is broken, that is, the majority opinion has a significant advantage over every other opinion.
Without loss of generality, we assume that the process starts at round~$0$.
We assume that opinion~$1$ is the majority opinion at round~$0$, but contrary to \Cref{sec:expected_growth}, we do not assume that the other opinions are sorted in decreasing order.
We write
\begin{equation*}
    \Delta_i(t) \coloneq  p_1(t) - p_i(t) \qquad \text{ and } \qquad \Delta(t) \coloneq  \min_{i > 1} \Delta_i(t)
\end{equation*}
to denote the {\em gaps} with respect to opinion 1.
Fix a sufficiently small absolute constant $c>0$ and then a sufficiently large absolute constant $C>0$ (depending on $c$).
We define the \emph{growth-rate parameter}~$\gamma(t)$ by
\begin{equation*}
    \gamma(t) \coloneq \min\{h^2p_1(t),1\}.
\end{equation*}
Under the hypotheses of \Cref{lemma:gap-drift}, the relevant conditional growth factor within an amplification phase is at least $1+c\gamma(0)$.
Accordingly, the time scale of a phase is proportional to $1/(c\gamma(0))$.
We say that the configuration $\bfp(t)$ at round~$t$ is \emph{good} if $p_1(t)>0$ and
\begin{equation}\label{eq:plurality-gap-assumption}
  \Delta(t) \ge C\sqrt{\frac{p_1(t)}{\gamma(t)}\cdot\frac{\ln n}{n}}. \tag{$\star$}
\end{equation}
Thus, the scale in the gap condition is expressed directly as $p_1(t)/\gamma(t)$.

\paragraph{Martingale notation.}
For a square-integrable martingale $(M_s)_{s\ge0}$ adapted to a filtration $(\calG_s)_{s\ge0}$, its \emph{increments} are $\xi_s=M_s-M_{s-1}$, for $s\ge1$.
Its \emph{predictable quadratic variation} is
\[
  V_s=\sum_{u=1}^s\expect*{\xi_u^2\mid\calG_{u-1}},
  \qquad V_0=0.
\]
Since the increments have conditional mean zero, $V_s$ is the sum of their conditional variances.
It measures the accumulated conditional variance and is the quantity appearing in Freedman's inequality (\cref{lem:tool:freedman}).

We will show the following result, which is just a reformulation of \cref{thm:plurality-consensus-intro}.

\begin{theorem}\label{thm:plurality-consensus}
  Let $h \ge 3$. If the configuration~$\bfp(0)$ is good in the sense of \eqref{eq:plurality-gap-assumption}, then plurality consensus is reached in
  \begin{equation*}
      O\left(\frac{\ln n}{\gamma(0)}\right)
  \end{equation*}
  rounds with probability at least $1-1/n$.
\end{theorem}

\subsection{Proof outline}

We start from a configuration at round~$0$ that is good in the sense of \Cref{eq:plurality-gap-assumption}.
We show that the density~$p_1(t)$ of the plurality opinion increases quickly until it reaches either twice its initial value~$p_1(0)$ or a sufficiently large constant (say, $3/5$).
We refer to the corresponding time period as an \emph{amplification phase}.
Once $p_1(t)$ reaches $3/5$, we show that the process quickly converges to consensus, and we refer to the corresponding time period as the \emph{final phase}.
Our analysis thus reduces to establishing high-probability bounds on the duration of each type of phase.
To this end, our main technique is to consider a fine-grained filtration in which the new opinions of the nodes are revealed one by one in each round.
We then control the process by applying Freedman's inequality to this ``node-exposure'' filtration.

To iterate the argument, we show that the configuration at the end of each amplification phase remains good. We also show that it satisfies an additional condition on the gap, namely $\Delta(t) \geq p_1(t)/32$. This property allows us to obtain tighter bounds on the duration of every amplification phase after the first.

Freedman's inequality (\cref{lem:tool:freedman}) applies to martingales.
That is why, in this section, we define martingales describing the evolution of \(p_1(t)\) and \(\Delta(t)\) by normalizing these quantities by their expected growth.
The martingales do not grow over time, implying that the two processes of interest \(p_1(t)\) and \(\Delta(t)\) accumulate growth factors.

\subsection{Amplification phases}

We say that the phase started at round 0 \emph{succeeds} at round $t$ when either of the following occurs:
\begin{enumerate}
  \item $p_1(t) \ge 3/5$, or
  \item $p_1(t) \ge 2p_1(0)$, $\Delta(t) \ge p_1(t)/32$, and the configuration $\bfp(t)$ is good (see \Cref{eq:plurality-gap-assumption}).
\end{enumerate}
\begin{restatable}{lemma}{phaselemma}
  \label{lemma:phase-lemma}
  If the configuration~$\bfp(0)$ is good, then the phase succeeds within
  \begin{center}
  \(
    \displaystyle
    T \coloneq  2\left\lceil \frac{32}{c\gamma(0)}
    \ln \left(\frac{2 \, p_1(0)}{\Delta(0)}\right)\right\rceil
  \)
  \end{center}
  rounds with probability at least $1-1/n^2$.
\end{restatable}

The remainder of the section is dedicated to the proof of \Cref{lemma:phase-lemma}, which is somewhat involved. We write~$p \coloneq  p_1(0)$, and re-use some notations from \Cref{sec:expected_growth}: let $\calF_t$ be the $\sigma$-algebra generated by the first $t$ rounds, and let
\begin{equation*} %
    \bfq(t) = \expecton{t}{\bfp(t+1)}
\end{equation*}
be the expected densities in round~$t+1$, conditional on the history up to round~$t$.
Furthermore, whenever $p_1(t) >0$, define
\begin{equation*}%
  \eta(t) \coloneq  \frac{q_1(t)}{p_1(t)},
\end{equation*}
with the convention that~$\eta(t) = 0$ if $p_1(t) = q_1(t) = 0$.
By \Cref{cor:multinomial:weak-comparison}, we have that $\eta(t)\ge 1$ whenever opinion 1 is the unique maximum opinion at round~$t$.
Finally, we define
\begin{equation*}
    H(t) \coloneq  \prod_{s=0}^{t-1} \eta(s)
\end{equation*}
the accumulated growth of $\eta$ (in other words, $H(0) = 1$ and $H(t+1) = \eta(t) \, H(t)$ for every~$t \geq 0$).
We will use the values~$H(\cdot)$ for normalisation.

\paragraph{Stopping time.}
We define the stopping time $\tau$ as the first round $t<T$ in which one of the following conditions holds:
\begin{enumerate}[label=(\roman*)]
  \item $p_1(t) \ge 3/5$,
  \item opinion 1 is not the unique maximum opinion, i.e., $p_1(t) \le p_i(t)$ for some $i > 1$, or
  \item $p_1(t)/H(t) \notin [p/2, 2p]$.
\end{enumerate}
If none of the above occur, we set $\tau = T$. Observe that $\tau$ is a stopping time.

\paragraph{Normalized densities and gaps.}
For $t \le \tau$, we define the \emph{normalized density} $P(t)$ of opinion 1, and the \emph{normalized gap} $D_i(\cdot)$ between opinion 1 and opinion $i$, as
\begin{equation*} %
  P(t) \coloneq  \frac{p_1(t)}{H(t)}  \qquad \text{ and } \qquad D_i(t) \coloneq  \frac{\Delta_i(t)}{H(t)}.
\end{equation*}
For $t > \tau$, we let  $P(t) \coloneq  P(\tau)$, and $D_i(t) \coloneq  D_i(\tau)$, that is, we freeze these processes to their values at round $\tau$. Analogously to $\Delta$, we define $D(t) \coloneq  \min_{i >1} D_i(t)$.
By definition of $\tau$,  for $t < \tau$ we have
\begin{NoHyper}
\begin{equation}\label{eq:useful-properties}
  H(t) \ge 1, \qquad
  P(t) \in [p/2, 2p], \qquad p_1(t) = H(t) P(t) \ge p/2, \qquad p_1(t) < 3/5,
\end{equation}
\end{NoHyper}
and opinion 1 is the unique maximum opinion, i.e., $D(t) > 0$.
Note that $H(\cdot)$ is non-decreasing up to round $\tau$.
For $t < \tau$, the following equalities that follow from the definition of $P$ and $H$ will be useful later on:
\begin{align}\label{eq:q1-to-p}
  P(t) = \frac{p_1(t)}{H(t)} =  \frac{q_1(t)}{H(t)\eta(t)} = \frac{q_1(t)}{H(t+1)}.
\end{align}

\subsubsection{Node-exposure filtration and probability theoretic tools}

To get enough control on the variance of the updates,
we define a more granular \emph{node-exposure filtration}, which reveals the updates made by the nodes one-by-one.
For each $0\le j\le n$, let $\calG_{t,j}$ be the sigma-algebra generated by the configurations up to round $t$, together with the opinions adopted by the first $j$ nodes in round $t+1$.
Thus, $\calG_{t,0}=\calF_t$ and
$\calG_{t,n}=\calF_{t+1}$.
Finally, let~$\calG$ be the filtration given by $\calG_{0,0},\ldots,\calG_{0,n}=\calG_{1,0}, \calG_{1,1},\ldots$.
In particular,~$\calG$ contains~$\calF$ as a subsequence.

The filtration $\calG$ can be seen as splitting a round into $n$ asynchronous updates.
Conditional on $\calF_t$, these updates are mutually independent because all
nodes sample their opinions from the configuration at the start of the round.
Given a process $X(t)$ adapted to $\calF_t$ whose round increments can be decomposed as
\[
  X(t+1)-X(t)=\sum_{j=1}^n\xi(t,j),
\]
we define its \emph{node-exposure interpolation} by
\[
  \widehat X(t,j)\coloneq X(t)-X(0)+\sum_{a=1}^j\xi(t,a),
  \qquad 0\le t<T,\quad 0\le j\le n.
\]
This considers the individual node contributions between consecutive rounds:
$\widehat X(t,0)=X(t)-X(0)$ and $\widehat X(t,n)=X(t+1)-X(0)$.
If each $\xi(t,j)$ is integrable, $\calG_{t,j}$-measurable, and satisfies
$\expect*{\xi(t,j)\mid\calG_{t,j-1}}=0$, then $\widehat X$ is a martingale with respect to the node-exposure filtration, starting at zero.
We refer to it as the \emph{interpolated martingale}.
We will make repeated use of the following concentration bound, obtained by applying Freedman's inequality to this martingale.

\begin{lemma}
  \label{lemma:new-interpolation-lemma}
  Let $(X(t))_{t=0}^T$ be adapted to $\calF$ such that
  \[
    X(t+1)-X(t)=\sum_{j=1}^n \xi(t,j),
  \]
  where $\xi(t,j)$ is $\calG_{t,j}$-measurable.
  Let $B,V>0$ be fixed constants and suppose that almost surely
  \[
    \expect*{\xi(t,j)\mid\calG_{t,j-1}}=0,
    \qquad |\xi(t,j)|\le B,
    \qquad \sum_{t=0}^{T-1}\sum_{j=1}^n
    \expect*{\xi(t,j)^2\mid\calG_{t,j-1}}\le V.
  \]
  Then, for every $a>0$, we have
  \[
    \pr*{\max_{0\le t\le T}|X(t)-X(0)|\ge a}
    \le 2\exp\left(-\frac{a^2}{2(V+Ba/3)}\right).
  \]
\end{lemma}
\begin{proof}
  Apply \Cref{lem:tool:freedman} to the interpolated martingale $\widehat X(t,j)$ defined above.
  Its increments are bounded by $B$, and its predictable quadratic variation is bounded by $V$.
  Since $\widehat X(t-1,n) - \widehat X(0,0) = X(t)-X(0)$ for each \(1 \le t \le T\), the resulting bound implies the claim.
\end{proof}

\subsubsection{Density remains bounded}

First we show that the process $P(\cdot)$ does not deviate too much from $p$ by using Freedman's inequality.

\begin{lemma}\label{lemma:density-band}
  With probability at least $1-1/n^4$, we have
  \begin{equation}
    p/2 \le P(t) \le 2p  \qquad \text{ for all } 0 \le t \le T.
  \end{equation}
\end{lemma}
\begin{proof}
  Let $1 \le j \le n$ be a node and $Z_j(t+1)$ be the indicator for whether node $j$ adopted opinion 1 in round $t+1$.
  Define, for \(t < \tau\),
  \[
    \xi(t,j) \coloneq  \ind{t < \tau} \cdot \frac{Z_j(t+1) - q_1(t) }{nH(t+1)},
  \]
  and set it to \(0\) for \(t \ge \tau\).
  Observe that
  \begin{align*}
    \sum_{j=1}^n Z_j(t+1) = np_1(t+1).
  \end{align*}
  For $t < \tau$,
  combining the above with \Cref{eq:q1-to-p} and rearranging yields
  \begin{align}
    \sum_{j=1}^n \xi(t,j) = \frac{1}{nH(t+1) }\sum_{j=1}^n \left( Z_j(t+1) - q_1(t) \right) = P(t+1) - P(t).
  \end{align}
  For $t \ge \tau$, both sides are zero.
  Observe that because $q_1(t)$ and $H(t+1)$ are $\calF_t$-measurable, and hence, $\calG_{t,j-1}$-measurable, we have by conditional independence that
  \[
    \expect{Z_j(t+1) \mid \calG_{t,j-1}} = q_1(t)  \quad \text{ and } \quad    \expect*{ \xi(t,j) \mid \calG_{t,j-1}} = 0.
  \]
  For $t<\tau$, we have $H(t+1)\ge1$, so $\abs{\xi(t,j)}\le1/n$ for all \(j \in [n]\).
  For $t\ge\tau$, \(\abs{\xi(t,j)} = 0 \le 1/n\) for all \(j \in [n]\).
  Linearity of expectation and $Z_j(t+1) = Z^2_j(t+1)$ gives
  \begin{align*}
    \expect{(Z_j(t+1) - q_1(t))^2 \mid \calG_{t,j-1}}
    = q_1(t)(1-q_1(t)) \le q_1(t).
  \end{align*}
  Furthermore, on the event $\{t<\tau\}$, which is
  $\calG_{t,j-1}$-measurable, because $H(t+1)$ is also
  $\calG_{t,j-1}$-measurable, we get
  \begin{align*}
    \expect*{\xi(t,j)^2 \mid \calG_{t,j-1}}
    &\le \frac{1}{n^2H(t+1)^2}
    \expect{(Z_j(t+1) - q_1(t))^2 \mid \calG_{t,j-1}} \\
    &\le \frac{q_1(t)}{n^2H(t+1)^2}
    \underset{(\ref{eq:q1-to-p})}{=}
    \frac{P(t)}{n^2H(t+1)} \le \frac{2p}{n^2}.
  \end{align*}
  On $\{t\ge\tau\}$, $\xi(t,j)$ is zero,
  so the same bound holds.
  In particular, we get that
  \begin{align*}
    \sum_{t=0}^{T-1} \sum_{j=1}^n \expect*{\xi(t,j)^2 \mid \calG_{t,j-1}} \le \frac{2Tp}{n}.
  \end{align*}
  Setting $B=1/n$, $V = 2Tp/n$ and $a=p/2$, and observing that~$Ba \le V$, yields
  \begin{align*}
    \frac{a^2}{2(V+Ba/3)} &\ge \frac{a^2}{4V}
    = \frac{np}{32T} = \Omega\left(
      \frac{cnp\gamma(0)}{\ln(2p/\Delta(0))}
    \right) & \text{(by definition of $T$)} \\
    &= \Omega\left(
      \frac{cn\Delta(0)^2\gamma(0)}{p}
    \right) &\text{(since $p \geq \Delta(0)$)} \\
    &= \Omega(cC^2 \ln n), & \text{(by \Cref{eq:plurality-gap-assumption})}
  \end{align*}
  Thus \Cref{lemma:new-interpolation-lemma} gives the claim with probability $1-1/n^4$ for a sufficiently large~$C$, after fixing~$c$.
\end{proof}

\subsubsection{One-round drift bounds}\label{sec:drift-bounds}

We now use \Cref{thm:multinomial:growth} to establish bounds on the expected drift in a single round $t<\tau$.

\begin{lemma}\label{lemma:one-round-growth}
  For $i >1$ and $t < \tau$, we have
  \begin{equation}
    q_1(t) - q_i(t) \ge  \eta(t) \Delta_i(t) \left(1 + \frac{128c\gamma(0)p_i(t)}{p_1(t)}\right).
  \end{equation}
\end{lemma}
\begin{proof}
  If $p_i(t)=0$, then $q_i(t) = 0$ and the claim is vacuous. Hence, assume $p_i(t) >0$.
  Since $t < \tau$, we have from \Cref{eq:useful-properties} that $p_1(t) \ge p/2$.
  For brevity, we will drop the round index $t$ and only consider random variables at round $t$.
  Since $p_1\ge p/2$, we have
  \[
    \min\{h^2p_1,1\}\ge \min\{h^2p/2,1\}
    \ge \frac{\min\{h^2p,1\}}{2}=\frac{\gamma(0)}{2}.
  \]
  Now further algebraic manipulations give
  \begin{align*}
    \frac{q_1}{q_i} &\ge 1 + \left(\frac{p_1}{p_i} - 1 \right)  \cdot \left(1 + \frac{ \min\{h^2 p_1, 1\} }{80}\right) && \text{(by \Cref{thm:multinomial:growth})} \\
    &= 1 + \frac{\Delta_i}{p_i} \cdot \left(1 + \frac{ \min\{h^2 p_1, 1\} }{80}\right) && \text{(using $p_1/p_i-1 = \Delta_i/p_i$)} \\
    &= \frac{p_1}{p_i} \left(1 + \frac{\Delta_i}{80p_1} \min\{h^2 p_1, 1\} \right) && \text{(by expanding and rearranging)} \\
    &\ge \frac{p_1}{p_i} \left(1 + \frac{\gamma(0)\Delta_i}{160p_1}\right) && \text{(using $\min\{ h^2 p_1, 1\} \ge \gamma(0)/2$ )}.
  \end{align*}
  Note that since $\Delta_i \le p_1$ and $\gamma(0) \le 1$, we have $\gamma(0)\Delta_i/(160p_1) \le 1/160$.
  Inverting the inequality and using $\eta = q_1/p_1$ gives
  \begin{align}\label{eq:cmp-qi}
    \frac{q_i}{\eta p_i} \le \frac{1}{1 + \gamma(0)\Delta_i/(160p_1)} \le 1- \frac{\gamma(0)\Delta_i}{320p_1} \le 1 - \frac{128c\gamma(0)\Delta_i}{p_1}
  \end{align}
  using $1/(1+x) \le 1 - x/2$ for $0 \le x \le 1$ and choosing $0<c\le 1/(128\cdot320)$.
  The claim of the lemma now follows from using $q_1 =  \eta \, p_1$ and $p_1-p_i = \Delta_i$, as
  \begin{align*}
    q_1-q_i
    &\ge \eta \, p_1-\eta \, p_i
    \left(1-\frac{128c\gamma(0)\Delta_i}{p_1}\right) \\
    &=\eta(p_1-p_i)+\frac{128c\gamma(0)\eta p_i\Delta_i}{p_1} \\
    &=\eta \, \Delta_i
    \left(1+\frac{128c\gamma(0)p_i}{p_1}\right). \quad\qedhere
  \end{align*}
\end{proof}

\begin{lemma}\label{lemma:gap-drift}
  Let $i > 1$ and $t < \tau$. Then we have
  \begin{enumerate}[label=(\alph*)]
    \item $\expect*{D_i(t+1)\mid\calF_t}\ge D_i(t)$, and \label{lemma:gap-drift-for-all-less-than-tau}
    \item if $D_i(t)<p/8$, then also
      \(
        \expect*{D_i(t+1)\mid\calF_t}\ge \left(1 + c\gamma(0) \right) D_i(t),
      \) \label{lemma:gap-drift-for-small-gap}
    \item
      if $\Delta(t) \ge p_1(t)/32$, then $\eta(t) \ge 1 + c\gamma(0)$.
  \end{enumerate}
\end{lemma}
\begin{proof}
  Consider a single round $t < \tau$.
  The definition of $D_i(t+1) = \Delta_i(t+1)/H(t+1)$, $H(t+1) = \eta(t)H(t)$, and using  \Cref{lemma:one-round-growth} gives
  \begin{align*}
    \expect*{D_i(t+1)\mid\calF_t} &= \expect*{ \frac{\Delta_i(t+1)}{\eta(t) H(t)} \mid \calF_t}
    = \frac{q_1(t) - q_i(t)}{\eta(t)H(t)}
    \ge D_i(t )\left(1 + \frac{128c\gamma(0)p_i(t)}{p_1(t)}\right).
  \end{align*}
  This gives the first claim.
  For brevity, we now omit writing $t$.
  If $D_i < p/8$, then using definitions of $D_i$ and $P$ and \Cref{eq:useful-properties} we have
  \begin{align*}
    1 - \frac{p_i}{p_1} = \frac{\Delta_i}{p_1} = \frac{D_i}{P} < \frac{p}{8P} \le \frac{1}{4},
  \end{align*}
  so $p_i/p_1 \ge 3/4$. Plugging this into the previous bound gives the second claim, as
  \[
    D_i \left(1 + \frac{128c\gamma(0)p_i}{p_1}\right) \ge
    D_i \left(1 + 96c\gamma(0) \right) \ge D_i \left(1 + c\gamma(0)\right).
  \]
  For the third claim, observe that \Cref{eq:cmp-qi} in \Cref{lemma:one-round-growth} and $\Delta_i \ge \Delta \ge p_1/32$ implies
  \[
    q_i \le \eta p_i\left(1 - \frac{128c\gamma(0)\Delta_i}{p_1} \right) \le \eta p_i \left(1 - 4c\gamma(0)\right).
  \]
  Now because $1-p_1 \ge 2/5 \ge 1/4$, we get
  \begin{align*}
    1 =q_1+\sum_{i>1}q_i \le \eta p_1+\eta(1-4c\gamma(0))(1-p_1) =\eta-4c\gamma(0)\eta(1-p_1) \le \eta - c\gamma(0)\eta.
  \end{align*}
  Using $\eta \ge 1$ and rearranging gives $\eta \ge 1 + c\gamma(0)$.
\end{proof}

\subsubsection{Gaps grow and stay large}

Let us write $\phi \coloneq  1 + c\gamma(0)$. Observe that
\begin{equation}\label{eq:phi-to-t2-lb}
  \phi^{T/2} \ge \left( \frac{2p}{\Delta(0)} \right)^{16},
\end{equation}
because
\(
  \ln \phi = \ln(1 +c\gamma(0)) \ge c\gamma(0)/2
\) and so
\( (T/2) \ln \phi \ge Tc\gamma(0)/4 \ge 16 \ln (2p/\Delta(0))\).
We now consider a fixed opinion $i > 1$ and define the following two processes
\begin{align}
  M(t) &\coloneq  \sum_{s=0}^{t-1}\left( D_i(s+1) - \expect{D_i(s+1) \mid \calF_s} \right), \label{eq:def-of-M} \\
  W(t) &\coloneq  \sum_{s=0}^{t-1} \phi^{-(s+1)} \left(  M(s+1) - M(s)\right).
\end{align}
It is easy to verify that these processes are martingales; in particular, $M(\cdot)$ is the martingale from the Doob decomposition of $D_i$.
They will be used to track the accumulated noise in the (normalized) gaps.

\begin{lemma}\label{lemma:m-w-bounds}
  With probability at least $1-1/n^3$, we have for each $0 \le t \le T$
  \[
    |M(t)| \le p/32 \qquad \text{ and } \qquad |W(t)| \le \Delta(0)/2. %
  \]
\end{lemma}
\begin{proof}
  For $1 \le j \le n$ and $t \ge 0$,
  define the random variable
  \[
    Y_j(t) \coloneq
    \begin{cases}
      1 & \text{if node } j \text{ adopts opinion 1 in round } t, \\
      -1 &  \text{if node } j \text{ adopts opinion } i \text{ in round } t, \\
      0 & \text{otherwise}.
    \end{cases}
  \]
  Observe that
  \[
    \sum_{j=1}^n Y_j(t+1) = np_1(t+1) - np_i(t+1) = n \Delta_i(t+1).
  \]
  For $t < \tau$, we define
  \[
    \xi(t,j) \coloneq  \frac{Y_j(t+1) - \expect{Y_j(t+1) \mid \calF_t} }{nH(t+1)}
  \]
  and set $\xi(t,j) \coloneq  0$ for $t \ge \tau$.
  Now observe that for $t < \tau$ we have
  \begin{align*}
    \sum_{j=1}^n \xi(t,j) &= \frac{1}{nH(t+1)} \cdot \sum_{j=1}^n \left( Y_j(t+1)  - \expect{Y_j(t+1) \mid \calF_t}\right) \\
    &= \frac{
      \Delta_i(t+1)
      -\expect*{\Delta_i(t+1)\mid\calF_t}
    }{H(t+1)}\\
    &= D_i(t+1)
    -\expect*{D_i(t+1)\mid\calF_t}\\
    &= M(t+1)-M(t).
  \end{align*}
  For $t \ge \tau$, we have that $\xi(t,j) = 0$ and $M(t+1)-M(t)=0$.
  Note that
  \begin{equation}\label{eq:moments-of-y}
    \expect{Y_j(t+1) \mid \calF_t} = q_1(t) - q_i(t) \qquad \text{ and } \qquad \expect{Y_j(t+1)^2 \mid \calF_t} = q_1(t) + q_i(t),
  \end{equation}
  because
  \[
    \expect*{Y_j(t+1)^2 \mid \calF_t} = 1^2 q_1 + (-1)^2 q_i + 0 \cdot (1-q_1(t)-q_i(t)).
  \]
  Clearly, $|\xi(t,j)| \le 2/n$, since $H(\cdot) \ge 1$ is non-decreasing until round $\tau$.
  Using \Cref{eq:moments-of-y} and conditional independence, we can bound the variance
  \begin{align*}%
    \expect{(Y_j(t+1) -  \expect{Y_j(t+1) \mid \calF_t})^2 \mid \calG_{t,j-1}} &= \expect{Y_j(t+1)^2 \mid \calG_{t,j-1}} - (\expect{Y_j(t+1) \mid \calG_{t,j-1}})^2 \\
    &\le \expect{Y_j(t+1)^2 \mid \calG_{t,j-1}} = q_1(t) + q_i(t) \le 2q_1(t).
  \end{align*}
  Since $H(t+1)$ is $\calF_t$-measurable, and hence, also $\calG_{t,j-1}$-measurable, gives
  \begin{align*}
    \expect*{ \xi(t,j)^2 \mid \calG_{t,j-1}} &\le \frac{2q_1(t)}{n^2H(t+1)^2} \\
    &\le \frac{2P(t)}{n^2 H(t+1)} && \text{(using \Cref{eq:q1-to-p})} \\
    &\le \frac{4p}{n^2} && \text{(since $P(t) \le 2p$ for $t < \tau$)}.
  \end{align*}
  The predictable quadratic variation of the interpolated martingale associated with $M$ is bounded as follows:
  \begin{align*}
    \sum_{t=0}^{T-1} \sum_{j=1}^n \expect*{\xi(t,j)^2 \mid \calG_{t,j-1}} &\le \frac{4Tp}{n} \\
    &\le \frac{8p}{n} \left\lceil \frac{32}{c\gamma(0)} \, \ln \left(\frac{2p}{\Delta(0)}\right)\right\rceil && \text{(by choice of $T$)}\\
    &\le O\left( \frac{p^3}{cn\gamma(0)\Delta(0)^2}\right) && (\text{$\ln(2x) \le x^2$ for $x = p/\Delta(0) \ge 1$}) \\
    &= O\left( \frac{p^2}{cC^2 \ln n} \right). && \text{(by \Cref{eq:plurality-gap-assumption} since~$\bfp(0)$ is good)}
  \end{align*}
  Let $a_M = p/32$ and $V_M \coloneq  4Tp/n$ and $B \coloneq  2/n$.
  Since $Ba_M \le V_M$, \Cref{lemma:new-interpolation-lemma} yields
  \[
    \pr*{\max_{0 \le t \le T} |M(t)| \ge p/32} \le 2\exp\left( - \frac{a_M^2}{3V_M}\right) \le \exp \left( - \Omega(cC^2 \log n)\right).
  \]
  Similarly, for $W$, using bounds on the geometric series and definition of $\phi$, we get
  \begin{align*}
    \sum_{t=0}^{T-1}\sum_{j=1}^n
    \expect*{(\phi^{-(t+1)}\xi(t,j))^2
    \mid\calG_{t,j-1}}
    &\le \frac{4p}{n}\sum_{t=0}^{T-1}\phi^{-2(t+1)} \\
    &\le \frac{4p}{n(\phi^2-1)}
    \le \frac{4p}{n(\phi-1)}
    = \frac{4p}{cn\gamma(0)} \le \frac{4\Delta(0)^2}{cC^2\ln n} =: V_W,
  \end{align*}
  since $n\Delta(0)^2\gamma(0)\ge C^2p\ln n$ by \Cref{eq:plurality-gap-assumption}.
  Applying \Cref{lemma:new-interpolation-lemma} with $a_W = \Delta(0)/2$ and $Ba_W \le V_W$
  gives
  \begin{align*}
    \pr*{\max_{0 \le t \le T} |W(t)| \ge \Delta(0)/2} &\le 2\exp\left( - \frac{a_W^2}{3V_W}\right) \\
    &\le 2 \exp\left( -\frac{\Delta(0)^2}{4 \cdot 12 \Delta(0)^2/(cC^2 \ln n)}\right)
    \le 2\exp \left( -  \frac{cC^2 \ln n}{48}\right).
  \end{align*}
  The claim now follows from the union bound and choosing $C$ sufficiently large after fixing $c$.
\end{proof}

\begin{lemma}\label{lemma:normalized-drifts}
  Let $i > 1$ be an opinion.
  With probability at least $1-1/n^3$, we have
  \begin{enumerate}
    \item $D_i(t) > 0 $ for each $0 \le t \le \tau$, and
    \item $D_i(t) \ge p/16$ for each $T/2 \le t \le \tau$.
  \end{enumerate}
\end{lemma}
\begin{proof}
  Assume the conditions of \Cref{lemma:m-w-bounds} hold.
  Let $\sigma \coloneq  \min\{ s : D_i(s) \ge p/8\}$.
  For any $0 \le s < \min\{ \sigma, \tau\}$, we have that
  \begin{align*}
    D_i(s+1) &= \expect{D_i(s+1) \mid \calF_s} + M(s+1) - M(s) && \text{(by rearranging \Cref{eq:def-of-M})} \\
  &\ge \phi D_i(s) + M(s+1) - M(s) && \text{(by \Cref{lemma:gap-drift} since $D_i(s) < p/8$))}.
\end{align*}
By dividing both sides by $\phi^{s+1}$, we get
\[
  \frac{D_i(s+1)}{\phi^{s+1}} - \frac{D_i(s)}{\phi^s} \ge \frac{M(s+1)-M(s)}{\phi^{s+1}}.
\]
Now by summing and telescoping we get for $0 \le t \le \min\{\sigma,\tau\}$ that
\begin{align*}
  \frac{D_i(t)}{\phi^t} - D_i(0) = \sum_{s=0}^{t-1} \left( \frac{D_i(s+1)}{\phi^{s+1}} - \frac{D_i(s)}{\phi^s} \right) \ge \sum_{s=0}^{t-1} \frac{M(s+1)-M(s)}{\phi^{s+1}} = W(t),
\end{align*}
which in turn gives us
\begin{align} \label{eq:delta-phit}
  D_i(t) \ge \phi^t \left( D_i(0) + W(t) \right) \ge
  \phi^t \left( D_i(0) - \Delta(0)/2 \right) \ge \frac{\phi^t  \Delta(0)}{2} > 0
\end{align}
by rearranging and using that $|W(t)| \le \Delta(0)/2$ by \Cref{lemma:m-w-bounds}.
For $\sigma \le t \le \tau$, we have
\begin{align*}
  D_i(t)
  &= D_i(\sigma)
  + \sum_{s=\sigma}^{t-1}
  \left(\expect*{D_i(s+1)\mid\calF_s}-D_i(s)\right)
  + M(t)-M(\sigma)\\
  &\ge D_i(\sigma)+M(t)-M(\sigma)
  \ge \frac p8-\frac p{32}-\frac p{32}
  = \frac{p}{16} > 0.
\end{align*}
This establishes the first claim, as $D_i(t) >0$ for $0 \le t \le \tau$.

For the second claim,
if $\tau < T/2$, then the claim is vacuous. Therefore, suppose $\tau \ge T/2$.
It suffices to show that  $\sigma \le T/2$. For the sake of contradiction,
suppose that $\sigma > T/2$.
Hence, $D_i(T/2) < p/8$ by definition of $\sigma$, but
\begin{align*}
  D_i(T/2) &\ge \frac{\phi^{T/2}\Delta(0)}{2} && \text{(using \Cref{eq:delta-phit})} \\
  &\ge \left(\frac{2p}{\Delta(0)}\right)^{16} \frac{\Delta(0)}{2} && \text{(using \Cref{eq:phi-to-t2-lb})} \\
  &\ge \left(\frac{2p}{\Delta(0)} \cdot \frac{\Delta(0)}{2} \right) = p && \text{(using $\Delta(0) \le p$)},
\end{align*}
which contradicts $D_i(T/2) < p/8$.
\end{proof}

\subsubsection{Proof of the phase lemma}

\phaselemma*
\begin{proof}
If $p \ge 3/5$, then the phase succeeds at round 0, so we restrict attention to the case~$p < 3/5$. We consider the following events:
\begin{align*}
    \calA_1 &\coloneq  \left\{ p/2 \le P(t) \le 2p \text{ for all } 0 \le t \le T  \right\}, \\
    \calA_2 &\coloneq  \left\{ D(t) > 0 \text{ for all } 0 \le t \le T  \right\}, \\
    \calA_3 &\coloneq  \left\{ D(t) \ge p/16 \text{ for all } T/2 \le t \le \tau  \right\}.
\end{align*}
Initially absent opinions remain absent.
Hence, for every initially absent opinion $i$, we have $D_i(t)=P(t)$ for all $0\le t\le T$.
On $\calA_1$, we have $P(t)\ge p/2$ for all $0\le t\le T$.
Therefore, $D_i(t)>0$ for all $0\le t\le T$, and
$D_i(t)\ge p/16$ for all $T/2\le t\le\tau$.
By \Cref{lemma:density-band,lemma:normalized-drifts} and a union bound over all initially present opinions other than opinion $1$ (of which there are at most $n-1$), we have $\pr*{\calA_1 \cap \calA_2 \cap \calA_3} \ge 1-1/n^2$.
In the remainder of the proof, we condition on these events.

\paragraph{}
First, we consider the case that~$\tau < T$. Then by definition of~$\tau$, one of the following holds:
\begin{enumerate}[label=(\roman*)]
  \item $p_1(\tau) \ge 3/5$,
  \item opinion 1 is not the unique maximum opinion, i.e., $D(\tau) \le 0$, or
  \item $p_1(\tau)/H(\tau) \notin [p/2, 2p]$.
\end{enumerate}
However, (ii) is incompatible with~$\calA_2$, and (iii) is incompatible with~$\calA_1$.
Thus, it must be the case that $p_1(\tau) \ge 3/5$, in which case the phase succeeds at round~$\tau$.

\paragraph{}
Then, we consider the case that~$\tau = T$. $\calA_1 \cap \calA_2 \cap \calA_3$ implies that for $T/2 \le t \le T$, we have
\[
  \frac{\Delta(t)}{p_1(t)} = \frac{D(t)}{P(t)} \ge \frac{p/16}{2p} = \frac{1}{32},
\]
and hence, $\eta(t) \ge \phi$ for $T/2 \le t < T$ by \Cref{lemma:gap-drift}.
By definition of $H$ and \Cref{eq:phi-to-t2-lb}, we get
\[
  H(T) \ge \phi^{T/2} \ge \left(\frac{2p}{\Delta(0)}\right)^{16} \ge 512,
\]
because $\Delta(0) \le p$. Furthermore, using the definition of \(\calA_1\), we get
\[
  p_1(T) = H(T)P(T) \ge \frac{H(T)p}{2} \ge 2p = 2p_1(0),
\]
so the density of opinion 1 doubles.
Observe that
\begin{align*}
  \frac{\Delta(T)^2\gamma(T)}{p_1(T)}
  &\ge \frac{H(T)^2p^2\min\{2H(T)h^2p,1\}}{16^2\cdot2H(T)p} && \text{(by $\calA_1$ and $\calA_3$)} \\
  &\ge \frac{H(T)p\min\{h^2p,1\}}{512} && \text{(since $H(T)\ge1$)} \\
  &\ge \frac{p^2\gamma(0)}{p} && \text{(since $H(T)\ge512$)} \\
  &\ge \frac{\Delta(0)^2\gamma(0)}{p} && \text{(since $\Delta(0)\le p$)} \\
  &\ge C^2\frac{\ln n}{n}. && \text{(by \Cref{eq:plurality-gap-assumption})}
\end{align*}
In the first inequality, we use that $x\mapsto \min\{h^2x,1\}/x=\min\{h^2,1/x\}$ is nonincreasing for $x>0$, together with $p_1(T)\leq 2H(T)p$ and $\Delta(T)\geq H(T)p/16$.
Rearranging now gives
\[
  \Delta(T)\ge C\sqrt{\frac{p_1(T)}{\gamma(T)}\cdot\frac{\ln n}{n}},
\]
so $\bfp(T)$ is good. Since we already established that $\Delta(T) \ge p_1(T)/32$ and~$p_1(T) \ge 2p_1(0)$, we conclude that the phase succeeds at round $T$.
\end{proof}

\subsection{Final phase}

After $O(\log n)$ amplification phases, we are at a configuration, where the plurality opinion has density at least $3/5$. From this configuration, it is easy to show that the process reaches consensus in $O(\log n)$ rounds with high probability.

\begin{lemma}\label{lemma:final-phase}
If $p_1(0) \ge 3/5$, then the process reaches consensus on opinion 1 in $O(\log n)$ rounds with probability at least $1-1/n^2$.
\end{lemma}
\begin{proof}
The technique is well-known~\cite{Doerr11}, but we nonetheless recall the arguments for the sake of completeness.
First, we consider the case $h=3$.
We condition throughout on the configuration at round~$t$, assuming that $p_1(t) \geq 3/5$.
Let $z = \sum_{i>1} p_i(t)$, let $Z = n \, z$ be the number of nodes that do not have the majority opinion in round~$t$, and let~$Z'$ be the same quantity in round~$t+1$.
The probability to not adopt the majority opinion in round~$t+1$ is at most $f(z) := z^2(3-2z)$, and in particular $\expect{Z'} \le n \, f(z)$.
Let~$A,B > 0$ be sufficiently large constants.

\begin{itemize}
    \item[(i)] Consider the case $Z \geq \sqrt{A \, n \log n}$.
    One can check that~$f(z) \leq \frac{22}{25} z$ for $z \leq \frac{2}{5}$, and for $A$ large enough the Chernoff bound gives
    \begin{equation*}
        \pr*{ Z' \geq \frac{23}{25} Z } \leq \exp\pa{ - \frac{2}{n} \pa{\frac{1}{25} Z}^2 } \leq \exp \pa{ -\frac{2A}{625} \log n} = O(n^{-3}).
    \end{equation*}
    \item[(ii)] Consider the case $Z \leq \sqrt{A \, n \log n}$, i.e., $z \leq \sqrt{(A \log n)/n}$. We have
    \begin{align*}
        \pr*{ Z' > B \log n} &\leq \binom{n}{B \log n} f(z)^{B \log n} \\
        &\leq \binom{n}{B \log n} \pa{\frac{3A \log n}{n}}^{B \log n} & \text{(using $f(z) \leq 3z^2$)} \\
        &\leq \pa{\frac{3e \, A}{B}}^{B \log n} & \text{(using $\binom{n}{k} \leq (en/k)^k$)} \\
        &= O(n^{-3}). & \text{(for $B$ large enough)}
    \end{align*}
    \item[(iii)] Consider the case $Z \leq B \log n$. Then
    \begin{equation*}
        \pr*{Z' > 0} \leq \expect{Z'} \leq 3n \, z^2 \leq \frac{3 B^2 \log^2 n}{n}.
    \end{equation*}
\end{itemize}
Using (i), (ii) and a union bound, we obtain that the number of nodes without the majority opinion gets below $B \log n$ in $O(\log n)$ rounds and remains below this threshold for another $\log n$ rounds, with high probability. The desired statement follows by using (iii) over a few rounds.
Finally, since $p_1(t)\geq 3/5>1/2$, opinion $1$ is the unique largest opinion.
After ordering the remaining opinions by density, we apply \cref{lem:sample-size-majorization} with \(b= 1\), and obtain
\[
    1-q_1^{(h)}(\bfp(t))\leq 1-q_1^{(3)}(\bfp(t))\leq f(z)
\]
for every $h>3$.
Therefore, the proof also applies to every $h>3$.
\end{proof}

\subsection{\texorpdfstring{Proof of \Cref{thm:plurality-consensus}}{Proof of the plurality-consensus theorem}}

\begin{proof}
Suppose we start with a good initial configuration and let~$K \coloneq  \lceil \log n\rceil$.
We iteratively apply \Cref{lemma:phase-lemma} to the initial configuration $K$ times.
Each phase starts where the preceding phase succeeds.
By the union bound, all applications succeed with probability at least $1-K/n^2$, which we assume from now on.
Let $T_k$ be the number of rounds it takes for the $k$th phase to succeed.
This means that after at most
\[
  S = T_1 + \ldots + T_K
\]
rounds we reach a configuration where opinion 1 has density at least $3/5$. By a union bound and \Cref{lemma:final-phase} applied to the configuration at round $S$, we get that plurality consensus is reached by round $S+O(\log n)$ with probability at least $1-1/n$.

It remains to bound $S$. By \Cref{lemma:phase-lemma}, we have that
\[
  T_1
  = O\left(
    \frac{1}{\gamma(0)}
    \log\frac{2 \, p_1(0)}{\Delta(0)}
  \right)
  = O\left(\frac{\log n}{\gamma(0)}\right).
\]
because $\Delta(0) > 0$ implies that $\Delta(0) \ge 1/n$.
When applying \Cref{lemma:phase-lemma} for the $(k+1)$-th time for $1 \le k < K$, if we have not reached density $3/5$ for the initial plurality opinion, then by definition of success at round $t = T_1 + \cdots + T_k$, we have
\(
  \Delta(t) \ge p_1(t)/32.
\)
Thus, we get that
\[
  T_{k+1} = O\left( \frac{1}{\gamma(t)} \log\left(\frac{2p_1(t)}{p_1(t)/32}\right)\right) = O\left(\frac{1}{\gamma(0)}\right),
\]
since the density at the start of each successful phase is at least its initial value, so $\gamma(t)\ge\gamma(0)$.
Since $K=O(\log n)$, combining the above with the definition of $\gamma(0)$ gives
\[
  S + O(\log n) = O\left(\frac{\log n}{\gamma(0)}\right). \qedhere
\]
\end{proof}
\section{Symmetry breaking}
\label{sec:symmetry_breaking}

The goal of this section is to prove \cref{thm:consensus-time-intro}. More specifically, we prove the following theorem.
\begin{theorem}
    \label{thm:proof of the upper bound on the consensus time}
    Let $C>0$ be a sufficiently large constant.
    The following bounds hold with high probability:
    \begin{align*}
        \tconsensus \leq
        \begin{dcases}
         \frac{C\log n}{\min\{h^2\pmax(0),1\}} & \text{if } \pmax(0)\geq C\max\qty{\frac{(\log n)^2}{n},\frac{\log n}{h\sqrt{n}}},\\
         C\qty(\log n+\frac{\sqrt{n}}{h})(\log n)^2 & \text{for any initial configuration}.
        \end{dcases}
    \end{align*}
\end{theorem}

To this end, we show that a unique dominant opinion emerges, even when the initial configuration is balanced.
In this setting, the index of the opinion that achieves the maximum proportion may change over time.
Recall that $\pmax(t)=\max_{i\in[k]} p_i(t)$ is the maximum proportion at the beginning of round $t$ and let
\begin{align*}
    I_t=\min\{i\in[k]:p_i(t)=\pmax(t)\}
\end{align*}
denote the index attaining the maximum proportion at round $t$, breaking ties by the smallest index.
In this section, we present the following lemma, which ensures that a unique strong opinion emerges as long as the initial opinion distribution has a sufficiently large maximum proportion.

\begin{lemma}[Emergence of the unique strong opinion]
    \label{lem:emergence of the unique strong opinion}
    Suppose
    \[\pmax(0)\geq C\max\qty{\frac{(\log n)^2}{n},\frac{\log n}{h\sqrt{n}}}\] for a sufficiently large constant $C>0$.
    Then, for some $T=O\qty(\frac{\log n}{\min\{h^2\pmax(0),1\}})$,
    \begin{align*}
       \pr*{\exists j\neq I_T: p_{j}(T) \geq \frac{3}{4}\pmax(T)} \leq n^{-10}.
    \end{align*}
\end{lemma}

This lemma implies that the gap $\Delta(T)$ between the largest and second-largest opinion proportions satisfies $\Delta(T)\geq \pmax(T)/4$ w.h.p.
Combining this result with the following lemma, which controls downward fluctuations of the maximum proportion, we obtain the initial-gap condition of the plurality consensus result (\cref{thm:plurality-consensus}).

\begin{lemma}[Maximum proportion remains large]
    \label{lem:taumaxdown}
    For any $T>0$,
    \begin{align*}
        \operatorname{Pr}{\left[\exists t\leq T: \pmax(t)\leq \frac{\pmax(0)}{2}\right]} \leq \exp\qty(-\Omega\qty(\frac{n\pmax(0)}{T})).
    \end{align*}
\end{lemma}

\subsection{Proof outline}
The key ingredient in the proof of \cref{lem:emergence of the unique strong opinion} is the expected decrease of a well-constructed potential function.
Let
\begin{align*}
    &G_j(t) = \frac{(\pmax(t)-p_j(t))^2}{\pmax(t)+p_j(t)} \text{ for } j\in [k] \text{ and } t\geq 0, %
    \\
    &\Phi_a(x) = \frac{1}{1+ax} \text{ for } a\geq 0 \text{ and } x\geq 0. %
\end{align*}
We prove the following one-round contraction property.
\begin{lemma}[Potential contraction lemma]
    \label{lem:drift for hmaj}
    Let $t\geq 0$ be an arbitrary round.
    Let $\rho$ be an arbitrary parameter that does not depend on $t$ and satisfies $0<\rho\leq \frac{1}{80}\min\{h^2\pmax(t),1\}$.
    Let $a=n\rho/28$.
    Then, we have the following:
    \begin{enumerate}
        \item \label{item:drift for single opinion}
        Let $j\in[k]$ be an arbitrary opinion such that $j\neq I_t$ and $p_j(t)\geq \pmax(t)/2$ hold. Then,
        \[\expecton{t}{\Phi_{a}(G_{j}(t+1))} \leq \qty(1-\frac{\rho}{420})\Phi_{a}(G_{j}(t)).\]
        \item \label{item:drift for a pair of non-weak opinions}
        Let $i$ and $j$ be an arbitrary pair of distinct opinions such that both $p_i(t)\geq \pmax(t)/2$ and $p_j(t)\geq \pmax(t)/2$ hold. Then,
        \[\expecton*{t}{\Phi_{a}(\max\{G_{i}(t+1),G_{j}(t+1)\})} \leq \qty(1-\frac{\rho}{420})\Phi_{a}\qty(\max\{G_{i}(t),G_{j}(t)\}).\]
    \end{enumerate}
\end{lemma}

In the proof of \cref{lem:emergence of the unique strong opinion}, we use \cref{lem:drift for hmaj} (\cref{item:drift for a pair of non-weak opinions}) with  $\rho=\frac{1}{160}\min\{h^2\pmax(0),1\}$.
We can choose this $\rho$ since $\pmax(t)\geq \pmax(0)/2$ holds w.h.p.~by \cref{lem:taumaxdown}.
Then, for any pair of opinions $i$ and $j$ satisfying $p_i(t)\geq \pmax(t)/2$ and $p_j(t)\geq \pmax(t)/2$, \cref{lem:drift for hmaj} (\cref{item:drift for a pair of non-weak opinions}) ensures the potential function $\Phi_{a}\qty(\max\{G_{i}(t),G_{j}(t)\})$ keeps decreasing with a factor of $1-\Omega(\rho)$ until one of the stopping conditions is met.
Consequently, with high probability, at least one of the two opinions becomes weak by round $T=O(\log n/\rho)$, unless the maximum proportion falls below half its initial value.

Once an opinion becomes weak, it is unlikely to become strong again within the time horizon under consideration.
We assure this formally by the following lemma.
\begin{lemma}[Weak cannot be strong]
    \label{lem:weak cannot be strong}
    Let $j$ be an arbitrary opinion such that $p_j(0)\leq \pmax(0)/2$.
    Then, for any $T>0$,
    \begin{align*}
        \operatorname{Pr}{\left[\exists t\leq T: p_j(t) \geq \frac{3}{4}\pmax(t)\right]} \leq \exp\qty(-\Omega\qty(\frac{n\pmax(0)}{T})).
    \end{align*}
\end{lemma}

Consequently, taking the union bound over all pairs of opinions, we can show that all but at most one opinion have become weak and cannot be strong at round $T=O(\log n/\rho)$.
This completes the proof outline of \cref{lem:emergence of the unique strong opinion}.

In the following, we first prove \cref{lem:drift for hmaj}.
Then, we prove \cref{lem:taumaxdown,lem:weak cannot be strong} using the same framework based on martingale concentration techniques.
Finally, we combine these and prove \cref{lem:emergence of the unique strong opinion}.

\subsection{\texorpdfstring{Proof of the potential contraction (\cref{lem:drift for hmaj})}{Proof of the potential contraction lemma}}
This section proves \cref{lem:drift for hmaj}.
First, we introduce the following basic lemma for the potential function $\Phi_a(x)=\frac{1}{1+ax}$.
Intuitively speaking, if a random process has both multiplicative and additive drift (\cref{lem:multi and additive drift}(\cref{item:multi and additive drift})) and its variance is sufficiently small (\cref{lem:multi and additive drift}(\cref{item:multi and additive drift variance})), the potential contracts multiplicatively in expectation.

\begin{lemma}
    \label{lem:multi and additive drift}
    Let $X$ be a non-negative random variable with $\expect{X^2}<\infty$.
    For parameters $x\geq 0, r\in (0,1], b>0$, and $C\geq 1$, suppose the following two conditions hold:
    \begin{enumerate}
        \item \label{item:multi and additive drift}
        $\expect{X}\geq (1+r)x+b$,
        \item \label{item:multi and additive drift variance}
        $\variance{X}\leq \frac{Cb}{r}(\expect{X}-x)$.
    \end{enumerate}
    Then, for $a=r/(2Cb)$,
    \begin{align*}
        \expect{\Phi_a(X)} \leq \qty(1-\frac{r}{6C})\Phi_a(x).
    \end{align*}
\end{lemma}
\begin{proof}
    Write $\mu=\expect{X}$ for convenience.
    We use the following exact expansion around $\mu$:
    \[
    \frac{1}{1+aX} = \frac{1}{1+a\mu} - \frac{a(X-\mu)}{(1+a\mu)^2} + \frac{a^2(X-\mu)^2}{(1+a\mu)^2(1+aX)}.
    \]
    Multiplying both sides by $1+ax$ and taking expectations, we obtain
    \begin{align}
        \expect*{\frac{1+ax}{1+aX}}
        &= \frac{1+ax}{1+a\mu} + \frac{(1+ax)a^2}{(1+a\mu)^2}\expect*{\frac{(X-\mu)^2}{1+aX}}  && (\text{by $\expect{X-\mu}=0$}) \nonumber\\
        &\leq \frac{1+ax}{1+a\mu} \qty(1+ \frac{a^2\variance{X}}{1+a\mu}) && (\text{by $1+aX\geq 1$}) \nonumber\\
        &=\qty(1-\frac{a(\mu-x)}{1+a\mu})\qty(1+\frac{a^2\variance{X}}{1+a\mu}) \nonumber\\
        &= 1 - \frac{\mu-x-a\variance{X}}{\mu+1/a}- \frac{a^3(\mu-x)\variance{X}}{(1+a\mu)^2}.
        \label{eq:multi and additive drift 1}
    \end{align}
    From \cref{lem:multi and additive drift}(\cref{item:multi and additive drift variance}), we have $\variance{X}\leq \frac{\mu-x}{2a}$. Hence,
    \begin{align*}
        \mu - x - a\variance{X} \geq \frac{\mu-x}{2}.
    \end{align*}
    Further, we have $\mu\geq (1+r)x+b>x$ (\cref{lem:multi and additive drift}(\cref{item:multi and additive drift})).
    Hence,
    \begin{align*}
        \mu+\frac{1}{a}
        &=\mu-x+x+\frac{1}{a}
        =(\mu-x)\qty(1+\frac{x+1/a}{\mu-x})\\
        &\leq (\mu-x)\qty(1+\frac{\frac{1}{r}rx+\frac{1}{ab}b}{rx+b})\\
        &\leq (\mu-x)\qty(1+\max\qty{\frac{1}{r},\frac{1}{ab}})
    \end{align*}
    Thus, from~\eqref{eq:multi and additive drift 1} with $a=r/(2Cb)$, we have
    \begin{align*}
        \frac{\expect{\Phi_a(X)}}{\Phi_a(x)}
        \leq 1-\frac{1}{2\qty(1+\max\qty{\frac{1}{r},\frac{1}{ab}})}
        =1-\frac{r}{2(2C+r)}
    \end{align*}
    holds. Since $r\leq 1\leq C$, we obtain the claim.
\end{proof}

To apply \cref{lem:multi and additive drift}, we need to ensure that some random variable $X$ satisfies both the multiplicative and additive drift conditions.
Our key observation is that $G_j(t) = \frac{(\pmax(t) - p_j(t))^2}{\pmax(t) + p_j(t)}$ satisfies both of these drift properties.
To estimate the expected value and variance of $G_j(t)$, we use $Z_j(t)$ defined below.
In the following, we use the notations of $x_+=\max\{x,0\}$ and $x_-=\min\{x,0\}$ for $x\in \bbR$.
\begin{equation*}
    Z_j(t) =
    \begin{dcases}
        \frac{(p_{I_{t-1}}(t) - p_j(t))_+^2}{p_{I_{t-1}}(t) + p_j(t)} & (\text{if } p_{I_{t-1}}(t) + p_j(t)\neq 0), \\
        0 & (\text{if } p_{I_{t-1}}(t) + p_j(t)=0).
    \end{dcases} \text{ for } j\in [k] \text{ and } t\geq 1. %
\end{equation*}
Technically, $Z_j(t)$ is much easier to handle than $G_j(t)$ because $I_{t-1}$ is a fixed index once $\calF_{t-1}$ is determined.
We now introduce the following lemma, which gives an appropriate estimation of the ratio of the sum of random variables.

\begin{lemma}
    \label{lem:sum of iids for drift lemma}
    Let $X_1, X_2, \ldots, X_n$ be $n$ independent random variables such that
    $\pr{X_i=1}=p$, $\pr{X_i=-1}=q$, and $\pr{X_i=0}=1-p-q$ for each $i\in [n]$.
    Suppose $0\leq q\leq p\leq 1$, $p+q\leq 1$, and $n(p+q)\geq 1$.
    Let $X=\sum_{i=1}^n X_i$ and $Y=\sum_{i=1}^n X_i^2$.
    Let
    \begin{align*}
        Z=\begin{dcases}
            \frac{X_+^2}{nY} = \frac{\qty(\frac{1}{n}\sum_{i=1}^nX_i)_+^2}{\frac{1}{n}\sum_{i=1}^nX_i^2} & (\text{if } Y\neq 0), \\
            0 & (\text{if } Y=0).
        \end{dcases}
    \end{align*}
    Then, we have the following:
    \begin{enumerate}
        \item \label{item:sum of iids for drift lemma} $\frac{(p-q)^2}{p+q}+\frac{pq}{n(p+q)^2} \leq \expect{Z} \leq \frac{(p-q)^2}{p+q}+\frac{4pq}{n(p+q)^2}$.
        \item \label{item:sum of iids for drift lemma variance} $\variance{Z} \leq \frac{7}{n}\qty(\frac{(p-q)^2}{p+q}+\frac{4pq}{n(p+q)^2})$.
    \end{enumerate}
\end{lemma}
\begin{proof}[Proof of \cref{lem:sum of iids for drift lemma}(\cref{item:sum of iids for drift lemma})]
    Let $\eta=\frac{p-q}{p+q}$, $s=\frac{4pq}{(p+q)^2}$, and $\nu=\frac{(p-q)^2}{p+q}$.
    First, we introduce the following random variables:
    \begin{itemize}
        \item For each $i\in [n]$, let $B_i\in \{0,1\}$ be a binary random variable with $\pr{B_i=1}=p+q$.
        \item For each $i \in [n]$, let $V_i\in \{-1,0,1\}$ be a random variable with $\pr{V_i=1}=\pr{V_i=-1}=\frac{q}{p+q}=\frac{1-\eta}{2}$ and $\pr{V_i=0}=\frac{p-q}{p+q}=\eta$. We have $\expect{V_i}=0$ and $\expect{V_i^2}=\variance{V_i}=\frac{2q}{p+q}=1-\eta$.
        \item For each $i \in [n]$, let $U_i=V_i+\ind{V_i=0}\in \{-1,1\}$. We have $\pr{U_i=1}=\frac{p}{p+q}=\frac{1+\eta}{2}$ and $\pr{U_i=-1}=\frac{q}{p+q}=\frac{1-\eta}{2}$.
        We have $\expect{U_i}=\frac{p-q}{p+q}=\eta$, $\expect{U_i^2}=1$ and $\variance{U_i}=1-\eta^2=s$.
    \end{itemize}
    Suppose all $(B_i)_{i=1}^n$ and $(V_i)_{i=1}^n$ are independent.
    By construction, $(X_1,\ldots,X_n)\stackrel{d}{=}(B_1U_1,\ldots,B_nU_n)$ and $(B_iU_i)^2=B_i$.
    Moreover, $(U_i)_{i=1}^n$ is an i.i.d.~family independent of $(B_i)_{i=1}^n$.
    Consequently, conditional on $\sum_iB_i=\ell$, the sum $\sum_iB_iU_i$ has the same distribution as
    $\sum_{i=1}^{\ell}U_i$.
    Hence, for any $\ell\in [n]$ with $\pr{Y=\ell}>0$, we have
    \begin{align}
        \expect{Z\mid Y=\ell}
        &= \expect*{\frac{\qty(\sum_{i\in [n]:X_i^2=1} X_i)_+^2}{n\sum_{i=1}^n X_i^2}\mid \sum_{i=1}^n X_i^2=\ell} \nonumber\\
        &= \frac{\expect*{\qty(\sum_{i\in [n]:B_i=1} B_iU_i)_+^2\mid \sum_{i=1}^n B_i=\ell}}{n\ell} \nonumber\\
        &= \frac{\expect*{\qty(\sum_{i=1}^{\ell} U_i)_+^2}}{n\ell}. \label{eq:Z cond on Y}
    \end{align}
    For $\ell\in[n]$, write $U^{(\ell)}=\sum_{i=1}^{\ell}U_i$.
    We next bound $\expect{(U^{(\ell)})_+^2}$ from above and below.
    Since $x_+^2\leq x^2$ for all $x\in \bbR$, we have $(U^{(\ell)})^2_+\leq (U^{(\ell)})^2$. Hence,
    \begin{align}
        \expect{(U^{(\ell)})^2_+}\leq \expect{(U^{(\ell)})^2}=\variance{U^{(\ell)}}+\expect{U^{(\ell)}}^2=\ell^2\eta^2+\ell s
        \label{eq:upper of Up}
    \end{align}
    Let $V^{(\ell)}=\sum_{i=1}^{\ell} V_i$.
    From definition, we have $U^{(\ell)}=V^{(\ell)}+\sum_{i=1}^{\ell}\ind{V_i=0}\geq V^{(\ell)}$.
    Hence, $(U^{(\ell)})^2_-\leq (V^{(\ell)})^2_-$ holds.
    Note that, for $x\leq y$, $x_-\leq y_-$ holds and hence $y_-^2\leq x_-^2$ holds.
    Further, from the symmetry of $V_i$, we have
    \begin{align*}
        \expect{(U^{(\ell)})^2_-}\leq
        \expect{(V^{(\ell)})^2_-} = \expect{(V^{(\ell)})^2_+} = \frac{1}{2}\expect{(V^{(\ell)})^2}=\frac{\ell (1-\eta)}{2}\leq \frac{\ell s}{2}.
    \end{align*}
    Note that $-(V^{(\ell)})_- \stackrel{d}{=} (V^{(\ell)})_+$ holds since $V^{(\ell)}$ is a sum of independent random variables symmetric about zero, it is also symmetric about zero.
    Further, $x^2_-+x^2_+=x^2$ holds for any $x\in \bbR$,
    \begin{align}
        \expect{(U^{(\ell)})^2_+} = \expect{(U^{(\ell)})^2} - \expect{(U^{(\ell)})^2_-} \geq  \ell^2\eta^2+\frac{\ell s}{2}.
        \label{eq:lower of Up}
    \end{align}
    Combining \cref{eq:Z cond on Y,eq:lower of Up,eq:upper of Up} and $\expect{Y}=n(p+q)$, we obtain
    \begin{align*}
        \expect{Z} = \sum_{\ell=1}^n\frac{\expect{(U^{(\ell)})^2_+}}{n\ell}\pr{Y=\ell}
        \leq \frac{1}{n}\sum_{\ell=1}^n\qty(\eta^2\ell + s)\pr{Y=\ell}
        =\frac{\eta^2\expect{Y}}{n}+\frac{s\pr{Y>0}}{n}
        \leq \nu+\frac{s}{n},
    \end{align*}
    and
    \begin{align}
        \expect{Z}
        \geq \frac{1}{n}\sum_{\ell=1}^n\qty(\eta^2\ell + \frac{s}{2})\pr{Y=\ell}
        =\frac{\eta^2\expect{Y}}{n}+\frac{s\pr{Y>0}}{2n}
        \geq \nu+\frac{s}{4n}
        \label{eq:Z lower}
    \end{align}
    In the last inequality, we use $\pr{Y=0}=(1-(p+q))^n\leq e^{-n(p+q)}\leq e^{-1}\leq 1/2$ since $n(p+q)\geq 1$.
\end{proof}

\begin{proof}[Proof of \cref{lem:sum of iids for drift lemma}(\cref{item:sum of iids for drift lemma variance})]
    We use the notation introduced in the proof of the first assertion.
    By the conditional-distribution representation used in \cref{eq:Z cond on Y}, for any $\ell\in [n]$, we have
    \begin{align*}
        \expect{Z^2\mid Y=\ell} = \frac{\expect{(X)_+^4\mid Y=\ell}}{n^2\ell^2}
        = \frac{\expect{(U^{(\ell)})^4_+}}{n^2\ell^2}
        \leq \frac{\expect{(U^{(\ell)})^4}}{n^2\ell^2}.
    \end{align*}
    Here, we use $x^4_+\leq x^4$ for any $x\in \bbR$.
    Recall $\expect{U_i}=\eta$ and $\expect{U_i^2}=1$.
    Hence, from $\expect{(U^{(\ell)})^2}=\ell^2\eta^2+\ell s$ (\cref{eq:upper of Up}), we have
    \begin{align*}
        \expect{(U^{(\ell)})^4}
        &= \sum_{\substack{a,b,c,d\in [\ell]: \\ \abs{\{a,b,c,d\}}=4}}\expect{U_aU_bU_cU_d}
        + \sum_{\substack{a,b,c,d\in [\ell]: \\ \abs{\{a,b,c,d\}}\leq 3}}\expect{U_aU_bU_cU_d}\\
        &\leq \sum_{\substack{a,b,c,d\in [\ell]: \\ \abs{\{a,b,c,d\}}=4}}\expect{U_a}\expect{U_b}\expect{U_c}\expect{U_d}  + \binom{4}{2}\sum_{a,b,c\in [\ell]}\expect{U_aU_bU_c^2}\\
        &= \sum_{\substack{a,b,c,d\in [\ell]: \\ \abs{\{a,b,c,d\}}=4}}\eta^4  + 6\ell\sum_{a,b\in [\ell]}\expect{U_aU_b}\\
        &\leq  \ell^4 \eta^4 + 6\ell \expect{(U^{(\ell)})^2}
        = \ell^4 \eta^4 + 6\ell (\ell^2\eta^2+\ell s).
    \end{align*}
    Recall $\expect{Y}=n(p+q)$ and $\expect{Y^2}\leq n^2(p+q)^2 + n(p+q)$.
    Hence,
    \begin{align*}
        \expect{Z^2}
        \leq \sum_{\ell=1}^n\frac{\ell^4 \eta^4 + 6\ell (\ell^2\eta^2+\ell s)}{n^2\ell^2}\pr{Y=\ell}
        =\frac{\eta^4\expect{Y^2}+6\eta^2\expect{Y}+6s\pr{Y>0}}{n^2}
        \leq \nu^2 + \frac{7\nu}{n} + \frac{6s}{n^2}.
    \end{align*}
    Thus, from $\expect{Z}\geq \nu \geq 0$ (\cref{eq:Z lower}), we obtain
    \begin{align*}
        \variance{Z}
        \leq \expect{Z^2} - \expect{Z}^2
        \leq \nu^2 + \frac{7\nu}{n} + \frac{6s}{n^2} - \nu^2
        \leq  \frac{7}{n}\qty(\nu+\frac{s}{n}).
    \end{align*}
\end{proof}

We apply \cref{lem:sum of iids for drift lemma} to $Z_j(t)$ to show the multiplicative and additive drift for $Z_j(t)$.
We use the following calculation, which is a natural consequence of \cref{thm:multinomial:growth,cor:multinomial:weak-comparison}.
\begin{lemma}\label{lem:drift for Z additive and multiplicative}
    Suppose that $p_1=\max_{i\in[k]}p_i\geq 1/n$.
    Let $j\ge 2$ be any opinion such that $p_j\geq p_1/2$.
    Let $\rho$ be a parameter satisfying $0<\rho \leq (1/80)\min\{h^2p_1,1\}$. Then,
    \begin{align*}
        \frac{(q_1-q_j)^2}{q_1+q_j} + \frac{q_1q_j}{n(q_1+q_j)^2}
        \geq \qty(1+\frac{\rho}{2}) \frac{(p_1-p_j)^2}{p_1+p_j} + \frac{1}{5n}.
    \end{align*}
\end{lemma}
\begin{proof}
According to \cref{thm:multinomial:growth}, we have $\frac{q_1}{q_j}-1 \geq (1+\rho)\left(\frac{p_1}{p_j}-1\right)$.
This implies that $\frac{q_j}{q_1} \leq \frac{p_j}{p_1+\rho(p_1-p_j)} \leq \frac{p_j}{p_1}$.
Additionally, $q_1\geq p_1$ holds from \cref{cor:multinomial:weak-comparison}.
Therefore,
\begin{align*}
    \frac{(q_1-q_j)^2}{q_1+q_j}
    =p_1 \frac{q_1}{p_1} \frac{\qty(1-\frac{q_j}{q_1})^2}{1+\frac{q_j}{q_1}}
    \geq p_1 \frac{\qty(1-\frac{p_j}{p_1+\rho(p_1-p_j)})^2}{1+\frac{p_j}{p_1}}
    =\qty(\frac{1+\rho}{1+\rho\qty(1-\frac{p_j}{p_1})})^2\frac{(p_1-p_j)^2}{p_1+p_j}.
\end{align*}
Here, in taking the above inequality, we used the bound $\frac{q_j}{q_1}\leq \frac{p_j}{p_1+\rho(p_1-p_j)}$ for the numerator, and $\frac{q_j}{q_1}\leq \frac{p_j}{p_1}$ for the denominator, respectively.

Since $p_j\geq p_1/2$ from the assumption, we have
$
\frac{1+\rho}{1+\rho\qty(1-p_j/p_1)} \geq \frac{1+\rho}{1+\rho/2} \geq 1+\frac{\rho}{3}
$.
Thus,
\begin{align}
    \frac{(q_1-q_j)^2}{q_1+q_j} - \frac{(p_1-p_j)^2}{p_1+p_j} \geq \qty(\qty(1+\frac{\rho}{3})^2-1)\frac{(p_1-p_j)^2}{p_1+p_j}\geq \frac{2\rho}{3}\cdot \frac{(p_1-p_j)^2}{p_1+p_j}.
    \label{eq:drift for Z multiplicative}
\end{align}

Next, combining $p_j\geq p_1/2$ and $q_1\geq p_1$ (\cref{cor:multinomial:weak-comparison}), we have $\frac{(p_1-p_j)^2}{p_1+p_j}\leq \frac{(1/2)^2p_1^2}{(3/2)p_1} \leq \frac{p_1}{6} \leq \frac{q_1+q_j}{6}$.
Further, we have $\frac{(q_1-q_j)^2}{q_1+q_j}
=(q_1+q_j)\qty(1-\frac{4q_1q_j}{(q_1+q_j)^2})$.
Hence,
\begin{align}
    \frac{(q_1-q_j)^2}{q_1+q_j} - \frac{(p_1-p_j)^2}{p_1+p_j}
    \geq \frac{1}{n(q_1+q_j)}\qty(\frac{(q_1-q_j)^2}{q_1+q_j} - \frac{(p_1-p_j)^2}{p_1+p_j})
    \geq \frac{1}{n}\qty(\frac{5}{6}-\frac{4q_1q_j}{(q_1+q_j)^2}).
    \label{eq:drift for Z additive}
\end{align}
The first inequality follows since \cref{eq:drift for Z multiplicative} implies $\frac{(q_1-q_j)^2}{q_1+q_j} - \frac{(p_1-p_j)^2}{p_1+p_j}\geq 0$ and $q_1\geq p_1 \geq 1/n$.
By taking a weighted average of \cref{eq:drift for Z multiplicative} and \cref{eq:drift for Z additive} with weights $3/4$ and $1/4$, respectively, we obtain
\begin{align*}
    \frac{(q_1-q_j)^2}{q_1+q_j} - \frac{(p_1-p_j)^2}{p_1+p_j}
    \geq \frac{\rho}{2} \cdot \frac{(p_1-p_j)^2}{p_1+p_j} + \frac{1}{4n}\qty(\frac{5}{6}-\frac{4q_1q_j}{(q_1+q_j)^2}).
\end{align*}
Adding $\frac{q_1q_j}{n(q_1+q_j)^2}$ to both sides, we obtain the claim.
\end{proof}

We now prove \cref{lem:drift for hmaj}.
\begin{proof}[Proof of \cref{lem:drift for hmaj}(\cref{item:drift for single opinion})]
    We outline the main steps of the proof as follows.
    First, we establish that $\Phi_a(G_j(t+1)) \leq \Phi_a(Z_j(t+1))$.
    Next, to utilize \cref{lem:multi and additive drift}, we check condition \cref{lem:multi and additive drift}(\cref{item:multi and additive drift}) by computing $\expecton{t}{Z_j(t+1)}$, and verify condition \cref{lem:multi and additive drift}(\cref{item:multi and additive drift variance}) by estimating $\varianceon{t}{Z_j(t+1)}$.

    Consider the function $f(x) = \frac{(x-b)^2_+}{x+b}$ for $x, b \geq 0$, where we set $f(x) = 0$ when $x = b = 0$.
    This function is non-decreasing in $x$:
    for $x \leq b$, $f(x) = 0$, and for $x > b$, we have $f(x) = \frac{(x-b)^2}{x+b}$ with derivative $f'(x) = \frac{(x-b)(3b+x)}{(x+b)^2} > 0$.
    Hence,
    \begin{align*}
    G_{j}(t+1)
    &=
    \frac{(\pmax(t+1)-p_j(t+1))^2_+}{\pmax(t+1)+p_j(t+1)}
    \geq
    \frac{(p_{I_t}(t+1)-p_j(t+1))^2_+}{p_{I_t}(t+1)+p_j(t+1)}
    =Z_j(t+1).
    \end{align*}
    Here, the first equality follows since $(\pmax(t+1)-p_j(t+1))^2=(\pmax(t+1)-p_j(t+1))^2_+$ always holds. The inequality is due to the fact that $\pmax(t+1)\geq p_{I_t}(t+1)$.
    Since $a\geq 0$, we obtain
    \begin{align}
    \Phi_{a}(G_{j}(t+1))\leq \Phi_{a}(Z_{j}(t+1)).
    \label{eq:inequality between G and Z}
    \end{align}

    Now, we show the following bound on $\expecton{t}{Z_j(t+1)}$:
    \begin{align}
        \expecton{t}{Z_j(t+1)} \geq \qty(1+\frac{\rho}{2})G_j(t)+\frac{1}{5n}.
        \label{eq:inequality for Z_j(t+1) and G_j(t)}
    \end{align}
    To see this, we condition on $\calF_t$ and apply \cref{lem:sum of iids for drift lemma} to $(X_v)_{v\in [n]}$ with $X_v = \ind{C_v(t+1)=I_t}-\ind{C_v(t+1)=j}$, where we recall that $C_v(t)\in[k]$ is the opinion of a node $v$ at the beginning of round $t$. %
    Then, we have $\frac{1}{n}\sum_{v\in [n]}X_v = p_{I_t}(t+1)-p_j(t+1)$, $\frac{1}{n}\sum_{v\in [n]}X_v^2 = p_{I_t}(t+1)+p_j(t+1)$, and hence $Z_j(t+1) = \frac{\qty(\frac{1}{n}\sum_{v\in [n]} X_v)_+^2}{\frac{1}{n}\sum_{v\in [n]} X_v^2}$.
    Further, let
    \[
        p_*\coloneq p_{I_t}(t)=\pmax(t), \qquad     q_*\coloneq\pron{t}{C_v(t+1)=I_t}=\pron{t}{X_v=1},
    \]
    and let
    \[
        q_j\coloneq\pron{t}{C_v(t+1)=j}  =\pron{t}{X_v=-1}.
    \]
    Then $n(q_*+q_j)\ge nq_*\ge1$ holds by
    \cref{cor:multinomial:weak-comparison}.
    Therefore, we can apply \cref{lem:sum of iids for drift lemma} and obtain
    \begin{align}
        \expecton{t}{Z_j(t+1)}
        \geq \frac{(q_*-q_j)^2}{q_*+q_j} +\frac{q_*q_j}{n(q_*+q_j)^2}.\label{eq:inequality for Z to q}
    \end{align}
    Hence, applying \cref{lem:drift for Z additive and multiplicative} with $p_*$ and $q_*$ in the roles of $p_1$ and $q_1$, respectively, and with $p_j=p_j(t)$, we obtain \cref{eq:inequality for Z_j(t+1) and G_j(t)}.
    Here, $\rho\le(1/80)\min\{h^2p_*,1\}$ and
    \[
        G_j(t)=\frac{(\pmax(t)-p_j(t))^2}{\pmax(t)+p_j(t)}
        =\frac{(p_*-p_j(t))^2}{p_*+p_j(t)}.
    \]

    Finally, we show the following relation between $\varianceon{t}{Z_j(t+1)}$ and $\expecton{t}{Z_j(t+1)}$:
    \begin{align}
        \varianceon{t}{Z_j(t+1)} \leq \frac{14}{n\rho}\qty(\expecton{t}{Z_j(t+1)} - G_j(t)).
        \label{eq:relation between Var and E for Z_j(t+1)}
    \end{align}
    To see this, we use \cref{eq:drift for Z multiplicative} in the proof of \cref{lem:drift for Z additive and multiplicative}.
    We have
    \begin{align*}
        \frac{(q_*-q_j)^2}{q_*+q_j} - \frac{(p_*-p_j)^2}{p_*+p_j}
        \geq \qty(1-\frac{1}{1+2\rho/3})\frac{(q_*-q_j)^2}{q_*+q_j} \geq \frac{\rho}{2}\frac{(q_*-q_j)^2}{q_*+q_j}.
    \end{align*}
    Note that $1-\frac{1}{1+2\rho/3}=\frac{2\rho}{3+2\rho}\geq \rho/2$ since $\rho\leq 1/80\leq 1/2$.
    Again, applying \cref{lem:sum of iids for drift lemma} with $\rho\leq 1/2$, we have
    \begin{align*}
        \expecton{t}{Z_j(t+1)}-\frac{(p_*-p_j)^2}{p_*+p_j}
        &\geq \frac{(q_*-q_j)^2}{q_*+q_j} + \frac{q_*q_j}{n(q_*+q_j)^2} -\frac{(p_*-p_j)^2}{p_*+p_j} &&(\text{by \cref{eq:inequality for Z to q}}) \nonumber \\
        &\geq \frac{\rho}{2}\frac{(q_*-q_j)^2}{q_*+q_j} + \frac{q_*q_j}{n(q_*+q_j)^2} \nonumber \\
        &\geq \frac{\rho}{2}\qty(\frac{(q_*-q_j)^2}{q_*+q_j}+\frac{4q_*q_j}{n(q_*+q_j)^2}) \nonumber \\
        &\geq \frac{n\rho}{14}\varianceon{t}{Z_j(t+1)}, && (\text{by \cref{lem:sum of iids for drift lemma}(\cref{item:sum of iids for drift lemma variance})})
    \end{align*}
    and we obtain \cref{eq:relation between Var and E for Z_j(t+1)}.

    Consequently, \cref{eq:inequality for Z_j(t+1) and G_j(t),eq:relation between Var and E for Z_j(t+1)} ensure that the conditions \cref{lem:multi and additive drift}(\cref{item:multi and additive drift,item:multi and additive drift variance}) are satisfied with parameters
    $x=G_j(t)$, $r=\rho/2$, $b=1/(5n)$, $C=35$, and $a=r/(2Cb)=n\rho/28$.
    We obtain
    \begin{align*}
        \expecton{t}{\Phi_a(Z_{j}(t+1))}\leq \qty(1-\frac{\rho}{420})\Phi_a(G_j(t)).
    \end{align*}
    Combining it with \cref{eq:inequality between G and Z}, we obtain the claim.
\end{proof}
\begin{proof}[Proof of \cref{lem:drift for hmaj}(\cref{item:drift for a pair of non-weak opinions})]
    Choose $\ell_t\in\{i,j\}\setminus\{I_t\}$ such that $G_{\ell_t}(t)=\max\{G_i(t),G_j(t)\}$, breaking ties by the smallest index.
    Such a choice is possible because $i\neq j$ implies that at least one of $i$ and $j$ differs from $I_t=\min\{i\in [k]: p_i(t) = \pmax(t)\}$, and $G_{I_t}(t)=0\leq G_l(t)$ for every $l\in[k]$.
    In particular, $\ell_t$ is $\calF_t$-measurable.
    Since $\Phi_a$ is nonincreasing, we obtain
    \begin{align*}
        &\expecton{t}{\Phi_a(\max\{G_{i}(t+1), G_j(t+1)\})} \\
        &\leq \expecton{t}{\Phi_a(G_{\ell_t}(t+1))} && (\text{since $\ell_t\in \{i,j\}$})\\
        &\leq \qty(1-\frac{\rho}{420})\Phi_a(G_{\ell_t}(t)) && (\text{by \cref{lem:drift for hmaj}(\cref{item:drift for single opinion})})\\
        &= \qty(1-\frac{\rho}{420})\Phi_a(\max\{G_i(t), G_j(t)\}).
    \end{align*}
\end{proof}

\subsection{
\texorpdfstring{Maximum proportion and weak opinions (\cref{lem:taumaxdown,lem:weak cannot be strong})}{Maximum proportion and weak opinions}
}
\label{sec:key properties for the maximum proportion and weak opinions}

This section shows \cref{lem:taumaxdown,lem:weak cannot be strong}.
Both are consequences of a martingale inequality based on the Bernstein condition (\cref{def:bernstein_condition}).
To begin with, we show the following common framework for the proofs of \cref{lem:taumaxdown,lem:weak cannot be strong}.
\begin{lemma}
    \label{lem:drift lemma for ratio}
    Fix an integer $T\geq 1$.
    Let $\calF_t$ be a natural filtration generated by $(\bfC(s))_{s\leq t}$.
    Let $(X(t))_{t\in\bbN_0}$ be a sequence of non-negative real-valued random variables adapted to $(\calF_t)_{t\in\bbN_0}$.
    Suppose that, for every $t\in[T]$, conditional on $\calF_{t-1}$, the following conditions hold:
    \begin{enumerate}
        \item \label{cond:drift lemma for ratio 1}
        $\frac{\expecton{t-1}{X(t)}}{\expecton{t-1}{p_{I_{t-1}}(t)}}\leq \frac{X(t-1)}{\pmax(t-1)}$,
        \item \label{cond:drift lemma for ratio 2}
        $X(t)-\expecton{t-1}{X(t)}$ satisfies the one-sided $\qty(\frac{b}{n},\frac{b\expecton{t-1}{p_{I_{t-1}}(t)}}{n})$-Bernstein condition for some $b\geq 1$.
    \end{enumerate}
    Let $\theta\in (0,1]$ and $I\in (0,\theta)$ be parameters and
    suppose $\frac{X(0)}{\pmax(0)}\leq\theta-I$. Then,
    \begin{align*}
        \pr*{
            \exists t\in\{0,\ldots,T\}:
            \frac{X(t)}{\pmax(t)}\geq\theta
        }
        \leq
        \exp\qty(
            -\frac{I^2n\pmax(0)}{4b(2T+I/3)}
        ).
    \end{align*}
\end{lemma}
\begin{proof}[Proof of \cref{lem:drift lemma for ratio}]
    Define the following parameters:
    \begin{itemize}
        \item For $t\geq 0$, let
        \[\theta_t = \theta - \frac{I}{2} + \frac{I\cdot t}{2T}.\]
        We have $0<\theta-\frac{I}{2} = \theta_0 \leq \theta_t$.
        Further, $\theta_t\leq \theta$ holds for all $t\in\{0,\ldots,T\}$.
        We will frequently use the fact that $\theta_t-\theta_{t-1}=\frac{I}{2T}$ in the proof.
        \item For $i\in[k]$ and $t\geq 0$, let
        \[Y_{i}(t) = \theta_t p_i(t) - X(t).\]
        In particular, write $Y(t) = \theta_t \pmax(t) - X(t)$.
        \item Let $\tau^+ = \inf\qty{t\geq 0: \frac{X(t)}{\pmax(t)} \geq \theta_t }$ and let $\tau=\min\{\tau^+, T\}$.
        \item Let $Z(t) = -Y(t\wedge \tau)$ and $M(t) = \exp(\lambda Z(t))$, where $\lambda = \frac{In/2}{2bT+Ib/3}$.
    \end{itemize}
    Recall $\expecton{t-1}{p_{I_{t-1}}(t)}\geq p_{I_{t-1}}(t-1)=\pmax(t-1)$ holds from \cref{cor:multinomial:weak-comparison}.
    Hence, for $\tau>t-1$, we have
    \begin{align}
    \expecton{t-1}{Y_{I_{t-1}}(t)}
    &=\qty(\theta_{t-1}+\frac{I}{2T})\expecton{t-1}{p_{I_{t-1}}(t)} - \expecton{t-1}{X(t)} \nonumber \\
    &= \frac{\expecton{t-1}{p_{I_{t-1}}(t)}}{\pmax(t-1)}
        \qty(\theta_{t-1} \pmax(t-1) - \frac{\pmax(t-1)\expecton{t-1}{X(t)}}{\expecton{t-1}{p_{I_{t-1}}(t)}}) + \frac{I}{2T}\expecton{t-1}{p_{I_{t-1}}(t)} \nonumber\\
        &\geq \frac{\expecton{t-1}{p_{I_{t-1}}(t)}}{\pmax(t-1)}
        \qty(\theta_{t-1}\pmax(t-1) - X(t-1)) + \frac{I}{2T}\expecton{t-1}{p_{I_{t-1}}(t)} \nonumber\\
        &\geq Y(t-1) + \frac{I}{2T}\expecton{t-1}{p_{I_{t-1}}(t)}.
        \label{eq:drift lemma for ratio Y difference}
    \end{align}
    The first inequality follows from the assumption \cref{lem:drift lemma for ratio}(\cref{cond:drift lemma for ratio 1}): $\frac{\expecton{t-1}{X(t)}}{\expecton{t-1}{p_{I_{t-1}}(t)}}\leq \frac{X(t-1)}{\pmax(t-1)}$.
    The second inequality follows from the fact that $\theta_{t-1}\pmax(t-1)-X(t-1)\geq 0$ for $\tau>t-1$ and $\expecton{t-1}{p_{I_{t-1}}(t)}/\pmax(t-1)\geq 1$.
    \Cref{eq:drift lemma for ratio Y difference} implies that $Y(t-1)\leq \expecton{t-1}{\theta_t p_{I_{t-1}}(t)-X(t)}-\frac{I}{2T}\expecton{t-1}{p_{I_{t-1}}(t)}$ for $\tau>t-1$.
    Additionally, we have $Y(t) = \theta_t \pmax(t) - X(t) \geq \theta_t p_{I_{t-1}}(t) - X(t)$.
    Hence,
    \begin{align*}
        &Z(t) - Z(t-1) \\
        &=\ind{\tau>t-1} \qty(Y(t-1)-Y(t)) \\
        &\leq \ind{\tau>t-1} \qty(\theta_t\qty(\expecton{t-1}{p_{I_{t-1}}(t)}-p_{I_{t-1}}(t))+\qty(X(t)-\expecton{t-1}{X(t)})-\frac{I}{2T}\expecton{t-1}{p_{I_{t-1}}(t)}).
    \end{align*}

    By \cref{lem:Basic properties of sync process} (\cref{item:BC of sync process,item:variance of sync process 1}), the centered random variable
    $\expecton{t-1}{p_{I_{t-1}}(t)}-p_{I_{t-1}}(t)$ satisfies a Bernstein condition.
    Since $\theta_t^2\leq 1$ for $\tau>t-1$, applying
    \cref{lem:basic properties of bernstein condition} (\cref{item:BS upper bounded,item:BS multiplied}) shows that
    \[
        \ind{\tau>t-1}\theta_t  \qty(\expecton{t-1}{p_{I_{t-1}}(t)}-p_{I_{t-1}}(t))
    \]
    satisfies the Bernstein condition with parameters
    \[
      \left(\frac{1}{n},\frac{\ind{\tau>t-1}\expecton{t-1}{p_{I_{t-1}}(t)}}{n}\right).
    \]
    Note that $\theta_t^2\leq 1$ for $\tau>t-1$.
    Hence, from the assumption \cref{lem:drift lemma for ratio}(\cref{cond:drift lemma for ratio 2}) and \cref{lem:basic properties of bernstein condition} (\cref{item:BS sum}),
    \begin{equation*}
    \ind{\tau>t-1} \qty(\theta_t\qty(\expecton{t-1}{p_{I_{t-1}}(t)}-p_{I_{t-1}}(t))+\qty(X(t)-\expecton{t-1}{X(t)}))
    \end{equation*}
    satisfies one-sided $\qty(\frac{2b}{n},\frac{4\ind{\tau>t-1}b\expecton{t-1}{p_{I_{t-1}}(t)}}{n})$-Bernstein condition.

    Now, we introduce the following basic claim:
    \begin{claim}
        \label{claim:bernstein condition and moment supermartingale}
        Let $D, s, R>0$ and $J\geq 0$ be parameters.
        Suppose $X$ satisfies one-sided $(D,sJ)$-Bernstein condition.
        Then, for $\lambda = \frac{R}{s/2+DR/3}$,
        $\expect{\mathrm{e}^{\lambda (X-JR)}}\leq 1$.
    \end{claim}
    \begin{proof}[Proof of \cref{claim:bernstein condition and moment supermartingale}]
        By the choice of $\lambda$, we have
        \begin{align*}
            1-\frac{\lambda D}{3}
            =\frac{s/2}{s/2+DR/3}>0 \text{ and }
            \frac{\lambda s}{2(1-\lambda D/3)}=R.
        \end{align*}
        Hence, by \cref{def:bernstein_condition},
        \begin{align*}
            \expect{\mathrm{e}^{\lambda(X-JR)}}
            \leq
            \exp\left(
                \frac{\lambda^2Js}{2(1-\lambda D/3)}
                -\lambda JR
            \right)
            =
            \exp\left(
                \lambda J\left(
                    \frac{\lambda s}{2(1-\lambda D/3)}-R
                \right)
            \right)
            =1.
        \end{align*}
    \end{proof}

    We condition on $\calF_{t-1}$ and apply
    \cref{claim:bernstein condition and moment supermartingale} with the
    following parameters:
    \[X=\ind{\tau>t-1} \qty(\theta_t\qty(\expecton{t-1}{p_{I_{t-1}}(t)}-p_{I_{t-1}}(t))+\qty(X(t)-\expecton{t-1}{X(t)})),\]
    $R=\frac{I}{2T}\expecton{t-1}{p_{I_{t-1}}(t)}$, $s=\frac{4b\expecton{t-1}{p_{I_{t-1}}(t)}}{n}$, $D=\frac{2b}{n}$, $J=\ind{\tau>t-1}$, and $\lambda = \frac{R}{s/2+DR/3}= \frac{In/2}{2bT+Ib/3}$.
    Then, we obtain
    \begin{align*}
        \frac{\expecton*{t-1}{M(t)}}{M(t-1)}
        &= \expecton{t-1}{\exp(\lambda(Z(t)-Z(t-1)))} \\
        &\leq \expecton{t-1}{\exp(\lambda(X-JR))} \\
        &\leq 1.
    \end{align*}
    i.e., $(M(t))_{t\in \bbN_0}$ is a supermartingale.
    Hence, we have
    \begin{align*}
        \expect{M(T)}
        \leq \expect{M(0)}
        =\exp(-\lambda \pmax(0) \qty(\theta_0-\frac{X(0)}{\pmax(0)}))
        \leq \exp(-\frac{\lambda \pmax(0)I}{2}).
    \end{align*}
    Here, we use the assumption of $\frac{X(0)}{\pmax(0)}\leq \theta-I$ and $\theta_0=\theta-I/2$.
    Further,
    \begin{align*}
        \pr*{\exists t\leq T: \frac{X(t)}{\pmax(t)}\geq \theta}
        &\leq \pr*{\exists t\leq T: \frac{X(t)}{\pmax(t)}\geq \theta_t}\\
        &=\pr{\tau^+\leq T}\\
        &\leq \pr{Z(T)\geq 0}\\
        &=\pr{M(T)\geq 1}\\
        &\leq \expect{M(T)}.
    \end{align*}
    The second inequality follows since $\tau^+\leq T \Rightarrow Y(\tau^+\wedge T)\leq 0$ and $Y(\tau^+\wedge T)=-Z(T)$ from definition.
\end{proof}

\begin{proof}[Proof of \cref{lem:taumaxdown}]
    Let $c_{\max}^\downarrow=1/2$.
    We apply \cref{lem:drift lemma for ratio} with
    $X(t)=(1-c_{\max}^\downarrow)\pmax(0)$,
    $b=1$,
    $\theta=1$, and
    $I=c_{\max}^\downarrow$.
    We have $\frac{X(0)}{\pmax(0)} = 1-c_{\max}^\downarrow=\theta-I$.
    From $\expecton{t-1}{p_{I_{t-1}}(t)}\geq p_{I_{t-1}}(t-1)=\pmax(t-1)$ (\cref{cor:multinomial:weak-comparison}), the condition of \cref{lem:drift lemma for ratio} (\cref{cond:drift lemma for ratio 1}) holds.
    Further, $X(t)-\expecton{t-1}{X(t)}$ satisfies the assumption of \cref{lem:drift lemma for ratio} (\cref{cond:drift lemma for ratio 2}) since $X(t)$ is a constant.
    Hence, applying \cref{lem:drift lemma for ratio}, we obtain
    \begin{align*}
        \operatorname{Pr}{\left[\exists t\leq T: \frac{X(0)}{\pmax(t)}\geq \theta\right]}
        \leq \exp\qty(-\Omega\qty(\frac{\pmax(0)}{\frac{T}{n}+\frac{1}{n}}))
        = \exp\qty(-\Omega\qty(\frac{n\pmax(0)}{T})).
    \end{align*}
\end{proof}

\begin{proof}[Proof of \cref{lem:weak cannot be strong}]
    Let $c_{\mathrm{weak}}=1/2$.
    We apply \cref{lem:drift lemma for ratio} with
    $X(t)=p_j(t)$, $b=1$,
    $\theta=1-\frac{c_{\mathrm{weak}}}{2}$,
    $I=c_{\mathrm{weak}}/2$.
    We have $\frac{X(0)}{\pmax(0)} = \frac{p_j(0)}{\pmax(0)}\leq 1-c_{\mathrm{weak}}=\theta-I$.
    We have
    \begin{enumerate}
        \item $\frac{\expecton{t-1}{p_j(t)}}{\expecton{t-1}{p_{I_{t-1}}(t)}}\leq \frac{p_j(t-1)}{p_{I_{t-1}}(t-1)}=\frac{p_j(t-1)}{\pmax(t-1)}$ from \cref{cor:multinomial:weak-comparison},
        \item $p_j(t)-\expecton{t-1}{p_j(t)}$ satisfies $\qty(\frac{1}{n},\frac{\expecton{t-1}{p_{I_{t-1}}(t)}}{n})$ Bernstein condition from \cref{lem:Basic properties of sync process} (\cref{item:BC of sync process,item:variance of sync process 1}).
    \end{enumerate}
    Note that $\expecton{t-1}{p_j(t)}\leq \expecton{t-1}{p_{I_{t-1}}(t)}$ holds from \cref{cor:multinomial:weak-comparison}.
    Hence, applying \cref{lem:drift lemma for ratio},
    \begin{align*}
        \operatorname{Pr}{\left[\exists t\leq T: \frac{X(t)}{\pmax(t)}\geq \theta\right]}
        \leq \exp\qty(-\Omega\qty(\frac{\pmax(0)}{\frac{T}{n}+\frac{1}{n}}))
        = \exp\qty(-\Omega\qty(\frac{n\pmax(0)}{T})).
    \end{align*}
\end{proof}

\subsection{
\texorpdfstring{Emergence of the unique strong opinion (Proof of \cref{lem:emergence of the unique strong opinion})}{Emergence of the unique strong opinion}}
\label{sec:emergence of the unique strong opinion}

\begin{proof}[Proof of \cref{lem:emergence of the unique strong opinion}]
    Set $c_{\mathrm{weak}}=c_{\max}^\downarrow=1/2$ and define
    \begin{align*}
        &\tau_{\max}^\downarrow = \inf\{t\geq 0: \pmax(t)\leq (1-c_{\mathrm{max}}^\downarrow)\pmax(0)\}, \\
        &\tau_i^{\mathrm{weak}} = \inf\{t\geq 0: p_i(t)\leq (1-c_{\mathrm{weak}})\pmax(t)\} \text{ for } i\in [k].
    \end{align*}
    For a pair of distinct opinions $i$ and $j$, let $\tau_{i,j} = \min\{\tau_i^{\mathrm{weak}},\tau_j^{\mathrm{weak}},\tau_{\max}^\downarrow\}$.
    Furthermore, set $\rho=\min\{h^2\pmax(0),1\}/160$ and $a=n\rho/28$.
    From definition, $\pmax(t)\geq \pmax(0)/2$ for all $t<\tau_{i,j}$.
    Hence, $\rho \leq \rho_t=(1/80)\min\{h^2\pmax(t),1\}$ for all $t<\tau_{i,j}$.

    Let $r=\rho/420$, $X(t)=(1-r)^{-t}\Phi_a(\max\{G_i(t),G_j(t)\})$, and $Y(t)=X(t\wedge \tau_{i,j})$.
    Set $T=\lceil C'\log n/r\rceil$ for a sufficiently large constant $C'>0$.
    $(Y(t))_{t\in \bbN_0}$ is a supermartingale since
    \begin{align*}
        &\expecton{t-1}{Y(t)-Y(t-1)}\\
        &=\ind{\tau_{i,j}>t-1}\expecton{t-1}{X(t)-X(t-1)}\\
        &=\ind{\tau_{i,j}>t-1}(1-r)^{-t}\qty(\expecton{t-1}{\Phi_a(\max\{G_i(t),G_j(t)\})}-(1-r)\Phi_a(\max\{G_i(t-1),G_j(t-1)\}))\\
        &\leq 0.
    \end{align*}
    The last inequality follows from \cref{lem:drift for hmaj} (\cref{item:drift for a pair of non-weak opinions}).
    Since $\Phi_a(x)=\frac{1}{1+ax}\geq \frac{1}{1+a}$ for any $0\leq x\leq 1$,
    \begin{align*}
        \expect{Y(T)}
        &\geq \expect{\ind{\tau_{i,j}>T}Y(T)}\\
        &=\expect{\ind{\tau_{i,j}>T}(1-r)^{-T}
        \Phi_a(\max\{G_i(T),G_j(T)\})}\\
        &\geq \pr{\tau_{i,j}>T}\frac{(1-r)^{-T}}{1+a}.
    \end{align*}
    Hence, we obtain
    \begin{align}
        \pr{\tau_{i,j}>T}
        &\leq (1+a)(1-r)^{T}\expect{Y(T)} \nonumber\\
        &\leq (1+a)(1-r)^{T}\expect{Y(0)} \nonumber\\
        &\leq 2n\exp\qty(-rT)\leq n^{-20}.
        \label{eq:probability of the event two weak}
    \end{align}
    Note that we have
    \begin{align*}
        &\frac{1}{\rho} = \frac{160}{\min\{h^2\pmax(0),1\}},
    \end{align*}
    and
    \begin{align}
        \pmax(0) \geq \frac{(\log n)^2}{\rho n}
        &\Leftrightarrow \pmax(0) \geq 160\frac{(\log n)^2}{n} \max\qty{1,\frac{1}{h^2\pmax(0)}} \nonumber\\
        &\Leftrightarrow \pmax(0) \geq \max\qty{\frac{160(\log n)^2}{n},\frac{\sqrt{160}\log n}{h\sqrt{n}}}. \label{eq:equivalence of the conditions}
    \end{align}

    Now, consider the following events:
    \begin{enumerate}
        \item Let $\calS_i(t)$ be the event that $p_i(t)\geq (1-\frac{c_{\mathrm{weak}}}{2})\pmax(t)$.
        \item Let $\calA_{i,j}$ be the event that $\tau_{\max}^\downarrow > T$ and $\tau_i^{\mathrm{weak}}>T$ and $\tau_j^{\mathrm{weak}}>T$.
        \item Let $\calB_{i,j}$ be the event that $\tau_{\max}^\downarrow > T$ and $\{\tau_i^{\mathrm{weak}}\leq T \text{ or } \tau_j^{\mathrm{weak}}\leq T\}$ and $\calS_i(T)$ and $\calS_j(T)$.
    \end{enumerate}
    Then,
    \begin{align}
        &\pr*{\exists \{i,j\}\in \binom{[k]}{2}:\calS_i(T) \text{ and } \calS_j(T)} \nonumber\\
        &\leq \pr*{\exists \{i,j\}\in \binom{[k]}{2}:\tau_{\max}^\downarrow \leq  T \text{ or } \calA_{i,j} \text{ or } \calB_{i,j}} \nonumber \\
        &\leq \pr*{\tau_{\max}^\downarrow\leq T} \nonumber %
        \\
        &+ \pr*{\tau_{\max}^\downarrow>T \text{ and } \exists \{i,j\}\in \binom{[k]}{2}: \tau_i^{\mathrm{weak}}>T \text{ and } \tau_j^{\mathrm{weak}}>T} \nonumber%
        \\
        &+\pr*{\tau_{\max}^\downarrow>T \text{ and } \exists i\in [k]: \tau_i^{\mathrm{weak}}\leq T \text{ and } \calS_i(T)} \label{eq:probability of the event two weak 2}
        .
    \end{align}
    Now, we estimate each term on the right-hand side of \cref{eq:probability of the event two weak 2}.
    First, from \cref{lem:taumaxdown}, we have
    \begin{align*}
        \pr*{\tau_{\max}^\downarrow\leq T}
        \leq \exp\qty(-\Omega\qty(\frac{n\pmax(0)}{T}))
        \leq \exp\qty(-\Omega\qty(\frac{n\pmax(0)\rho}{\log n}))
        \leq n^{-15}.
    \end{align*}
    Second, applying the union bound and \cref{eq:probability of the event two weak}, we have
    \begin{align*}
        &\pr*{\tau_{\max}^\downarrow>T \text{ and } \exists \{i,j\}\in \binom{[k]}{2}: \tau_i^{\mathrm{weak}}>T \text{ and } \tau_j^{\mathrm{weak}}>T} \\
        &\leq \sum_{i,j\in [k]: i\neq j} \pr*{\tau_{\max}^\downarrow>T \text{ and } \tau_i^{\mathrm{weak}}>T \text{ and } \tau_j^{\mathrm{weak}}>T} \leq n^{-15}.
    \end{align*}
    Third,
    we have
    \begin{align}
        &\pr*{\tau_{\max}^\downarrow > T \text{ and } \tau_j^{\mathrm{weak}}\leq T \text{ and } \calS_j(T)}\nonumber\\
        &\leq \pr*{
            \tau_j^{\mathrm{weak}}\leq T
            \text{ and } \pmax(\tau_j^{\mathrm{weak}})>(1-c_{\max}^\downarrow)\pmax(0) \text{ and } \calS_j(T)}\nonumber\\
        &=\expect*{\ind{\tau_j^{\mathrm{weak}}\leq T
        \text{ and } \pmax(\tau_j^{\mathrm{weak}})>(1-c_{\max}^\downarrow)\pmax(0) } \pr{\calS_j(T)\mid \calF_{\tau_j^{\mathrm{weak}}}}}\nonumber\\
        &\leq \expect*{\ind{\tau_j^{\mathrm{weak}}\leq T
        \text{ and } \pmax(\tau_j^{\mathrm{weak}})>(1-c_{\max}^\downarrow)\pmax(0) } \exp\qty(-\Omega\qty(\frac{n\pmax(\tau_j^{\mathrm{weak}})}{T}))} \nonumber\\
        &\leq n^{-20}. \nonumber %
    \end{align}
    Here, we use \cref{lem:weak cannot be strong} in the last inequality.
    Taking the union bound, we obtain the claim.
\end{proof}

\subsection{\texorpdfstring{Proof of \cref{thm:consensus-time-intro}}{Proof of the upper bound on the consensus time}}
\label{sec:proof of the upper bound on the consensus time}
\begin{proof}[Proof of \cref{thm:proof of the upper bound on the consensus time}]
    Let $p_*=C_1\max\qty{\frac{(\log n)^2}{n},\frac{\log n}{h\sqrt{n}}}$ for a sufficiently large constant $C_1>0$.
    From \cref{eq:equivalence of the conditions},
    \begin{align*}
        \pmax(0)\geq p_* \Rightarrow
        \pmax(0)\geq \frac{C_2(\log n)^2}{n\min\{h^2\pmax(0),1\}}
        \Leftrightarrow \pmax(0)\geq \sqrt{C_2\frac{\pmax(0)}{\min\{h^2\pmax(0),1\}}\cdot\frac{(\log n)^2}{n}}
    \end{align*}
    for a sufficiently large constant $C_2>0$.

    First, consider the case where $\pmax(0)\geq p_*$.
    Then, we apply \cref{lem:emergence of the unique strong opinion}:
    For some $T_{\textrm{L}}=O\qty(\frac{\log n}{\min\{h^2\pmax(0),1\}})$, $p_{j}(T_{\textrm{L}})\leq \frac{3}{4}\pmax(T_{\textrm{L}}) \Leftrightarrow \pmax(T_{\textrm{L}})-p_j(T_{\textrm{L}})\geq \frac{\pmax(T_{\textrm{L}})}{4}$ holds for all $j\neq I_{T_\textrm{L}}$ w.h.p.
    Further, from \cref{lem:taumaxdown}, we have $\pmax(T_{\textrm{L}}) \geq \pmax(0)/2\geq p_*/2$ w.h.p.
    Thus, we can apply \cref{thm:plurality-consensus-intro} and the process reaches consensus within additional $O\qty(\frac{\log n}{\min\{h^2\pmax(T_{\textrm{L}}),1\}})$ rounds w.h.p.
    We obtain the claim from $\pmax(T_{\textrm{L}}) \geq \pmax(0)/2$.

    Next, consider the case where $\pmax(0)<p_*$.
    Then, set $T_{\textrm{S}}=\lceil C_3p_*n\log n\rceil$ for a sufficiently large constant $C_3>0$.
    Applying \cref{cor:majorization-largest-opinion}, $\pmax(T_{\textrm{S}})\geq p_*$ with high probability.
    Hence, starting from $T_{\textrm{S}}$, we can proceed in the same way as in the first case.
    The total time for this case is bounded by
    $O\qty(T_{\textrm{S}}+\frac{\log n}{\min\{h^2p_*,1\}}) = O\qty((\log n)^3+\frac{\sqrt{n}(\log n)^2}{h})$ and we obtain the claim.
\end{proof}

\begin{proof}[Proof of \cref{thm:consensus-time-intro}]
We claim that $\tconsensus \leq C\qty(1+\frac{C}{\min\{h^2\pmax(0),1\}})(\log n)^3$ w.h.p.~for any initial configuration.
If this is the case, then \cref{thm:consensus-time-intro} follows from \cref{thm:proof of the upper bound on the consensus time}.

Let $p_*=C\max\qty{\frac{(\log n)^2}{n},\frac{\log n}{h\sqrt{n}}}$ for a sufficiently large constant $C>0$.
If $\pmax(0)\geq p_*$, then the claim follows from \cref{thm:proof of the upper bound on the consensus time}.
We now consider the case where $\pmax(0)<p_*$.
In this case, $\pmax(0)\leq C\frac{(\log n)^2}{n} \Leftrightarrow \frac{\sqrt{n}}{h}\leq \frac{\sqrt{C}\log n}{h\sqrt{\pmax(0)}}$ or $\pmax(0)\leq C\frac{\log n}{h\sqrt{n}} \Leftrightarrow \frac{\sqrt{n}}{h}\leq \frac{C\log n}{h^2\pmax(0)}$ hold.
Further, since $\sqrt{x}\leq \max\{1,x\}$ holds for any $x\geq 0$, we have $\frac{1}{h\sqrt{\pmax(0)}}\leq \max\qty{1,\frac{1}{h^2\pmax(0)}}=\frac{1}{\min\{h^2\pmax(0),1\}}$.
Thus,
\begin{align*}
    \frac{\sqrt{n}}{h} \leq \max\qty{\frac{\sqrt{C}\log n}{h\sqrt{\pmax(0)}},\frac{C\log n}{h^2\pmax(0)}}
    \leq \frac{C\log n}{\min\{h^2\pmax(0),1\}},
\end{align*}
i.e., $C\qty(\log n +\frac{\sqrt{n}}{h})(\log n)^2 \leq C\qty(1+\frac{C}{\min\{h^2\pmax(0),1\}})(\log n)^3$ holds.
We obtain the claim from \cref{thm:proof of the upper bound on the consensus time}.
\end{proof}

\section{\texorpdfstring{Lower bound on neutral $h$-input dynamics}{Lower bound on neutral h-input dynamics}}
\label{sec:lowerbound on h-input dynamics}

We prove the consensus-time lower bound for rules that may use both the sampled opinions and the updating vertex's current opinion.

\subsection{Dynamics and statement}

In this section, we write $\mathcal{L}_X$ for the distribution of a random variable $X$.

We describe an update algorithm by a fixed distribution $\mathcal{L}_{\calW}$ over functions from $[k]\times[k]^h$ to $[k]$.
Each such function takes a vertex's current opinion and an ordered vector of $h$ sampled opinions and returns its new opinion.
We refer to the current opinion together with the $h$ sampled opinions as the $h+1$ inputs to the rule.
We write $\calW$ for a generic random function with this distribution.
The distribution $\mathcal{L}_{\calW}$ is the same for all vertices and rounds; it specifies the algorithm, including any internal randomness.
A deterministic update rule corresponds to a distribution concentrated on a single function.

We consider synchronous updates.
At the beginning of each round $t$, every vertex $v$ draws a fresh independent copy $\calW_v(t)$ of $\calW$.
Independently of this draw, it samples $h$ vertices uniformly with replacement and records their current opinions in the ordered vector $\bfZ^{(v)}(t)$.
The sampling is independent across vertices, and vertex $v$ updates to
\[
  C_v(t+1)=\calW_v(t)(C_v(t),\bfZ^{(v)}(t)).
\]
For a permutation $\sigma:[k]\to[k]$ and $\bfz\in[k]^h$, write $\sigma\bfz=(\sigma(z_1),\ldots,\sigma(z_h))$.

\begin{definition}[$h$-input rules and neutrality]
\label{def:h-input rule}
Fix integers $k\ge2$ and $h\ge1$.
An \emph{$h$-input rule} is a distribution $\mathcal{L}_{\mathcal{W}}$ of a random variable $\calW$ taking values in the set of functions from $[k]\times[k]^h$ to $[k]$ such that, for every $c\in[k]$ and $\bfz=(z_1,\ldots,z_h)\in[k]^h$, $\pr{\calW(c,\bfz)\in\{c,z_1,\ldots,z_h\}}=1$.
The rule is \emph{neutral} if, for every permutation $\sigma:[k]\to[k]$, every $c,i\in[k]$, and every $\bfz\in[k]^h$,
$\pr{\calW(c,\bfz)=i}=\pr{\calW(\sigma(c),\sigma\bfz)=\sigma(i)}$.
\end{definition}

The neutrality condition in \cref{def:h-input rule} concerns opinion labels, while input positions can be exploited by the specific update rule.
This includes rules that adopt the first repeated opinion in the sample sequence, as in $h$-\dejavu; see \cref{ex:extended-lb-dejavu}.
We also introduce the following parameter, which allows us to refine the lower bound for dynamics such as \dejavu.
\begin{definition}
    \label{def:ignore}
    For $\pignore\in[0,1]$, an $h$-input rule $\mathcal{L}_{\mathcal{W}}$ is \emph{$\pignore$-ignoring} if, for every input $(c,x_1,\ldots,x_{h})\in[k]^{h+1}$ with distinct entries, it holds that
    \begin{equation*}
        \pr{\calW(c,x_1,\ldots,x_{h})=c}\geq \pignore %
    \end{equation*}
\end{definition}
Intuitively, $\pignore$ is a lower bound on the probability that a vertex ignores the sampled opinions and retains its current opinion when all $h+1$ inputs are distinct.
Note that any $h$-input dynamics is $ 0$-ignoring.
\dejavu capped at $h$ samples, defined in \cref{ex:extended-lb-dejavu}, is $1$-ignoring.

\begin{definition}[$\calW$-dynamics]
\label{def:h-input dynamics}
Let $\mathcal{L}_{\mathcal{W}}$ be an $h$-input rule as in \cref{def:h-input rule}, and let $n\ge2$.
The \emph{$\calW$-dynamics} starts from a fixed configuration $\bfC(0)\in[k]^n$ and, in each round $t\ge0$, updates all vertices $v\in[n]$ simultaneously according to
$C_v(t+1)=\calW_v(t)(C_v(t),\bfZ^{(v)}(t))$.
Here, at the beginning of round $t$, conditional on $\bfC(0),\ldots,\bfC(t)$, the vertices independently choose pairs $(\calW_v(t),\bfZ^{(v)}(t))$.
For each vertex $v$, $\calW_v(t)$ has the same distribution as $\calW$, and, independently of $\calW_v(t)$, the vector $\bfZ^{(v)}(t)=(Z_1^{(v)}(t),\ldots,Z_h^{(v)}(t))$ is obtained by sampling $h$ vertices uniformly with replacement from $[n]$, possibly including $v$, and taking their opinions at round $t$.
\end{definition}

As in the \hmaj setting, we use $\bfC(t)$, $\bfp(t)$, $\pmax(t)$, and $\tconsensus$ for the configuration, opinion densities, maximum density, and consensus time.
We use the filtration $\calF_t$ and the notation $\pron{t}{\cdot}$,
$\expecton{t}{\cdot}$, $\varianceon{t}{\cdot}$, and $\bfq(t)$ from
\cref{sec:preliminaries}, now for this process.

Conditional on $\calF_t$, the sample vectors have distribution $\bfp(t)^{\otimes h}$ and are independent across vertices.
By \cref{def:h-input rule}, consensus is absorbing and an opinion absent initially can never appear.

We now establish a general lower bound on the consensus time of neutral $h$-input dynamics.
\begin{theorem}%
    \label{thm:general lower bound on h-input dynamics}
    Suppose $\pmax(0)<1$ and $h\geq 1$.
    For any neutral $h$-input dynamics, the consensus time is at least $T$ with probability at least $1-1/n$, where
    \begin{equation*}
      T=\Omega\left(
        1+\min\left\{
          \frac{1}{\min\{h^2\pmax(0),1\}},
          \frac{\sqrt n}{h\sqrt{\log n}}
        \right\}
      \right).
    \end{equation*}
    Moreover, if the dynamics is $\pignore$-ignoring, then the lower bound can be improved to
    \begin{align*}
      T=\Omega\left(
        1+\min\left\{
          \frac{1}{\min\{h^2\pmax(0),1\}},
          \frac{n}{h^2\log n},
          \frac{\sqrt n}{h\sqrt{(1-\pignore)\log n}}
        \right\}
      \right),
    \end{align*}
    where the last term is omitted when $\pignore=1$.
    In particular, for $\pignore=1$, one may take
    \begin{equation}
      T=\Omega\left(
        1+\min\left\{
          \frac{1}{\min\{h^2\pmax(0),1\}},
          \frac{n}{h^2\log n}
        \right\}
      \right).
      \label{eq:lower bound separated with ignoring 1}
    \end{equation}
\end{theorem}

\begin{example}[Majority and minority]
For $\bfz\in[k]^h$, let $\mathsf{Top}(\bfz)$ be the set of opinions with the highest frequency in $\bfz$.
Then $\pr{\calW(c,\bfz)=i}=\frac{\ind{i\in\mathsf{Top}(\bfz)}}{|\mathsf{Top}(\bfz)|}$ corresponds to the \hmaj protocol.
Similarly, let $\mathsf{Bot}(\bfz)$ be the set of opinions with the lowest frequency among those present in $\bfz$.
Then $\pr{\calW(c,\bfz)=i}=\frac{\ind{i\in\mathsf{Bot}(\bfz)}}{|\mathsf{Bot}(\bfz)|}$ corresponds to the $h$-\textsc{minority} protocol, studied for $k=2$ in~\cite{BecchettiCPTVZ24}.
Both are neutral $h$-input rules in the sense of \cref{def:h-input rule} for every number of opinions.
\end{example}

\begin{example}[\twochoices]
For $h=2$, the (deterministic) rule
\[
  \calW(c,(z_1,z_2))=
  \begin{cases}
    z_1,&z_1=z_2,\\
    c,&z_1\ne z_2
  \end{cases}
\]
is a neutral $2$-input rule, even when the retained opinion is absent from the sample.
Its $\calW$-dynamics is \twochoices~\cite{SS25}.
\end{example}

\begin{example}[\dejavu in \pullmodel{h}]
\label{ex:extended-lb-dejavu}
\dejavu in \pullmodel{h} works as follows: each node reads samples in order and stops at the first repeated opinion or after $h$ samples, whichever occurs first \cite{DDGGN26}.
In the standard \pullmodel{h}, the node must still sample all $h$ opinions, even if its next opinion is determined earlier.
In case of \emph{adaptive sampling}, the node may stop sampling as soon as it sees a duplicate opinion.
If $r\in\{2,\ldots,h\}$ is the smallest index for which $z_r\in\{z_1,\ldots,z_{r-1}\}$, the rule returns $z_r$; if no such index exists, it returns $c$.
We may generate the entire vector $\bfz=(z_1,\ldots,z_h)$ in advance and leave its unused suffix unread.
This representation has the same update distribution as sampling until the stopping condition is met, and it defines a neutral $h$-input rule in the sense of \cref{def:h-input rule}.
The case $h=2$ is exactly \twochoices.

\cref{thm:proof of the upper bound on the consensus time,thm:general lower bound on h-input dynamics} imply that, for $h\ge3$, when $\pmax\ll 1/(h\sqrt n)$, \hmaj can be asymptotically faster, with the following ratio between their convergence times holding with high probability:
\[
\frac{T_{\mathrm{DV}}}{T_{\mathrm{HM}}}
=\widetilde{\Omega}\left(\frac{1}{h\sqrt n\,\pmax}\right),
\]
where $T_{\mathrm{DV}}$ and $T_{\mathrm{HM}}$ denote the consensus times for \dejavu and \hmaj, respectively, starting from the same initial configuration.

For a dynamics with consensus time $T$, we define its sample complexity as the random variable \(S\) that is the average over all nodes of the total number of samples drawn by a node before consensus.
In the standard \pullmodel{h}, every node draws exactly $h$ samples in every round, so the sample complexity is exactly $hT$.
Consequently, with high probability, the sample complexities of \hmaj and \dejavu satisfy, respectively,
\[
  S_{\mathrm{HM}} =\widetilde{O}\left( h+\min\left\{\frac{1}{h\pmax},\sqrt n\right\} \right)
\]
and
\[
  S_{\mathrm{DV}} = \widetilde{\Omega}\left( h+\frac{1}{h\pmax} \right).
\]
In particular, in this regime, \hmaj uses at most
$\widetilde{O}(\sqrt n)$ samples per node, whereas \dejavu requires $\widetilde{\Omega}(1/(h\pmax))$ samples per node.
Thus, with high probability,
\[
  \frac{S_{\mathrm{DV}}}{S_{\mathrm{HM}}}   =\widetilde{\Omega}\left(\frac{1}{h\sqrt n\,\pmax}\right).
\]
This gives a polynomial separation whenever $\pmax$ is polynomially smaller than $1/(h\sqrt n)$.

In case of adaptive sampling, \dejavu can reduce its sample complexity.
Fix any constant $\varepsilon>0$.
Suppose that the initial configuration satisfies \(np_1=\omega(\log^2 n)\)
and, for a sufficiently large constant $\lambda>0$,
\[
  p_1-p_2 \ge \lambda\sqrt{\max\left\{\frac{1}{nh^2},\frac{p_1}{n}\right\}\log n}.
\]
Suppose further that
\[
  h=\Omega\left(\min\left\{\frac{n^{-1/4+\varepsilon}}{p_1},\frac{1}{\sqrt{p_1}}\right\}\right).
\]
Then, \cite[Theorem~3]{DDGGN26} gives, with high probability,
\[
  \frac{S_{\mathrm{HM}}^{\mathrm{adaptive}}}{S_{\mathrm{DV}}^{\mathrm{adaptive}}}=\widetilde{\Omega}\left(\max\{1,h\norm{\bfp}_2\}\right).
\]
Thus, when $h\norm{\bfp}_2$ is sufficiently large, \dejavu uses fewer samples than \hmaj, even though their convergence times are comparable up to logarithmic factors.

In the standard \pullmodel{h}, the expression $h+1/(hp_1)$ is minimized when $h^2p_1=\Theta(1)$.
The upper bound on the adaptive sample complexity of \dejavu in \cite{DDGGN26} is also minimized up to constant factors by this choice of $h$.

Applying \cref{thm:general lower bound on h-input dynamics}
(\cref{eq:lower bound separated with ignoring 1})
to \twochoices recovers the consensus-time lower bound in \cite{Archivio26}, whereas it establishes optimality (up to polylogarithmic factors) of the upper bound of \dejavu for plurality consensus under the assumptions of \cite{DDGGN26}.
\end{example}

\subsection{Proof}
We first bound the probability that a neutral rule selects a specified opinion from independent inputs.

\begin{lemma}[Selection probability]
\label{lem:upper bound on probability of winner opinion}
Let $\calW$ be a neutral $h$-input rule as in \cref{def:h-input rule}, and let $\bfp=(p_1,\ldots,p_k)$ be a probability vector.
Let $\bfZ=(Z_1,\ldots,Z_{h+1})\sim\bfp^{\otimes(h+1)}$ be independent of $\calW$.
Then, for every $i\in[k]$,
\[
  \pr{\calW(Z_1,(Z_2,\ldots,Z_{h+1}))=i}
  \le p_i\left(1+h(h-1)\norm{\bfp}_2^2
                   +\binom{h+1}{2}p_i\right).
\]
\end{lemma}

\begin{proof}
Set $m=h+1$ and write
$\calW(\bfz)=\calW(z_1,(z_2,\ldots,z_m))$.
Let $S_i$ be the set of vectors in $[k]^m$ containing exactly one $i$,
and let $\mathsf{Sing}(\bfy)$ count the opinions appearing exactly once
in a vector $\bfy$.

We first bound the probability of selecting $i$ when $\bfZ\in S_i$.
Let $\tau_{ij}$ swap labels $i$ and $j$, with $\tau_{ii}$ the identity.
On $S_i\cap S_j$, this map is a bijection preserving both
$\pr{\bfZ=\bfz}$ and $\mathsf{Sing}(\bfz)$, since $i$ and $j$ each
occur once.  Neutrality gives
$\pr{\calW(\bfz)=i}=\pr{\calW(\tau_{ij}\bfz)=j}$.
Consequently,
\begin{align*}
  \pr{\calW(\bfZ)=i,\ \bfZ\in S_i}
  &=\sum_{j=1}^k\sum_{\bfz\in S_i\cap S_j}
    \frac{\pr{\calW(\bfz)=i}\pr{\bfZ=\bfz}}
         {\mathsf{Sing}(\bfz)}\\*
  &=\sum_{j=1}^k\sum_{\bfz\in S_i\cap S_j}
    \frac{\pr{\calW(\tau_{ij}\bfz)=j}\pr{\bfZ=\tau_{ij}\bfz}}
         {\mathsf{Sing}(\tau_{ij}\bfz)}\\*
  &=\sum_{j=1}^k\sum_{\bfz\in S_i\cap S_j}
    \frac{\pr{\calW(\bfz)=j}\pr{\bfZ=\bfz}}
         {\mathsf{Sing}(\bfz)}\\*
  &\le\sum_{\bfz\in S_i}
    \frac{\pr{\bfZ=\bfz}}{\mathsf{Sing}(\bfz)}.
\end{align*}
The first equality uses
$\sum_j\ind{\bfz\in S_j}=\mathsf{Sing}(\bfz)$;
the third is the change of variables $\bfz\mapsto\tau_{ij}\bfz$.
The last inequality uses $\sum_j\pr{\calW(\bfz)=j}=1$.

Let $\bfU=(U_1,\ldots,U_h)\sim\bfp^{\otimes h}$ and write
$S=\mathsf{Sing}(\bfU)$.  Summing over the $m$ possible positions
of the unique $i$ gives
\begin{align*}
  \sum_{\bfz\in S_i}\frac{\pr{\bfZ=\bfz}}{\mathsf{Sing}(\bfz)}
  &=mp_i\sum_{\bfy\in([k]\setminus\{i\})^h}
    \frac{\pr{\bfU=\bfy}}{1+\mathsf{Sing}(\bfy)}\\*
  &\le mp_i\sum_{\bfy\in[k]^h}
    \frac{\pr{\bfU=\bfy}}{1+\mathsf{Sing}(\bfy)}\\*
  &=p_i\expect*{\frac{m}{1+S}}
  \le p_i\bigl(m-\expect{S}\bigr).
\end{align*}
Here $0\le S\le h=m-1$, and $m/(1+s)\le m-s$ for such values of \(S\)
because $(m-s)(1+s)-m=s(m-1-s)\ge0$.

To bound $\expect{S}$, let
$N_j=\sum_{r=1}^h\ind{U_r=j}$ be the number of occurrences
of opinion $j$.  Since $\sum_jN_j=h$ and each opinion with $N_j=1$
accounts for exactly one entry, $h-S$ counts the entries whose
opinion appears at least twice.  Thus
\begin{align*}
  h-S
  &=\sum_{j:N_j\ge2}N_j\\*
  &\le\sum_{j:N_j\ge2}N_j(N_j-1)\\*
  &=2\sum_{j=1}^k\binom{N_j}{2}\\*
  &=2\sum_{1\le r<s\le h}\ind{U_r=U_s}.
\end{align*}
The inequality uses $N_j-1\ge1$ when $N_j\ge2$.
The last equality counts pairs of positions with equal opinions:
opinion $j$ occurs at $N_j$ different positions and therefore gives
$\binom{N_j}{2}$ such pairs.
Taking expectations and using
$\pr{U_r=U_s}=\sum_jp_j^2=\norm{\bfp}_2^2$ yields
\[
  \expect{S}\ge h-2\binom h2\norm{\bfp}_2^2
  =h-h(h-1)\norm{\bfp}_2^2.
\]
This also holds for $h=1$, when the sum over pairs is empty.  Thus
\begin{equation}
  \label{eq:selection probability singleton bound}
  \pr{\calW(\bfZ)=i,\ \bfZ\in S_i}
  \le p_i\left(1+h(h-1)\norm{\bfp}_2^2\right).
\end{equation}

By \cref{def:h-input rule}, the rule can select $i$ only if $i$
occurs among its inputs.  Therefore, if $\calW(\bfZ)=i$ and
$\bfZ\notin S_i$, at least two entries of $\bfZ$ equal $i$.
A union bound gives
\begin{equation}
  \label{eq:selection probability multiple-occurrence bound}
  \pr{\calW(\bfZ)=i,\ \bfZ\notin S_i}
  \le\sum_{1\le r<s\le m}\pr{Z_r=Z_s=i}
  =\binom m2p_i^2.
\end{equation}
Adding \cref{eq:selection probability singleton bound,eq:selection probability multiple-occurrence bound}
and substituting $m=h+1$ proves the claim.
\end{proof}

Then, we bound the one-round expected growth and variance of any opinion density.

\begin{lemma}
  \label{lem:expectation and variance of h input dynamics}
  For any neutral $h$-input dynamics that is $\pignore$-ignoring, the following holds.
  Fix $t\geq 1$.
  Conditional on $\calF_{t-1}$, for every $i\in [k]$, we have
  \begin{enumerate}
      \item \label{item:expectation of h input dynamics} $\expecton{t-1}{p_i(t)} \leq p_i(t-1)+2h^2\pmax(t-1)^2$.
      \item \label{item:variance of h input dynamics}
      $\varianceon{t-1}{p_i(t)}\leq \frac{\qty(2(1-\pignore)+5h^2\pmax(t-1))\pmax(t-1)}{n}$.
  \end{enumerate}
\end{lemma}
\begin{proof}
Throughout the proof, we condition on $\calF_{t-1}$ and omit the
round index $t-1$ from the density notation: we write
$p_j=p_j(t-1)$ for every $j\in[k]$, $\bfp=\bfp(t-1)$,
and $\pmax=\pmax(t-1)$.
In particular,
\[
  p_i\le \pmax,
  \qquad \norm{\bfp}_2^2=\sum_jp_j^2
  \le \pmax\sum_jp_j=\pmax.
\]

For \cref{lem:expectation and variance of h input dynamics}(\cref{item:expectation of h input dynamics}), choose a vertex $V$
uniformly from $[n]$, independently of the dynamics.
Its current opinion has law $\bfp$, and conditional on any choice
of $V$, its $h$ sampled opinions have law $\bfp^{\otimes h}$,
independently of its update rule.
Thus the entire input vector has law $\bfp^{\otimes(h+1)}$ and is
independent of the rule \(\calW\).  Averaging over $V$ and applying
\cref{lem:upper bound on probability of winner opinion}, we obtain
\begin{align*}
  \expecton{t-1}{p_i(t)}
  &=\frac1n\sum_{v=1}^n\pron{t-1}{C_v(t)=i}
  =\pron{t-1}{C_V(t)=i}\\*
  &\le p_i\left(1+h(h-1)\norm{\bfp}_2^2
    +\binom{h+1}{2}p_i\right)\\*
  &\le p_i+\left(h(h-1)+\binom{h+1}{2}\right)\pmax^2
  \le p_i+2h^2\pmax^2.
\end{align*}
The last inequality uses
$h(h-1)+\binom{h+1}{2}=(3h^2-h)/2\le2h^2$.

For \cref{lem:expectation and variance of h input dynamics}(\cref{item:variance of h input dynamics}), let
$\bfX=(X_1,\ldots,X_h)\sim\bfp^{\otimes h}$ independently of
$\calW$, and set
\[
  a_i=\pr{\calW(i,\bfX)\ne i}.
\]
This is the conditional probability that a vertex currently holding
$i$ leaves that opinion.  There are $np_i$ such vertices, so
\cref{lem:Basic properties of sync process}(\cref{item:variance of sync process 2})
and \cref{lem:expectation and variance of h input dynamics}(\cref{item:expectation of h input dynamics}) give
\begin{align}
  n\varianceon{t-1}{p_i(t)}
  &\le\expecton{t-1}{p_i(t)}-p_i
    +\frac2n\sum_{\substack{v\in[n]\\C_v(t-1)=i}}a_i\nonumber\\
  &\le2h^2\pmax^2+2p_i a_i.
  \label{eq:variance of h input dynamics 1}
\end{align}

Let $D$ be the event that $i,X_1,\ldots,X_h$ are pairwise distinct.
By \cref{def:ignore} and independence of the rule and the samples,
\[
  \pr{\calW(i,\bfX)\ne i,\ D}
  \le(1-\pignore)\pr{D}
  \le1-\pignore.
\]
A collision occurs precisely when some $X_r=i$ or some
$X_r=X_s$ with $r<s$.  Hence
\begin{align*}
  \pr{D^{\mathsf c}}
  &\le\sum_{r=1}^h\pr{X_r=i}
    +\sum_{1\le r<s\le h}\pr{X_r=X_s}\\*
  &=hp_i+\binom h2\sum_jp_j^2
  =hp_i+\binom h2\norm{\bfp}_2^2.
\end{align*}
Splitting according to $D$ and $D^{\mathsf c}$ therefore gives
\begin{equation}
  \label{eq:variance of h input dynamics 2}
  a_i\le1-\pignore+hp_i+\binom h2\norm{\bfp}_2^2.
\end{equation}
Combining \cref{eq:variance of h input dynamics 1,eq:variance of h input dynamics 2}
and using $p_i,\norm{\bfp}_2^2\le \pmax$, we obtain
\begin{align*}
  n\varianceon{t-1}{p_i(t)}
  &\le2h^2\pmax^2+
    2p_i\left(1-\pignore+hp_i+
      \binom h2\norm{\bfp}_2^2\right)\\*
  &\le2(1-\pignore)\pmax+
    \left(2h^2+2h+h(h-1)\right)\pmax^2\\*
  &\le\left(2(1-\pignore)+5h^2\pmax\right)\pmax.
\end{align*}
Here $2h^2+2h+h(h-1)=3h^2+h\le5h^2$ for $h\ge1$.
Dividing by $n$ proves the variance bound.
\end{proof}

We are now ready to complete the proof of \cref{thm:general lower bound on h-input dynamics}.
\begin{proof}[Proof of \cref{thm:general lower bound on h-input dynamics}]
We prove the parameterized bound for $\pignore$-ignoring dynamics with
\begin{equation}
  T=\left\lceil
  \frac{1}{32\min\left\{
    h^2\pmax(0)+\frac{32h^2\log n}{n}
    +h\sqrt{\frac{32(1-\pignore)\log n}{n}},1
  \right\}}
  \right\rceil
  \label{eq:lower bound on consensus time with ignoring}
\end{equation}
Write $P_{\mathrm{comply}}=1-\pignore$ and set
\[
  L=\pmax(0)+\frac{32\log n}{n}
    +\frac1h\sqrt{\frac{32P_{\mathrm{comply}}\log n}{n}}.
\]
Then the time in the theorem is
$T=\left\lceil\frac{1}{32\min\{h^2L,1\}}\right\rceil$.
If $32h^2L\ge1$, then $T=1$, and the claim follows from
$\pmax(0)<1$.  Henceforth assume $32h^2L<1$, so that
\begin{equation}
  \label{eq:extended lower bound on T}
  T\le\frac1{32h^2L}+1\le\frac1{16h^2L},
  \qquad 2L<1.
\end{equation}

Let $A=\{i\in[k]:p_i(0)>0\}$.  By \cref{def:h-input rule},
absent opinions cannot appear, and $\abs{A}\le n$.
Define
\[
  \tau=\inf\{t\ge0:\pmax(t)\ge2L\}.
\]
Since $\pmax(0)\le L$, we have $\tau\ge1$.
For each $i\in A$, define a stopped martingale by
\[
  Y^{(i)}_t=\sum_{s=1}^t\ind{\tau>s-1}
    \left(p_i(s)-\expecton{s-1}{p_i(s)}\right),
  \qquad Y^{(i)}_0=0.
\]
To check that it is a martingale, note that, for every $s\ge1$,
\[
  Y^{(i)}_s-Y^{(i)}_{s-1}
  =\ind{\tau>s-1}
    \left(p_i(s)-\expecton{s-1}{p_i(s)}\right).
\]
This expression has conditional mean zero given $\calF_{s-1}$ because
$\{\tau>s-1\}\in\calF_{s-1}$.
Using
\cref{lem:expectation and variance of h input dynamics}(\cref{item:expectation of h input dynamics}) and the telescoping sum gives
\begin{align}
  p_i(T\wedge\tau)-p_i(0)
  &=Y^{(i)}_T+\sum_{s=1}^T\ind{\tau>s-1}
    \left(\expecton{s-1}{p_i(s)}-p_i(s-1)\right)\nonumber\\
  &\le Y^{(i)}_T+\sum_{s=1}^T\ind{\tau>s-1}
    2h^2\pmax(s-1)^2\nonumber\\
  &\le Y^{(i)}_T+8h^2L^2T
  \le Y^{(i)}_T+\frac L2.
  \label{eq:extended lower bound stopped drift}
\end{align}
Here in the last two inequalities we have used that $\pmax(s-1)<2L$ on $\{\tau>s-1\}$ and \cref{eq:extended lower bound on T}.

By \cref{lem:Basic properties of sync process}(\cref{item:BC of sync process}) and \cref{lem:expectation and variance of h input dynamics}(\cref{item:variance of h input dynamics}),
conditional on $\calF_{s-1}$, $Y^{(i)}_s-Y^{(i)}_{s-1}$ satisfies
the $(1/n,s_*)$-Bernstein condition, where
\[
  s_*=\frac{4P_{\mathrm{comply}}L+20h^2L^2}{n}.
\]
Indeed, on $\{\tau>s-1\}$, $Y^{(i)}_s-Y^{(i)}_{s-1}$ satisfies the
$(1/n,s_*)$-Bernstein condition because
\[
  \frac{\left(2P_{\mathrm{comply}}+
    5h^2\pmax(s-1)\right)\pmax(s-1)}{n}
  \le\frac{(2P_{\mathrm{comply}}+10h^2L)2L}{n}=s_*;
\]
on $\{\tau\le s-1\}$, $Y^{(i)}_s-Y^{(i)}_{s-1}=0$.
Thus, in \cref{cor:Freedman moment}, we may take $s_{s-1}=s_*$
for every $s\in[T]$.  For every $t\le T$, this gives
\[
  \sum_{s=1}^t s_{s-1}=t s_*\le T s_*,
\]
so we may choose $W=T s_*$ in \cref{cor:Freedman moment}.

We first bound the denominator $W+(hD)/3$ in the exponent of
\cref{cor:Freedman moment}.  Substituting $h=L/2$, $D=1/n$,
and $W=Ts_*$ into that corollary, this denominator becomes
$Ts_*+L/(6n)$.  Here $h$ denotes the parameter of the corollary,
not the number of sampled opinions.
The definition of $L$ gives
\[
  \frac{P_{\mathrm{comply}}}{h^2}\le
    \frac{nL^2}{32\log n},
  \qquad
  L\le\frac{nL^2}{32\log n}.
\]
Together with \cref{eq:extended lower bound on T}, these imply
\begin{align*}
  Ts_*+\frac{L}{6n}
  &\le\frac1n\left(
    \frac{P_{\mathrm{comply}}}{4h^2}
    +\frac{5L}{4}+\frac L6\right)\\*
  &=\frac1n\left(
    \frac{P_{\mathrm{comply}}}{4h^2}+\frac{17L}{12}\right)\\*
  &\le\left(\frac14+\frac{17}{12}\right)
    \frac{L^2}{32\log n}
  =\frac{5L^2}{96\log n}
  \le\frac{L^2}{16\log n}.
\end{align*}
Apply \cref{cor:Freedman moment} with $h=L/2$,
$D=1/n$, and $W=Ts_*$.  Since the event at
round $T$ is contained in the corresponding crossing event, we obtain
\begin{equation}
  \label{eq:extended lower bound Freedman tail}
  \pr*{Y^{(i)}_T\ge\frac L2}
  \le\exp\left(-\frac{L^2/8}{Ts_*+L/(6n)}\right)
  \le\exp(-2\log n)=n^{-2}.
\end{equation}

If $\tau\le T$, some $i\in A$ satisfies
$p_i(T\wedge\tau)=p_i(\tau)\ge2L$.  Since $p_i(0)\le L$,
\cref{eq:extended lower bound stopped drift} then gives
$Y^{(i)}_T\ge L/2$.  A union bound and
\cref{eq:extended lower bound Freedman tail} therefore yield
\[
  \pr{\tau\le T}
  \le\sum_{i\in A}\pr*{Y^{(i)}_T\ge\frac L2}
  \le\abs{A}n^{-2}\le\frac1n.
\]
On $\{\tau>T\}$, we have $\pmax(t)<2L<1$ for every
$0\le t\le T$, so consensus has not occurred by round $T$.
Consequently,
\[
  \pr{\tconsensus\ge T}\ge\pr{\tau>T}\ge1-\frac1n,
\]
as claimed.

It remains to express \eqref{eq:lower bound on consensus time with ignoring}
in the form stated in the theorem. Set
\[
  a=h^2\pmax(0),\qquad
  b=\frac{32h^2\log n}{n},\qquad
  c=h\sqrt{\frac{32(1-\pignore)\log n}{n}}.
\]
Writing $M=\max\{\min\{a,1\},\min\{b,1\},\min\{c,1\}\}$, we have
\begin{align}
  M
  &\le \min\{a+b+c,1\}\nonumber\\
  &\le \min\{a,1\}+\min\{b,1\}+\min\{c,1\}\nonumber\\
  &\le 3M.
  \label{eq:truncated sum comparison}
\end{align}
Taking reciprocals in \eqref{eq:truncated sum comparison} and using
$\lceil u\rceil=\Theta(1+u)$ for $u>0$, the choice of $T$ in
\eqref{eq:lower bound on consensus time with ignoring} satisfies
\begin{equation}
  \label{eq:lower bound reciprocal scales}
  T=\Theta\left(1+\frac1M\right)
   =\Theta\left(1+\min\left\{
       \frac1{\min\{a,1\}},\frac1b,\frac1c
     \right\}\right).
\end{equation}
Here a reciprocal with zero denominator is omitted; removing the truncations
from $b$ and $c$ changes the expression by at most a constant factor because
of the additive $1$. Substituting $a,b,c$ into
\eqref{eq:lower bound reciprocal scales} proves the parameterized bound in
\cref{thm:general lower bound on h-input dynamics}. When
$\pignore=1$, we have $c=0$, so the same substitution gives
\eqref{eq:lower bound separated with ignoring 1} directly.

Every neutral $h$-input dynamics is $0$-ignoring. In this case, let
\[
  x=\sqrt{\frac{32h^2\log n}{n}},
\]
so that $b=x^2$ and $c=x$. Since $x^2\le x$ when $x\le1$, while
both terms are absorbed by the truncation at $1$ when $x\ge1$, we have
\begin{equation}
  \label{eq:absorb quadratic fluctuation term}
  \min\{a+x+x^2,1\}=\Theta\bigl(\min\{a+x,1\}\bigr).
\end{equation}
Consequently, setting $\pignore=0$ in
\eqref{eq:lower bound on consensus time with ignoring} and applying
\eqref{eq:absorb quadratic fluctuation term}, followed by
\eqref{eq:truncated sum comparison} with the three terms $a,x,0$, gives
\begin{align*}
  T=\Omega\left(
    1+\min\left\{
      \frac{1}{\min\{h^2\pmax(0),1\}},
      \frac{\sqrt n}{h\sqrt{\log n}}
    \right\}
  \right).
\end{align*}
This is the first bound in the theorem.

\end{proof}
 \newpage
\section*{Acknowledgements}
The authors are grateful to Niccolò D’Archivio for facilitating the collaboration that led to this publication. %
\section*{AI methodology}
We used GPT-5.4, GPT-5.5, GPT-5.6, and GPT-6 to assist with proofs and exposition.
The key ideas originated with the authors.
The tool materially affected the proofs in \cref{sec:expected_growth,sec:plurality-consensus,sec:symmetry_breaking,sec:lowerbound on h-input dynamics}.

For the multinomial analysis (\cref{sec:expected_growth}), we identified three regimes and attempted to formulate transport lemmas inspired by~\cite{h-Majority-DISC25}.
Given these regimes and our preliminary attempts, GPT-5.4 and later GPT-5.5 produced suitable transport lemmas and proofs by May 2026;
earlier attempts without this guidance had been unsuccessful.
We subsequently identified how to merge two of these regimes, further simplifying the analysis.

For the plurality consensus section (\cref{sec:plurality-consensus}), we proposed applying Freedman's inequality~\cite{Freedman1975} to the multinomial estimates.
GPT-5.5 produced a complicated proof with several phases, giving a first draft proof in June.
Through subsequent iterations with GPT-5.6 and GPT-6, we substantially simplified and improved the argument to a single amplification phase.

For the symmetry breaking section (\cref{sec:symmetry_breaking}), we also used GPT-6 to simplify proofs, including the design of potential functions and expectation calculations, and to help make a symmetry argument rigorous.
The authors had independently developed a proof of the symmetry breaking result (\cref{lem:emergence of the unique strong opinion}) before discussions with AI led to the simpler potential-based argument presented in \cref{lem:drift for hmaj}.

As for the lower bound section (\cref{sec:lowerbound on h-input dynamics}), the authors developed a weaker version of \cref{thm:neutral-lower-bound-intro} which held only for dynamics where nodes adopt an opinion that is actually sampled.
GPT-6 proposed a simple extension to dynamics where a node can retain its own opinion, which we further generalized by introducing the parameter \(\pignore\).

The authors verified the correctness and originality of all content, including references.
This manuscript has been written by humans, for humans.
 \begingroup
\emergencystretch=1em
\printbibliography
\endgroup
\appendix
\section{Tools} \label{sec:tools}

\begin{definition}[Bernstein condition (Definition 3.3 of \cite{SS25})]
    \label{def:bernstein_condition}
    Let $D$ and $s$ be non-negative parameters.
    A random variable $X$ satisfies $(D,s)$-Bernstein condition if for all $\lambda \in \bbR$ with $\abs{\lambda}D<3$,
    \[
    \expect{\mathrm{e}^{\lambda X}}\leq \exp\qty(\frac{\lambda^2s/2}{1-(\abs{\lambda}D)/3}).
    \]
    It satisfies one-sided $(D,s)$-Bernstein condition if the same bound holds for all $\lambda\geq 0$ with $\lambda D<3$.
\end{definition}

\begin{lemma}[Basic properties of Bernstein condition (Lemma 3.4 of \cite{SS25} and Lemma 3.8 of \cite{USD_PODC26})]
    \label{lem:basic properties of bernstein condition}
    Let $X$ and $Y$ be random variables.
    We have the following:
    \begin{enumerate}
    \item If $\expect{X}=0$ and $\abs{X}\leq D$, then $X$ satisfies $(D,\variance{X})$-Bernstein condition.
    \item \label{item:BS upper bounded} If $X$ satisfies $(D,s)$-Bernstein condition, then $X$ satisfies $(D',s')$-Bernstein condition for any $D'\geq D$ and $s'\geq s$ (same for the one-sided version).
    \item \label{item:BS multiplied} If $X$ satisfies $(D,s)$-Bernstein condition, then $aX$ satisfies $(\abs{a}D,a^2s)$-Bernstein condition for any $a\in \bbR$ (same for the one-sided version if $a\geq 0$).
    \item If $X$ satisfies one-sided $(D,s)$-Bernstein condition and $Y\leq X$ almost surely, then $Y$ satisfies one-sided $(D,s)$-Bernstein condition.
    \item \label{item:BC for sum of independent random variables} If $X_1,\ldots,X_n$ are independent random variables such that each $X_i$ satisfies $(D,s_i)$-Bernstein condition, then $\sum_{i=1}^n X_i$ satisfies $(D,\sum_{i=1}^ns_i)$-Bernstein condition (same for the one-sided version).
    \item \label{item:BS sum} If $X_1,\ldots,X_n$ are random variables such that each $X_i$ satisfies $(D,s_i)$-Bernstein condition, then $\sum_{i=1}^n X_i$ satisfies $(nD,n\sum_{i=1}^ns_i)$-Bernstein condition (same for the one-sided version).
    \end{enumerate}
\end{lemma}

We use the notation of \cref{sec:preliminaries} for a general opinion process.
\begin{lemma}
    \label{lem:Basic properties of sync process}
    Suppose that, for every $t\geq 0$, the random variables $(C_v(t+1))_{v\in[n]}$ are conditionally independent given $\calF_t$.
    Then, for every $t\geq 0$ and $i\in[k]$, the following hold conditional on $\calF_t$:
    \begin{enumerate}
        \item \label{item:expectation of sync process} If $\pron{t}{C_v(t+1)=i}$ does not depend on $v\in [n]$, then $\expecton{t}{p_i(t+1)}=\pron{t}{C_v(t+1)=i}$.
        \item \label{item:variance of sync process 1} $\varianceon{t}{p_i(t+1)}\leq \frac{\expecton{t}{p_i(t+1)}}{n}$.
        \item \label{item:variance of sync process 2} $\varianceon{t}{p_i(t+1)}\leq \frac{1}{n}\qty(\expecton{t}{p_i(t+1)}-p_i(t)+\frac{2}{n}\sum_{v\in[n]:C_v(t)=i}\pron{t}{C_v(t+1)\neq i})$.
        \item \label{item:BC of sync process} $p_i(t+1)-\expecton{t}{p_i(t+1)}$ satisfies $\qty(\frac{1}{n},\varianceon{t}{p_i(t+1)})$-Bernstein condition.
    \end{enumerate}
\end{lemma}
\begin{proof}[Proof of \cref{lem:Basic properties of sync process}(\cref{item:expectation of sync process})]
    Since $\pron{t}{C_v(t+1)=i}$ does not depend on $v\in [n]$, we obviously obtain
    \begin{align*}
        \expecton{t}{p_i(t+1)}
        =\frac{1}{n}\sum_{v\in[n]}\expecton{t}{\ind{C_v(t+1)=i}}
        =\frac{1}{n}\sum_{v\in[n]}\pron{t}{C_v(t+1)=i}
        =\pron{t}{C_v(t+1)=i}.
    \end{align*}
\end{proof}
\begin{proof}[Proof of \cref{lem:Basic properties of sync process}(\cref{item:variance of sync process 1})]
    From the independence of $(C_v(t+1))_{v\in[n]}$, we have
    \begin{align*}
        \varianceon{t}{p_i(t+1)}
        =\frac{1}{n^2}\sum_{v\in[n]}\varianceon{t}{\ind{C_v(t+1)=i}}
        \leq \frac{1}{n^2}\sum_{v\in[n]}\expecton*{t}{\ind{C_v(t+1)=i}^2}
        =\frac{\expecton{t}{p_i(t+1)}}{n}.
    \end{align*}
    Note that $\expecton{t}{p_i(t+1)}=\frac{1}{n}\sum_{v\in[n]}\expecton{t}{\ind{C_v(t+1)=i}}$.
\end{proof}
\begin{proof}[Proof of \cref{lem:Basic properties of sync process}(\cref{item:variance of sync process 2})]
    We have
    \begin{align*}
        \varianceon{t}{p_i(t+1)}
        &=\varianceon{t}{p_i(t+1)-p_i(t)}\\
        &=\frac{1}{n^2}\sum_{v\in [n]}\varianceon*{t}{\ind{C_v(t+1)=i}-\ind{C_v(t)=i}}\\
        &\leq \frac{1}{n^2}\sum_{v\in [n]}\expecton*{t}{\qty(\ind{C_v(t+1)=i}-\ind{C_v(t)=i})^2}\\
        &=\frac{1}{n^2}\sum_{v\in [n]}\expecton*{t}{\qty(\ind{C_v(t+1)=i}-\ind{C_v(t)=i})+2\ind{C_v(t)=i}\qty(1-\ind{C_v(t+1)=i})}\\
        &=\frac{1}{n}\qty(\expecton{t}{p_i(t+1)}-p_i(t) + \frac{2}{n}\sum_{v\in [n]:C_v(t)=i}\pron{t}{C_v(t+1)\neq i}).
    \end{align*}
    Note that $(a-b)^2=a-b+2b(1-a)$ holds for any $a, b\in \{0,1\}$.
\end{proof}
\begin{proof}[Proof of \cref{lem:Basic properties of sync process}(\cref{item:BC of sync process})]
    Write $Y_v=\frac{\ind{C_v(t+1)=i}-\expecton{t}{\ind{C_v(t+1)=i}}}{n}$ for convenience.
    Then, $p_i(t+1)-\expecton{t}{p_i(t+1)}=\sum_{v\in[n]}Y_v$.
    Note that $(Y_v)_{v\in[n]}$ are independent random variables.
    Applying \cref{lem:basic properties of bernstein condition} (\cref{item:BC for sum of independent random variables}) shows that $p_i(t+1)-\expecton{t}{p_i(t+1)}$ satisfies $\qty(\frac{1}{n},\sum_{v\in[n]}\varianceon{t}{Y_v})$-Bernstein condition.
    Since
    \[\sum_{v\in[n]}\varianceon{t}{Y_v}
    =\varianceon*{t}{\sum_{v\in[n]}Y_v}
    =\varianceon*{t}{p_i(t+1)-\expecton{t}{p_i(t+1)}}
    =\varianceon{t}{p_i(t+1)},\]
    we obtain the claim.
\end{proof}

   \begin{theorem}[Theorem 2.6 of \cite{FGL15}]
    \label{thm:Freedman moment}
    Let $(X_t)_{t\in \bbN_0}$ be a real-valued supermartingale associated with a filtration $(\calF_t)_{t\in \bbN_0}$.
    Assume that $V_{t-1}$, $t\in [T]$ are positive and $\calF_{t-1}$-measurable random variables.
    Suppose that $\expecton{t-1}{\exp(\lambda(X_t-X_{t-1}))} \leq \exp(f(\lambda) V_{t-1})$ holds for all $t\in [T]$ and for a positive function $f$ for some $\lambda\in (0,\infty)$.
    Then, for all $h,W>0$,
    \begin{align*}
        \pr*{\exists t\leq T, X_t-X_0\geq h \text{ and }\sum_{i=1}^tV_{i-1} \leq W} \leq \exp\qty(-\lambda h + f(\lambda)W).
    \end{align*}
    \end{theorem}
    \begin{corollary}
        \label{cor:Freedman moment}
        Let $(X_t)_{t\in \bbN_0}$ be a real-valued supermartingale associated with a filtration $(\calF_t)_{t\in \bbN_0}$.
        Suppose that, for every $t\geq 1$, conditional on
        $\calF_{t-1}$, the difference $X_t-X_{t-1}$ satisfies the
        one-sided $(D,s_{t-1})$-Bernstein condition.
        Then, for any $h,W>0$,
        \begin{align*}
            \pr*{\exists t\leq T, X_t-X_0\geq h \text{ and }\sum_{i=1}^t s_{i-1} \leq W} \leq \exp\qty(-\frac{h^2/2}{W+(hD)/3}).
        \end{align*}
    \end{corollary}
    \begin{proof}
        We apply \cref{thm:Freedman moment} with the following parameters: $\lambda=\frac{h}{W+(hD)/3}$, $f(\lambda)=\frac{\lambda^2/2}{1-(\lambda D)/3}$, $V_t=s_t$.
        Note that $\expecton{t-1}{\exp(\lambda(X_t-X_{t-1}))} \leq \exp(f(\lambda) s_{t-1})$ holds from the assumption that $X_t-X_{t-1}$ satisfies one-sided $(D,s_{t-1})$-Bernstein condition.
        Then, we obtain
        \begin{align*}
            \pr*{\exists t\leq T, X_t-X_0\geq h \text{ and }\sum_{i=1}^t s_{i-1} \leq W}
            &\leq \exp\qty(-\lambda h + \frac{\lambda^2/2}{1-(\lambda D)/3}W)\\
            &=\exp\qty(-\frac{h^2/2}{W+(hD)/3}).
        \end{align*}
    \end{proof}

We obtain the following two-sided form of Freedman's inequality~\cite{Freedman1975}.

\begin{corollary}[Freedman's inequality]
\label{lem:tool:freedman}
Let \((M_t)_{t=0}^N\) be a martingale with respect to a filtration
\((\calG_t)_{t=0}^N\), and assume \(M_0=0\).
Let \(\xi_t=M_t-M_{t-1}\) for \(1\le t\le N\).
Assume \(\abs{\xi_t}\le B\) almost surely for every \(1\le t\le N\).
Let
\begin{align*}
  V_t=\sum_{s=1}^t\expect*{\xi_s^2\mid\calG_{s-1}}.
\end{align*}
Then, for every \(r>0\) and every \(V>0\),
\begin{align*}
  \pr*{\exists t\le N: |M_t| \ge r \text{ and }V_t\le V}
  \le
  2\exp\left(-\frac{r^2}{2(V+Br/3)}\right).
\end{align*}
\end{corollary}
\begin{proof}
Since $M_t$ is a martingale, $\expect*{\xi_t\mid\calG_{t-1}}=0$.
By the first item of \cref{lem:basic properties of bernstein condition}, conditional on $\calG_{t-1}$, both $\xi_t$ and $-\xi_t$ satisfy the one-sided $(B,s_{t-1})$-Bernstein condition, where
\[
  s_{t-1}=\expect*{\xi_t^2\mid\calG_{t-1}}.
\]
Extend \(M_t\) to be constant after time \(N\).
Apply \cref{cor:Freedman moment} to $M_t$ and $-M_t$ with $D=B$, $h=r$, $W=V$, and $T=N$.
For both signs, $\sum_{i=1}^t s_{i-1}=V_t$.
A union bound over the two signs gives the result.
\end{proof}

\end{document}